\documentclass[11pt,letterpaper]{article}

\usepackage{jheppub}

\usepackage{amsfonts,amssymb,amsmath,latexsym,graphics,graphicx,epsfig,multirow,comment,feyn,slashed,color,xcolor,afterpage,makecell,appendix}
\usepackage{booktabs}
\usepackage{slashbox}
\usepackage{tabularx,afterpage}

\newcolumntype{C}[1]{>{\centering\let\newline\\\arraybackslash\hspace{0pt}}m{#1}}

\newcommand{\be}{\begin{equation}}
\newcommand{\ee}{\end{equation}}
\newcommand{\bea}{\begin{equation}\begin{aligned}}
\newcommand{\eea}{\end{aligned}\end{equation}}
\newcommand{\bal}{\begin{align}}
\newcommand{\eal}{\end{align}}
\newcommand{\nn}{\nonumber\\}
\newcommand{\bi}{\begin{itemize}}
\newcommand{\ei}{\end{itemize}}
\newcommand{\ben}{\begin{enumerate}}
\newcommand{\een}{\end{enumerate}}
                   
\newcommand{\Eq}[1]{Eq.~(\ref{#1})}
\newcommand{\Fig}[1]{Fig.~\ref{#1}}
\newcommand{\Refc}[1]{Ref.~\cite{#1}}
\newcommand{\Refcs}[1]{Refs.~\cite{#1}}

\newcommand{\cW}{c_{\scriptscriptstyle W}}
\newcommand{\sW}{s_{\scriptscriptstyle W}}

\newcommand{\mW}{m_{\scriptscriptstyle W}}
\newcommand{\pW}{p_{\scriptscriptstyle W}}

\newcommand{\EW}{E_{\scriptscriptstyle W}}

\newcommand{\mZ}{m_{\scriptscriptstyle Z}}
\newcommand{\EZ}{E_{\scriptscriptstyle Z}}

\newcommand{\EH}{E_{\scriptscriptstyle H}}
\newcommand{\mH}{m_{\scriptscriptstyle H}}
\newcommand{\gW}{g_{\scriptscriptstyle W}}
\newcommand{\qW}{q_{\scriptscriptstyle W}}

\newcommand{\IW}{I_{\scriptscriptstyle W}}

\newcommand{\eN}{e_{\scriptscriptstyle N}}
\newcommand{\mN}{m_{\scriptscriptstyle N}}
\newcommand{\mV}{m_{\scriptscriptstyle V}}
\def\pd{\partial}

\begin{document}

\title{Dark Matter Spectra from the Electroweak to the Planck Scale}

\author[1,2]{\small Christian W. Bauer,}
\author[1,2]{\small Nicholas L. Rodd,}
\author[3]{\small Bryan R. Webber}
\affiliation[1]{\footnotesize Berkeley Center for Theoretical Physics, University of California, Berkeley, CA 94720, USA}
\affiliation[2]{\footnotesize Theoretical Physics Group, Lawrence Berkeley National Laboratory, Berkeley, CA 94720, USA}
\affiliation[3]{\footnotesize University of Cambridge, Cavendish Laboratory, J.J.~Thomson Avenue, Cambridge, UK}

\abstract{
We compute the decay spectrum for dark matter (DM) with masses above the scale of electroweak symmetry breaking, all the way to the Planck scale.
For an arbitrary hard process involving a decay to the unbroken standard model, we determine the prompt distribution of stable states including photons, neutrinos, positrons, and antiprotons.
These spectra are a crucial ingredient in the search for DM via indirect detection at the highest energies as being probed in current and upcoming experiments including IceCube, HAWC, CTA, and LHAASO.
Our approach improves considerably on existing methods, for instance, we include all relevant electroweak interactions.}

\maketitle
\setcounter{page}{2}

\section{Introduction}

If the dark matter (DM) of our universe is a particle with a mass between the electroweak and Planck scales, then it could be discovered via indirect detection of standard model (SM) particles produced from its decay.
Such decays can be initiated by an underlying hard process where the DM decays to two SM states, $\chi \to X \bar{X}$.
The SM states, injected with virtuality $\mu \sim m_{\chi}$, will shower and eventually hadronize, evolving down to on-shell stable particles such as photons, neutrinos, positrons, and anti-protons.

\begin{figure}[t]
\leavevmode
\vspace{-0.2cm}
\begin{center}
\includegraphics[width=.6\textwidth]{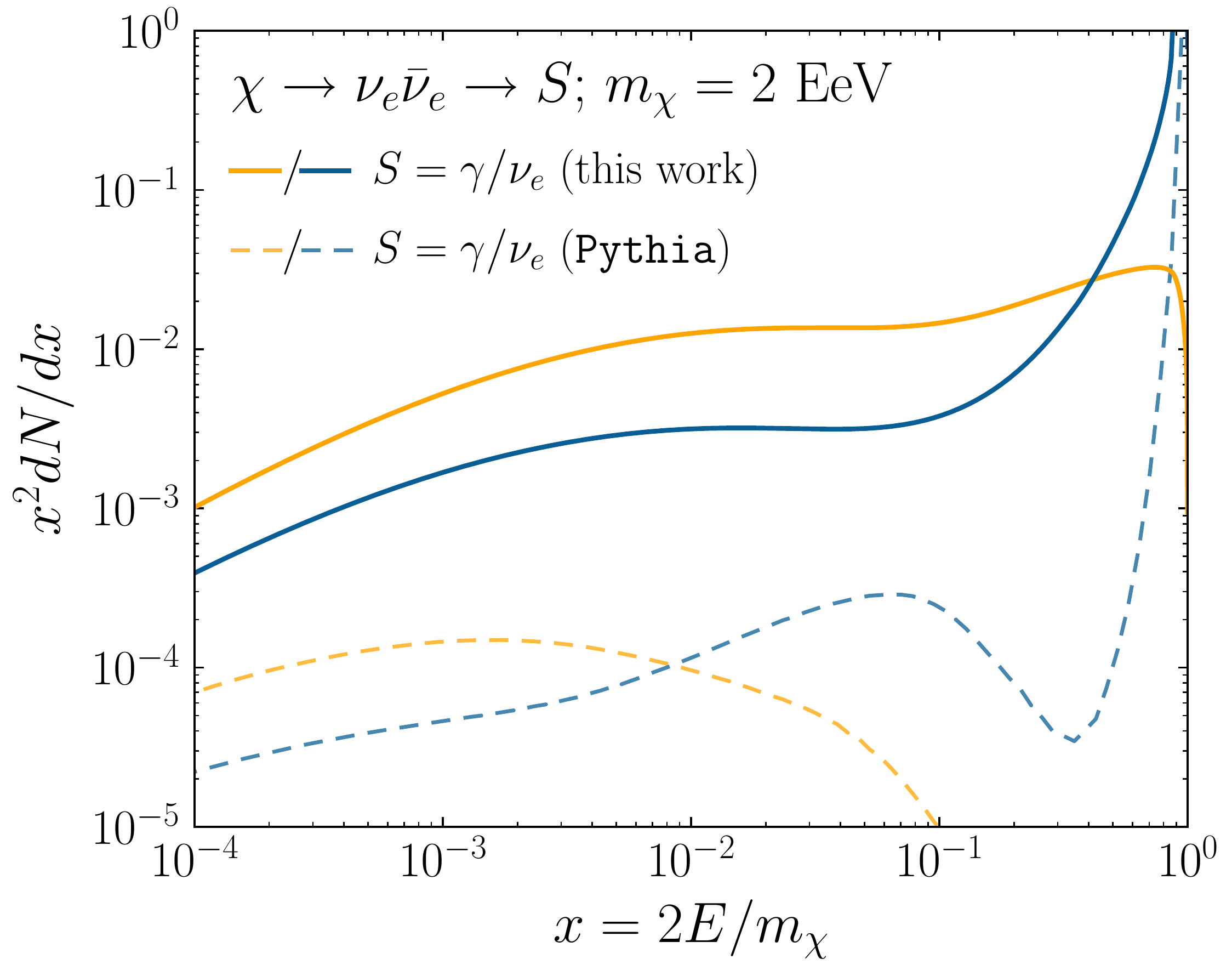}
\end{center}
\vspace{-0.5cm}
\caption{The prompt electron neutrino and photon spectrum resulting from the decay of a 2 EeV DM particle to $\nu_e \bar{\nu}_e$, as currently being searched for at IceCube~\cite{Aartsen:2018mxl}.
Solid curves represent the results of this work, and predict orders of magnitude more flux at certain energies than the dashed results of \texttt{Pythia} 8.2, one of the only existing methods to generate spectra at these masses.
In both cases energy conservation is satisfied: there is a considerable contribution to a $\delta$-function at $x=1$, associated with events where an initial $W$ or $Z$ was never emitted and thus no subsequent shower developed.
Large disagreements are generically observed at these masses for electroweak dominated channels, while the agreement is better for colored initial SM states.
}
\label{fig:PythiaComparison}
\end{figure}

Calculation of the resulting prompt spectra is a central ingredient in testing the hypothesis of heavy DM.
At present, a common approach is to simulate these events using \texttt{Pythia}~\cite{Sjostrand:2006za,Sjostrand:2007gs,Sjostrand:2014zea}, which accurately reproduces most of the relevant physics up to $\sim$TeV scales.
\texttt{Pythia} is not, however, at present designed to operate well above these scales, for example it is missing interactions such as triple gauge couplings in the electroweak sector that can become increasingly important.
In this paper, we propose an alternative approach, which in certain channels can produce spectra that differ significantly from existing results, as demonstrated in \Fig{fig:PythiaComparison}.
We make our full results publicly available~\cite{PubCode}.

Heavy decaying DM can be realized in a number of different scenarios, including Wimpzillas~\cite{Chung:1998zb,Benakli:1998ut,Kolb:1998ki,Blasi:2001hr,Kolb:2017jvz,Alcantara:2019sco}, glueballs~\cite{Faraggi:2000pv,Boddy:2014yra,Forestell:2016qhc,Halverson:2016nfq,Cohen:2016uyg,Forestell:2017wov}, gravitinos~\cite{Pagels:1981ke,Steffen:2006hw,Ishiwata:2008cu}, superstring relics~\cite{Chang:1996vw,Faraggi:1999iu,Coriano:2001mg,DelleRose:2017vvz}, and more recent proposals, see for example~\cite{Contino:2018crt,Babichev:2018mtd,Kim:2019udq,Dudas:2020sbq,Kramer:2020oqi,Hambye:2020lvy,Garcia:2020hyo}.
Independent of UV motivations, there is a clear reason to consider searching for such DM: the robust experimental program to probe astrophysical messengers at higher energies.
Many instruments can probe heavy DM, including HAWC~\cite{Abeysekara:2017jxs}, IceCube~\cite{Abbasi:2011eq,Esmaili:2013gha,Rott:2014kfa,Aartsen:2018mxl,Bhattacharya:2019ucd,Chianese:2019kyl,Liu:2020ckq}, ANTARES~\cite{ANTARES:2019svn,Aartsen:2020tdl}, Pierre Auger Observatory~\cite{ThePierreAuger:2015rma,Esmaili:2012us,Kuznetsov:2016fjt,Aab:2019auo}, Telescope Array~\cite{Verzi:2017hro,Abbasi:2018ywn}, and in the future CTA~\cite{CTAConsortium:2018tzg,Silverwood:2014yza}, LHAASO~\cite{Bai:2019khm,He:2019bcr}, IceCube-Gen2~\cite{Aartsen:2014njl}, and KM3NET~\cite{Adrian-Martinez:2016fdl,Ng:2020ghe}.
Taken together, these experiments demonstrate that in the coming years we will continue to probe the universe at higher energies and to greater sensitivities.
Prompt spectra with a full treatment of SM interactions are required to discover DM with this data.

The remainder of this paper presents our approach to obtaining these.
We begin by describing how the calculation of DM spectra can be mapped onto fragmentation functions (FFs), which can be evolved from the UV scale, $\mu \sim m_{\chi}$, down to the IR, $\mu \sim 0$.
The computation can be performed in three stages: 1) Evolution from $m_{\chi}$ to the weak scale $\qW\sim 100$ GeV; 2) Matching through $\qW$; and 3) Continued evolution down to 0.
In the main text we outline the methods used at each step, providing full technical details in related appendices.

\section{Framework}

The flux of an observable particle $S$ produced from DM decay depends centrally on the prompt spectrum, defined as\footnote{The discussion is couched in the language of DM decay due to the Kamionkowski-Griest bound~\cite{Griest:1989wd} providing a naive obstruction to DM annihilation at these masses.
The bound can be evaded, see e.g.~\cite{Berlin:2016vnh,Harigaya:2016nlg,Berlin:2016gtr,Cirelli:2018iax}, and our results can be readily ported to annihilation with the simple identification $m_{\chi}^{\rm dec.} = 2 m_{\chi}^{\rm ann.}$.}
\be
\frac{dN_S}{dx} = \frac{1}{\Gamma_0} \frac{d\Gamma}{dx} (\chi \to S + \ldots)\,.
\label{eq:dNSdx}
\ee
Here $\Gamma$ is the inclusive decay rate of $\chi$ to $S$, $\Gamma_0 = 1/\tau$ is the inverse lifetime, and we use dimensionless variables $x = 2E/m_{\chi}$.
If the decay is seeded by an underlying process $\chi \to X \bar{X}$, for an arbitrary SM state $X$, the process begins with each particle at a virtuality scale $m_{\chi}/2$.
The problem is then to determine the probability that $X$ and $\bar{X}$ evolve to produce $S$ carrying a fraction $x$ of the initial energy.
This process is described by a fragmentation function (FF) $D_a^b(x;\,\mu_Q, \mu_0)$, which determines the probability of an initial particle $a$ at a scale $\mu_Q$ evolving to produce a particle $b$ at $\mu_0$ carrying a momentum fraction $x$; in the absence of any evolution we would have $D_a^b(x;\,\mu_Q, \mu_0) = \delta_a^b \delta(1-x)$.
In this language, we can write the spectrum as\footnote{\Eq{eq:spec2ff} applies for a hard two-body decay.
The formalism can be extended to $(n>2)$-body decays, as described in the App.~\ref{sec:DMDecAnn}.}
\be
\frac{dN_S}{dx} = D_X^S(x;\, m_{\chi}/2, 0) + D_{\bar{X}}^S(x;\, m_{\chi}/2, 0)\,.
\label{eq:spec2ff}
\ee

\begin{figure}[t]
\leavevmode
\begin{center}
\includegraphics[width=.45\textwidth]{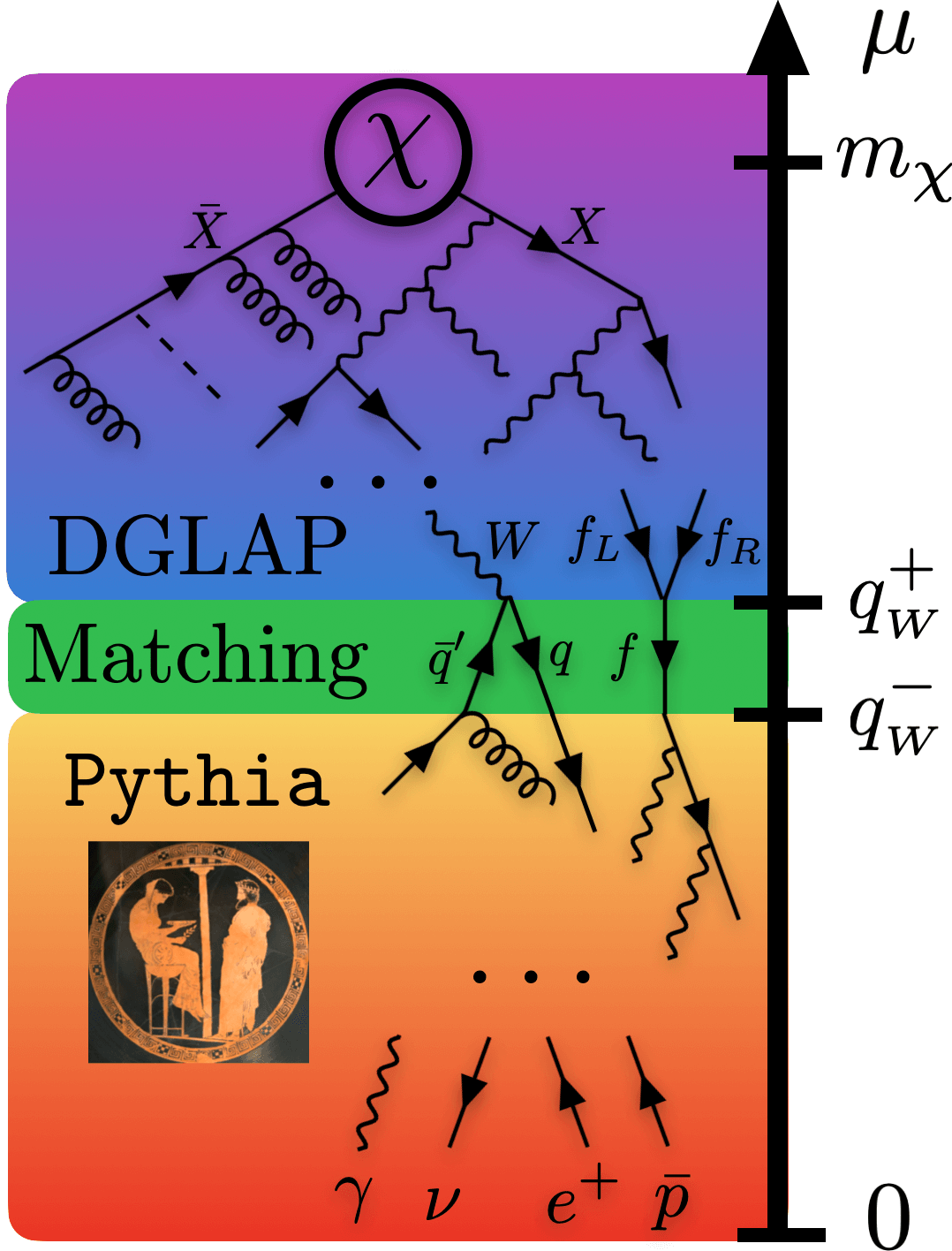}
\end{center}
\vspace{-0.3cm}
\caption{A cartoon of the three steps used to calculate the UV to IR evolution.
An initial 2-body final state resulting from a hard interaction of $\chi \to X \bar{X}$ is DGLAP evolved down to $\qW^+$, just above the weak scale.
At $\qW^+$, states with masses above the electroweak scale are integrated out.
Finally, at $\qW^-$ these results are matched onto \texttt{Pythia} which handles the subsequent evolution and hadronization effects.
}
\label{fig:ThreeSteps}
\end{figure}

At this stage, we have simply rephrased the problem.
The power of \Eq{eq:spec2ff} is that it allows us to bring to bear the considerable formalism of FFs to the calculation of DM spectra.
In particular, the full evolution in virtuality can be decomposed into easier to compute segments, and then convolved together.
For the present work we will exploit this result to break the calculation up as follows,
\begin{align}
&\hspace{-0.15cm}D_X^S(x;\, m_{\chi}/2, 0)
=\sum_{M,N} \int_x^1 \frac{dy}{y}\,\int_{x/y}^1 \frac{dz}{z}
\underbrace{D_X^M(y;\, m_{\chi}/2, \qW^+)}_{\rm DGLAP} \nn
&\hspace{1.5cm}
\times\, \underbrace{D_M^N(z;\, \qW^+, \qW^-)}_{\rm Matching}\, \times \,
\underbrace{D_N^S(x/(yz);\, \qW^-, 0)}_{\texttt{Pythia}}\,.
\label{eq:threesteps}
\end{align}
The three pieces to be calculated are as follows.
Firstly we evolve from the scale of the DM mass down to just above the weak scale, $\qW^+$, using the DGLAP equations~\cite{Gribov:1972ri,Dokshitzer:1977sg,Altarelli:1977zs} and in particular an implementation using all interactions in the unbroken SM, as well as a partial treatment of soft-coherence effects~\cite{Chudakov:1955aa,Ermolaev:1981cm,Mueller:1981ex,Bassetto:1982ma,Bassetto:1984ik}.
We next perform a matching by evolving across a parametrically small region through the weak scale, removing all particles with electroweak scale masses.
Finally, these results are matched onto \texttt{Pythia} below $\qW$, where it is used to calculate the subsequent showering, hadronization, and light particle decays in a regime where it has been extensively vetted.
A simplified depiction of the full evolution is given in \Fig{fig:ThreeSteps}, and we next flesh out the details involved at each stage.

\section{High Scale Evolution and Soft Coherence}

The first step of our calculation is to take the two body spectrum at $\mu = m_{\chi}/2$, and evolve this down to just above the weak scale, $\mu = \qW^+$.\footnote{In practice, to improve numerical stability, the DGLAP equations are solved by evolving from $\qW$ to $m_{\chi}/2$, rather than the other way around.
See App.~\ref{sec:HighScale} for details.}
To do so we include the dominant effects associated with the leading collinear and collinear-soft divergences in the theory, both of which are described by the unregulated Altarelli-Parisi splitting functions $\hat{P}(z)$.
The evolution of the FFs under the $1 \to 2$ splitting interaction $I$ encoded in $\hat{P}(z)$ is described by the DGLAP evolution equations, which take the schematic form (suppressing the fixed high scale)
\be
\left[\mu \frac{\pd}{\pd \mu} D(x;\, \mu)\right]_I
= - \frac{\alpha_I}{\pi} \int_0^1 dz\, \hat{P}(z) \left[ \frac{1}{z} D(x/z;\, \mu) - D(x;\,\mu) \right],
\label{eq:SchematicDGLAP}
\ee
where $\alpha_I$ is the corresponding SM coupling.
The distribution of momenta amongst the particles, described by the FF $D(x, \mu)$, can only evolve in two ways: contributions are received by particles with momentum fractions greater than $x$ splitting to exactly that value, and they are lost if a particle with fraction $x$ splits at all.
This describes the two terms in square brackets above, and they are associated with real and virtual emissions in the theory respectively.
The evolution down to a desired scale is achieved by solving \Eq{eq:SchematicDGLAP}, accounting for all interactions in the unbroken SM: SU(3), SU(2)$_L$, U(1)$_Y$, and Yukawa.

The actual problem is more complex.
In the unbroken SM there are 58 states of interest, implying \Eq{eq:SchematicDGLAP} is in truth 3364 coupled equations, where the $\hat{P}(z)$ now account for all interactions in the SM.
Further, flavor changing interactions in the electroweak sector can lead to an incomplete cancellation of the real and virtual contributions, and the subsequent development of electroweak double logarithms~\cite{Ciafaloni:2000df}.
To solve this problem, we use the results of~\Refcs{Manohar:2018kfx,Bauer:2018xag,Bauer:2018arx}, which included a treatment of polarization effects.
These polarization effects are critical.
Even if in the UV we start with an unpolarized initial state, the chiral nature of the electroweak interaction will generate a polarization, which can be considerable over the evolution scales we consider here.
As the spectra of particles produced by decays of the electroweak states $t$, $W$, and $Z$ depend on their polarizations, we will need to keep track of this through to the matching step described below.

The DGLAP evolution can effectively be described as a semi-classical shower that develops via consecutive $1 \to 2$ splittings, each occuring with probability determined by the appropriate $\hat{P}(z)$.
Soft physics breaks this picture.
As an example, consider a splitting $W \to d \bar{u}$, and then the subsequent emission of a gluon off either the $\bar{u}$ or $d$
\begin{center}
\includegraphics[width=.18\textwidth]{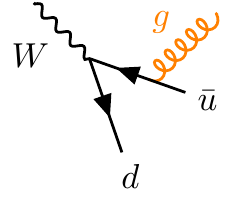}\hspace{1.0cm}
\includegraphics[width=.18\textwidth]{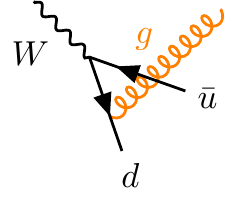}
\vspace{-0.3cm}
\end{center}
If the gluon is emitted at a wide angle with respect to the $d \bar{u}$ pair, then in the soft limit, its wavelength will be such that it cannot resolve the individual quarks.
As they came from an uncolored $W$, the gluon will now see 0 net color charge and cannot be emitted.
In particular, there is destructive interference between the two diagrams, and consequently a large suppression of small-$x$ states.
The result is the well known angular-ordering effect, and in the context of QCD Monte Carlo simulations, it is understood how to include this effect~\cite{Marchesini:1983bm,Marchesini:1987cf}.
As a modification to the full SM DGLAP equations, it is not.
A complete treatment of the problem is beyond the scope of the present work, however, we introduce an identity in order to capture the largest effect of soft coherence: the reduction of real radiation at small-$x$.
As we derive in the App.~\ref{sec:HighScale}, the equation that describes the leading effect of angular ordering, can be rewritten as a DGLAP equation, but evolved from a scale $x \times m_{\chi}/2$, rather than $m_{\chi}/2$.
This allows us to take our solution to the full form of \Eq{eq:SchematicDGLAP}, and augment them with the substitution $m_{\chi} \to x\, m_{\chi}$.\footnote{Technically this substitution is only appropriate for the single logarithmic terms associated with the isosinglet evolution.
As such we need to factor out the non-cancelling electroweak double logs from this change of variables.
See App.~\ref{sec:HighScale} for details.}

This substitution allows for a simple inclusion of the soft physics, but it is not perfect.
Soft-coherence not only reduces the real emission, it also increases the associated virtual no-emission probability, and our result only accounts for the former.
This deficiency manifests itself as a failure of momentum conservation generally at the level of $\sim$$1-3\%$, although for particular states and masses it can be as large as 10\%.
Given the large impact of the effect on the spectra at small-$x$, we choose to accept this shortcoming, leaving the complete treatment as an open problem.

\section{Weak Matching}

We now take our DGLAP evolved, soft-coherence corrected, FFs and evolve them across the electroweak threshold.
Formally we evolve across $\qW^{\pm} = \qW (1\pm \epsilon)$, with $\epsilon \ll 1$, a parametrically small separation of scales, ensuring this step cannot generate large logs from the evolution.
Instead, the point of this step is a matching from the unbroken to broken SM, where we integrate out the electroweak mass states $t$, $W$, $Z$, and $h$.
For $h$ we let \texttt{Pythia} handle the decay.
For $t$, $W$, and $Z$, we instead need to account for the fact that our evolution at the first step can generate a significantly polarized spectrum for each state.
In order to ensure this physics persists into our final results, we decay each of these states analytically, accounting for the polarization.

For our purposes, the details of the polarized decays are sufficiently described by the differential spectra obtained from the tree level diagrams, such as depicted below.
\begin{center}
\includegraphics[width=.3\textwidth]{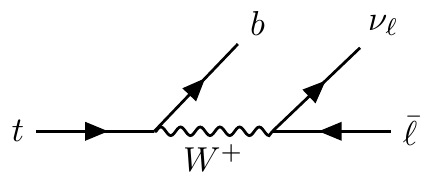}
\vspace{-0.3cm}
\end{center}
The calculations are straightforward, although for the top slightly involved, and so we postpone the full details to the appendices.
In each case, we obtain an analytic or simple parameterization for $D_{W_{0,\pm}}^f(x;\,\qW^+,\qW^-)$, where $f$ represents the fermions that can be produced from a $W$ decay, and the equivalent functions for $Z_{0,\pm}$ and $t_{\pm}$, where $0,\pm$ correspond to the longitudinal and $\pm$ transverse polarizations, respectively.
Full details are given in App.~\ref{sec:Matching}.

For states without $m \sim \qW$, other than the contribution they receive from decaying particles, the threshold is uneventful.
We remove polarization at this stage by, for example, combining $f_L + f_R = f$, because its effects in evolution below the electroweak scale are small and \texttt{Pythia} does not take account of them.
This step also represents the first appearance of electroweak masses, which are neglected in the initial evolution.
For $m_{\chi} \sim \qW$, we will accordingly underestimate the phase space suppression of electroweak states, and therefore do not quote results for $m_{\chi} <~{\rm TeV}$.
There are many options available at these scales, including \texttt{Pythia} or PPPC4DMID~\cite{Cirelli:2010xx,Ciafaloni:2010ti}.
Another possibility would be to match our results to fixed-order matrix elements without double counting, in the way proposed in~\Refc{Bauer:2017bnh}.\footnote{For neutrino states and $m_{\chi} \sim {\rm TeV}$ we observe $\mathcal{O}(1)$ differences between our results and PPPC4DMID.
In the absence of the fixed order corrections we take this as a theoretical uncertainty, although it will be suppressed significantly as we increase in mass.}

\section{Low Scale Evolution with \texttt{Pythia}}

Our final step is to take each particle that resulted from the weak matching, and continue its evolution down to lower scales using \texttt{Pythia}, which will include the remaining showering, soft coherence, light particle decays, and importantly the non-perturbative hadronization.
For each state, starting at a scale $\qW$, the spectra of stable states, $S \in \{\gamma, e^{\pm}, p^{\pm},\nu_{e,\mu,\tau},\bar{\nu}_{e,\mu,\tau}\}$, are determined.
When convolved with the earlier steps the full DM result is obtained.

For our purposes there is one deficiency with a full \texttt{Pythia} treatment at this step.
\texttt{Pythia} models the final state radiation (FSR) emission of photons off charged particles, $f \to f \gamma$, only down to an isolation or $p_T$ cut.
Photons that are highly collinear with the parent charged particle cannot be separated in the environment of a collider like the LHC.
If they travel over galactic or cosmological distances, they certainly can however.
As such, we want to include the photons that \texttt{Pythia} deliberately excludes at small-$x$.
To do so, we turn photon FSR off in \texttt{Pythia} and instead include it analytically to first order using the appropriate result given by,
\be
[D_N^{\gamma}(x;\,\qW,0)]_{\rm FSR} 
= \frac{\alpha \eN^2}{2\pi} \frac{1+(1-x)^2}{x} \left[ \ln \left( \frac{4\qW^2(1-x)}{\mN^2} \right) - 1 \right].
\label{eq:photonFSR}
\ee
Here $\eN$ is the charge of the emitter, and $\mN$ its mass.
If $N$ indexes a quark, we take $m_q = {\rm max}(m_q, \Lambda_{\rm QCD})$, as the evolution will stop at the higher of the two scales.
This treatment means we neglect processes like subsequent splitting of
the FSR photons into charged fermions but as these are formally higher order and not enhanced by large logs, the above treatment is sufficient for our purposes.

In addition, we also need to account for the fact that turning off photon FSR in \texttt{Pythia} means the charged particles that would have emitted photons now have a higher momentum fraction.
In other words, including only \Eq{eq:photonFSR} is inconsistent with momentum conservation.
However, it is straightforward to derive the required modifications to the charged particle FFs, simply from momentum conservation, and these are applied.
The full expressions are provided in App.~\ref{sec:LowScale}.

\begin{figure}[t]
\leavevmode
\vspace{-0.2cm}
\begin{center}
\includegraphics[width=.6\textwidth]{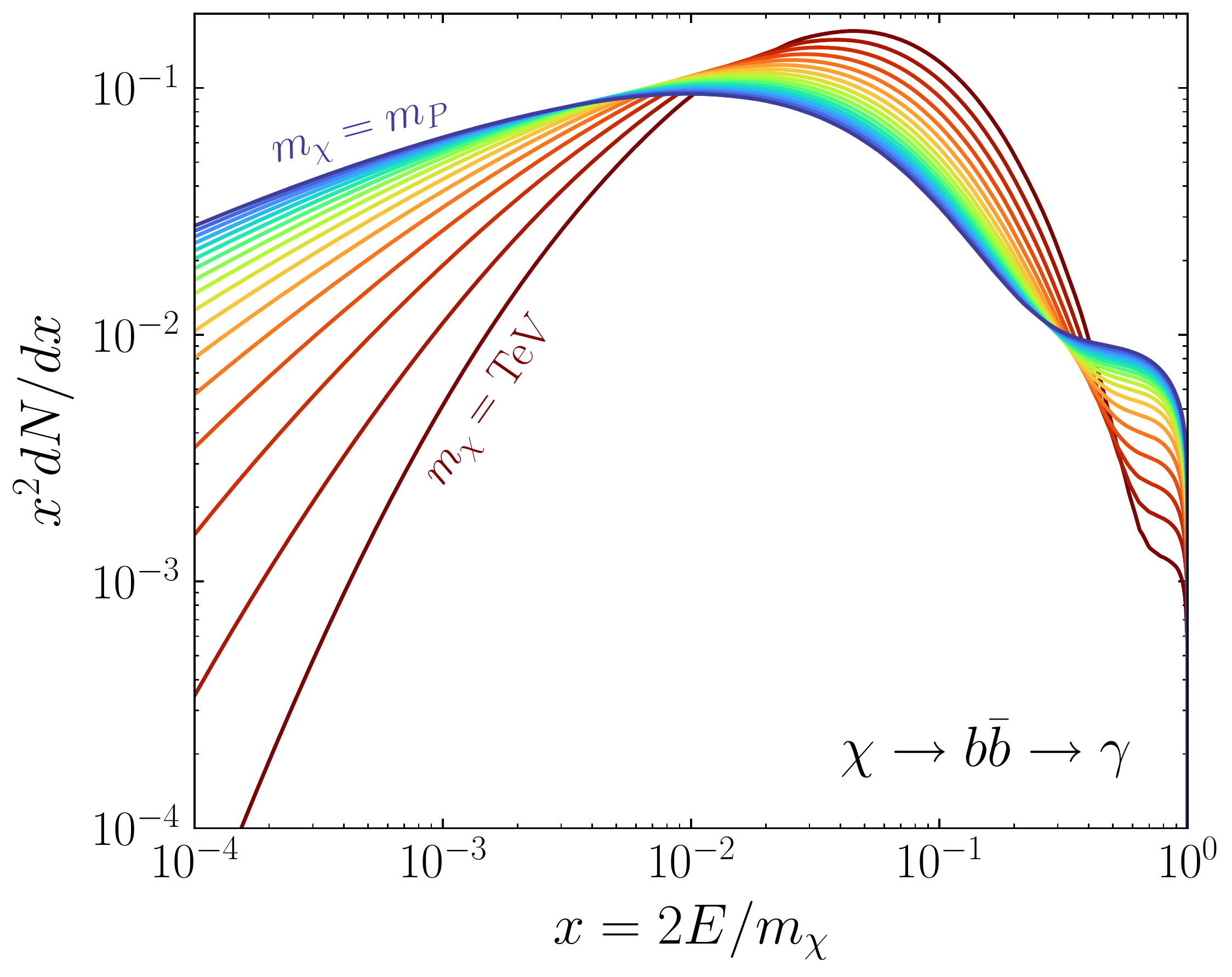}
\end{center}
\vspace{-0.5cm}
\caption{The photon spectrum resulting from $\chi \to b \bar{b}$, shown across the entire mass range considered in this work. 
A curve is shown for every decade in mass between a TeV and the Planck scale.
}
\label{fig:ballM}
\end{figure}

\section{Discussion}

Combining these three FFs, with the aid of Eqs.~(\ref{eq:spec2ff},\ref{eq:threesteps}), we obtain the desired prompt DM spectra.\footnote{We emphasize the qualifier {\it prompt}.
The spectrum at Earth depends on additional propagation effects, see e.g.~\cite{Murase:2012xs,Esmaili:2015xpa,Batista:2016yrx,Blanco:2018bbf}.}
An example is shown in \Fig{fig:PythiaComparison}.
The difference between our results and those of \texttt{Pythia} is driven by the lack of the full electroweak interactions in the latter.
In \texttt{Pythia}, $Z$ emission from the hard neutrino leads to an endpoint contribution to the neutrino spectrum, whilst the low energy bumps are associated with QCD states.
However, a full electroweak shower cannot form in \texttt{Pythia} as it can in our results, and by $m_{\chi} \sim {\rm EeV}$, the shower can involve a large number of electroweak bosons that produce significantly more emission across the allowable energy fractions.
For the same reason, in other channels where electroweak showers contribute significantly, such as $\chi \to e^- e^+$, large disagreements are also observed.
Smaller differences occur for channels where QCD showers dominate.
Yet in all cases, for $m_{\chi}$ well above the electroweak scale, running \texttt{Pythia} will eventually become impractical given the size of the generated showers.
Our approach encounters no such obstacle, as exemplified in \Fig{fig:ballM} where we show the photon spectrum from $\chi \to b \bar{b}$ for DM masses all the way to the Planck scale.
The spectrum of hard photons is seen to increase with $m_{\chi}$, whilst the lower energy peak associated predominantly with neutral pion decays progressively softens.
These are only a few representative examples.
In total we compute 638 FFs $D_X^S$: $X$ indexes the 58 states in the unbroken SM, and $S \in \{\gamma, e^{\pm}, p^{\pm}, \nu_{e,\mu,\tau},\bar{\nu}_{e,\mu,\tau}\}$.
The full spectra are available~\cite{PubCode}.\footnote{We release 56 rather than 58 values for $X$, and therefore only 616 FFs. We exclude the two states associated with interference of $B$ and $W^3$, although they are included in the evolution.}
An exploration of the impact these results have on searches for DM with neutrino final states is provided in \Refc{Liu:2020ckq}.

Our results are not the final word on heavy DM spectra.
There are a number of directions in which our results can be systematically improved, including:
\bi
\item A rigorous treatment of the relevant soft physics as part of a full next-to-leading-logarithmic (NLL) calculation.
\item Matching to fixed-order electroweak matrix elements using the method of~\Refc{Bauer:2017bnh}.
\item Inclusion of estimated theoretical uncertainties associated with both the DGLAP and parton shower evolution, for the latter see~\Refcs{Cembranos:2013cfa,Amoroso:2018qga,Niblaeus:2019ldk}.
\item Polarized decays of the $\mu$ and $\tau$. 
\item We assume no additional thresholds are crossed between $m_{\chi}$ and $\qW$.
Results going beyond this have been considered for supersymmetric QCD~\cite{Berezinsky:2000up,Berezinsky:2002hq,Aloisio:2003xj} and also the minimal supersymmetric SM~\cite{Barbot:2002ep,Barbot:2002gt}.
\ei
At present, our spectra carry quantifiable theoretical uncertainties.
Our calculation is performed with double logarithmic accuracy, and therefore single logs are not resummed.
Accordingly, our results carry a global uncertainty of order $\exp [\alpha_I L]/(1+\alpha_I L)$ with $L=\ln(m_{\chi}/\qW)$ and $I=2,3$.
At an EeV this corresponds to an $\mathcal{O}(10\%)$ error, growing to $\mathcal{O}(20\%)$ by the Planck scale.
For $x \lesssim 10^{-3}$ there are even larger uncertainties associated with an incomplete treatment of the soft physics associated with coherent emission of gauge bosons.
We can obtain an uncertainty estimate by comparing our default procedure of implementing the $m_{\chi} \to x\, m_{\chi}$ substitution at the high scale to two alternative approaches: 1. an underestimate of soft-coherence, where no accounting for the effect is performed; and 2. an overestimate where we apply the substitution to the FFs combined across all three scales, which will double count the soft-coherence already present in \texttt{Pythia}.
Doing so, we conclude that our treatment can produce $\mathcal{O}(1)$ errors at small-$x$ values for an EeV, although the effect at a given $x$ decreases with mass.
A fuller discussion of uncertainties is given in App.~\ref{sec:Limitations}.

Even with these uncertainties, our spectra represent a manifest improvement over existing treatments: we include effects that are demonstrably important, and our formalism extends all SM states to arbitrarily high masses.
Combining these results with astrophysical probes of the high energy universe, the heavy DM hypothesis will be put to the test in the coming years.
In the event of an excess, we may finally begin to unravel the particle nature of DM.

\begin{acknowledgments}

Our work benefited from discussions with Carlos Arg\"{u}elles, Marco Cirelli, Marat Freytsis, Pat Harding, Qinrui Liu, Carsten Rott, Filippo Sala, Torbj\"{o}rn Sj\"{o}strand, Juri Smirnov, Varun Vaidya, and members of the Cambridge Pheno Working Group.
This work was supported by the Miller Institute for Basic Research in Science at the University of California, Berkeley (NLR), the Director, Office of Science, Office of High Energy Physics of the U.S. Department of Energy under the Contract No. DE-AC02-05CH11231 (CWB), and partially supported by
U.K. STFC consolidated grant ST/P000681/1 (BRW).
BRW is grateful for the hospitality of Kavli IPMU while part of this work was performed.
Kavli IPMU is supported by the World Premier International Research Center Initiative (WPI Initiative), MEXT, Japan.
This work made use of resources provided by the National Energy Research Scientific Computing Center, a U.S. Department of Energy Office of Science User Facility supported by Contract No. DE-AC02-05CH11231.

\end{acknowledgments}

\appendix
\addcontentsline{toc}{section}{\protect\numberline{}Appendices}%
\addtocontents{toc}{\protect\setcounter{tocdepth}{1}}
\section*{Appendix}

As mentioned, we have postponed many of the technical aspects of our calculation to the appendices, and we will organize that discussion as follows.
In App.~\ref{sec:DMDecAnn} we begin by providing an expanded discussion of the connection between DM spectra and FFs, and a comparison between the approach to the problem presented in this work with ideas discussed previously.
In Apps.~\ref{sec:HighScale},~\ref{sec:Matching}, and~\ref{sec:LowScale} we provide an unabridged discussion of the three steps of our evolution.
Afterwards, in App.~\ref{sec:Limitations} we provide an estimate of the theoretical uncertainties associated with our spectra.
Appendix~\ref{sec:AdditionalResults} contains additional results highlighting the physics inherent in our spectra, and finally App.~\ref{sec:PubCode} outlines the details of the public spectra and code.

\section{Dark Matter Spectra and Fragmentation Functions}
\label{sec:DMDecAnn}

This section expands upon the connection between DM spectra and FFs.
In particular, we detail the steps between \Eq{eq:dNSdx} and \Eq{eq:spec2ff}, as well as outlining the corresponding result for DM annihilation.
In doing so, it will become clear what steps are needed to modify our approach when the hard interactions is $(n>2)$-body.

To begin with, let us establish our conventions for indirect detection.
We follow \Refc{Lisanti:2017qoz}, and refer there for additional details.
The DM differential energy flux into an observable state $S$ for decay (dec.) annihilation (ann.), is given by\footnote{The factorization of particle physics and astrophysics this expression is predicated upon is an assumption.
For instance, the annihilation cross-section could depend on the relative DM velocity, which varies between astrophysical systems.
In such a case, however, an effective factorization can still be achieved and so the discussion of this sections ports over directly, although the velocity dependence of the cross-section is now placed into the astrophysics factor, see e.g.~\cite{Boddy:2018ike}.}
\bea
\frac{d\Phi}{dE} 
= \Bigg\{\begin{array}{l}
d\Phi_{\rm pp}^{\rm dec.}/dE \times D\,, \\
d\Phi_{\rm pp}^{\rm ann.}/dE \times J\,.
\end{array}
\eea
In both cases, $d\Phi/dE$ gives the number of $S$ states, per detector area, per observation time, per energy interval, carrying units $[{\rm particles}/{\rm cm}^2/{\rm s}/{\rm TeV}]$.
The $D$ and $J$-factors dictate the variation in flux across the celestial sphere for decay and annihilation, controlled by the DM density or density squared, respectively.
The focus of the present work is the result of fundamental DM-SM interaction, and this enters into the particle physics (pp) factor.
For a single annihilation or decay channel (for multiple channels simply sum over each weighted by the appropriate branching fractions), this can be written as
\bea
\frac{d\Phi_{\rm pp}^{\rm dec.}}{dE} &= \frac{1}{4\pi m_{\chi} \tau} \frac{dN_S}{dE}\,,\\
\frac{d\Phi_{\rm pp}^{\rm ann.}}{dE} &= \frac{\langle \sigma v \rangle}{8\pi m_{\chi}^2} \frac{dN_S}{dE}\,.
\label{eq:PPfactor}
\eea
These expressions expose the prompt spectrum per decay or annihilation, which represented our starting point in \Eq{eq:dNSdx}.
We note that this expression neglects propagation effects such as oscillation, redshifting, and other interactions which will transform the spectrum from the point of production to detection.
For details of how these effects can be incorporated, see e.g.~\cite{Murase:2012xs,Batista:2016yrx,Blanco:2018bbf}.

Although \Eq{eq:PPfactor} is how the particle physics factor is usually represented, the spectra should really be thought of as emerging from the differential decay width or annihilation cross-section, so that instead we have
\bea
\frac{d\Phi_{\rm pp}^{\rm dec.}}{dE} &= \frac{1}{4\pi m_{\chi}} \frac{d\Gamma}{dE}(\chi \to S + \ldots)\,,\\
\frac{d\Phi_{\rm pp}^{\rm ann.}}{dE} &= \frac{1}{8\pi m_{\chi}^2} \frac{d\langle \sigma v \rangle}{dE} (\chi\chi \to S + \ldots)\,.
\eea
The spectra can then be defined as
\bea
\frac{dN_S}{dE} &= \frac{1}{\Gamma_0} \frac{d\Gamma}{dE}(\chi \to S + \ldots)\,, \\
\frac{dN_S}{dE} &= \frac{1}{\langle \sigma v \rangle_0} \frac{d\langle \sigma v \rangle}{dE} (\chi\chi \to S + \ldots)\,,
\eea
from which \Eq{eq:PPfactor} follows, with the identification of the lifetime and cross sections appearing there as taking the tree-level or Born values, indicated by the $0$ subscript.
As we will consider spectra over a wide range of masses, it is convenient to move to dimensionless variables.
Specializing to the case of decay from now on, we take $x = 2E/m_{\chi}$, and then the above becomes exactly \Eq{eq:dNSdx}.
From now we will discuss the case of decay exclusively, although results for annihilation follow by rescaling $m_{\chi} \to 2 m_{\chi}$.

To calculate the decay spectrum we need to determine $d\Gamma/dx$.
The decay will be initiated by a hard process occurring at a scale $\mu \sim \chi$, where the DM decays to a set of SM particles each denoted by $I$.
We can then approximate each of these SM particles evolving separately down in virtuality to a set of stable observable SM particles, labelled by $S$, which is exactly described by a FF as described in the main body.
Corrections to this picture are described in the next subsection.
For the moment, however, the spectrum is given by the convolution of the hard process with each FF,
\be
\frac{d\Gamma}{dx}(\chi \to S + \ldots) = \sum_I \int_x^1 \frac{dz}{z}\, \frac{d\Gamma(\chi \to I)}{dz}\, D_I^S(x/z;\,m_{\chi}/2,0)\,.
\label{eq:spec2ffgeneral}
\ee
In the particular case where the hard interaction is a simple two body decay $\chi \to X \bar{X}$, then
\be
\frac{d\Gamma(\chi \to X \bar{X})}{dz} = \Gamma_0\,\delta(1-z)\,,
\label{eq:hardspec2body}
\ee
so that the convolution is trivial, and
\be
\frac{dN_S}{dE} = D_X^S(x/z;\,m_{\chi}/2,0) + D_{\bar{X}}^S(x/z;\,m_{\chi}/2,0)\,,
\ee
exactly as in \Eq{eq:spec2ff}.
If the initial hard process is more complicated than a two body decay, then it can be incorporated with a simple modification.
In detail, the appropriate generalisation of \Eq{eq:hardspec2body} should be substituted into \Eq{eq:spec2ffgeneral}, and combined with the FFs we provide~\cite{PubCode}.
We emphasize that the individual FFs are made available, and can be used directly.
As an additional example of how the FFs could be utilized, one signature of the formation of DM bound state formation could be the emission of a photon carrying away the binding energy, see e.g.~\cite{Asadi:2016ybp,Smirnov:2019ngs,Mahbubani:2019pij}.
For heavy DM, this photon can be sufficiently energetic that higher order processes become relevant, and a the photon should be replaced by the appropriate FF, thereby modifying the observable signature.
Further details are provided in App.~\ref{sec:PubCode}.

\begin{figure}[t]
\leavevmode
\vspace{-0.2cm}
\begin{center}
\includegraphics[width=.47\textwidth]{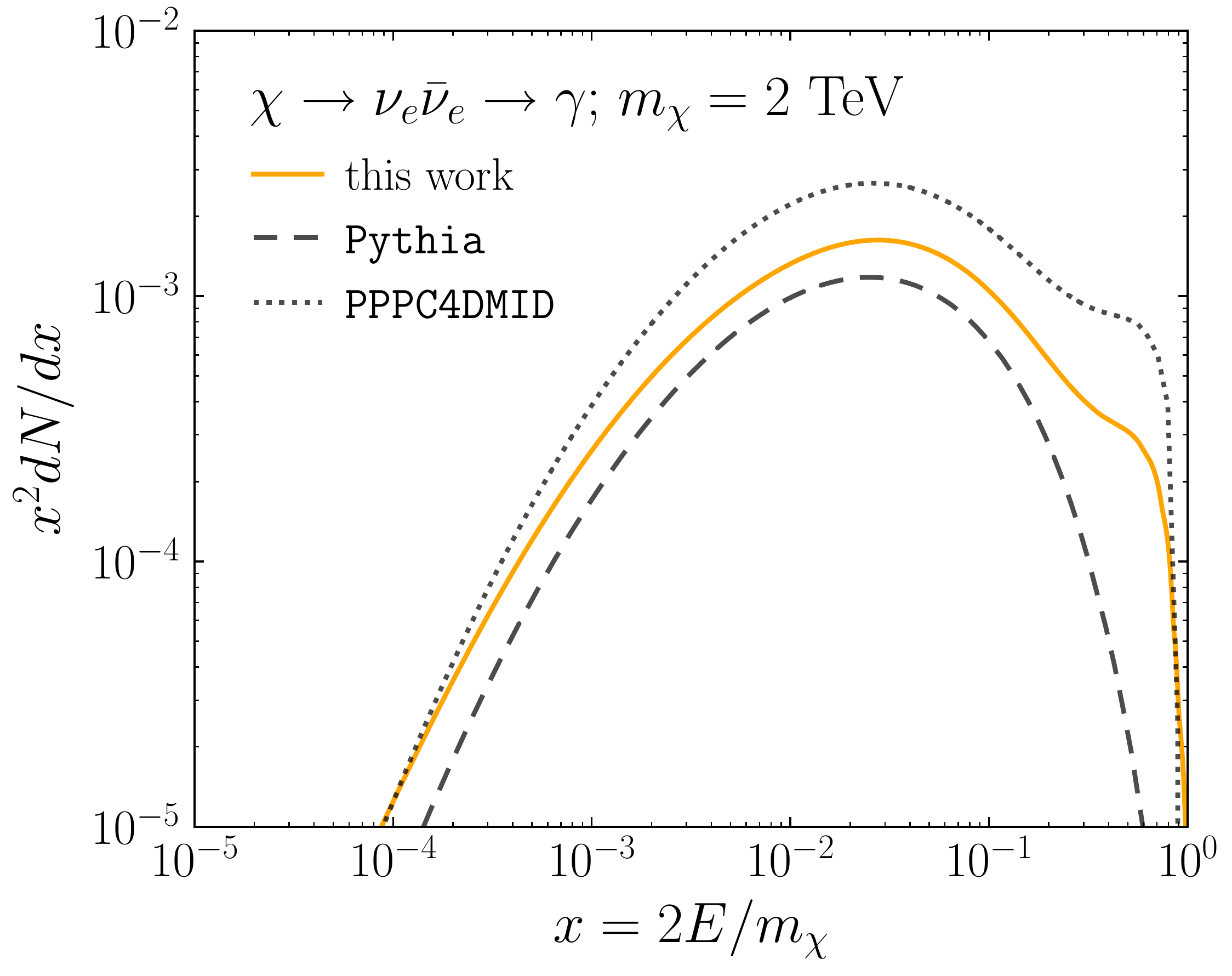}\hspace{0.1cm}
\includegraphics[width=.47\textwidth]{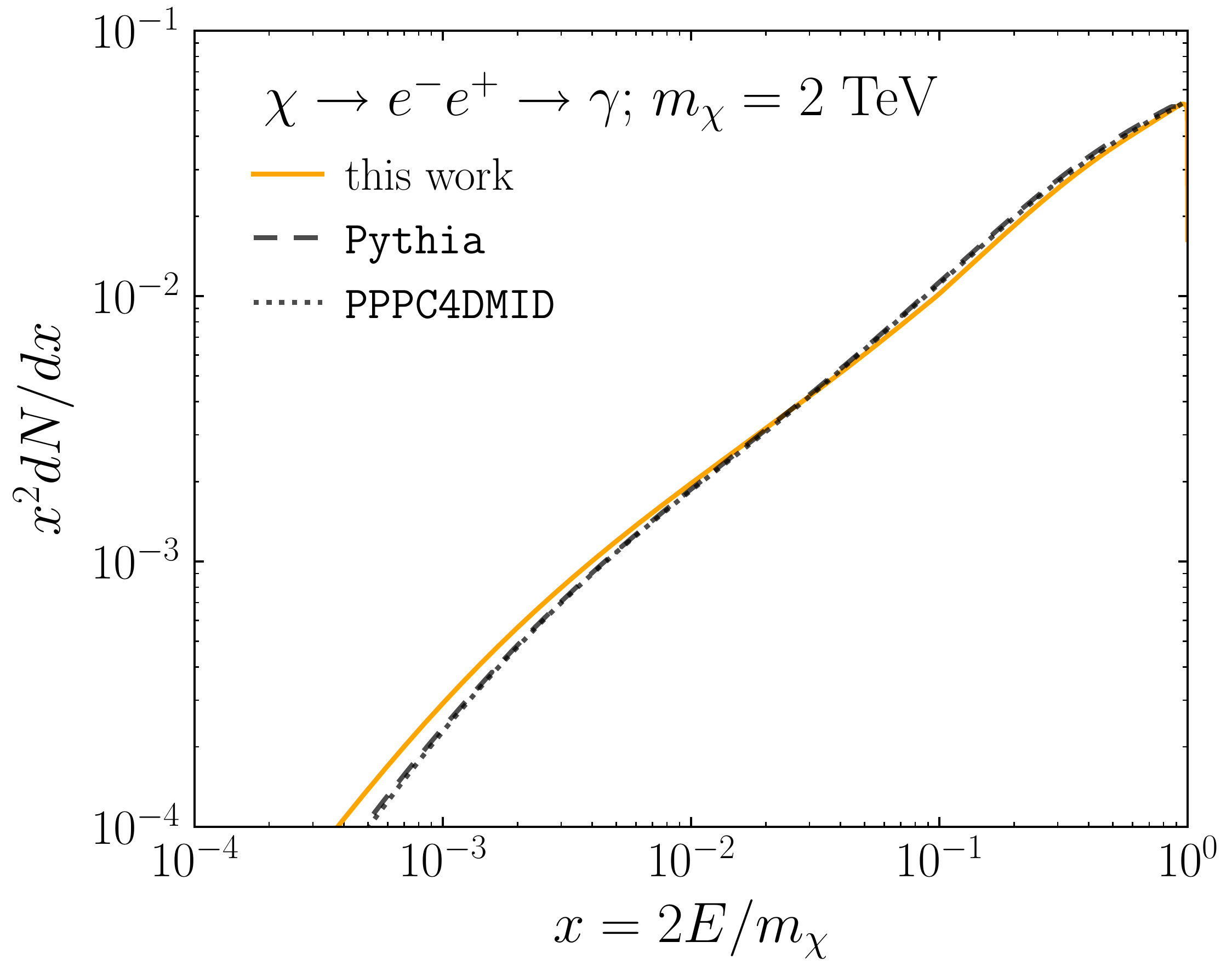}\vspace{0.5cm}\\
\includegraphics[width=.47\textwidth]{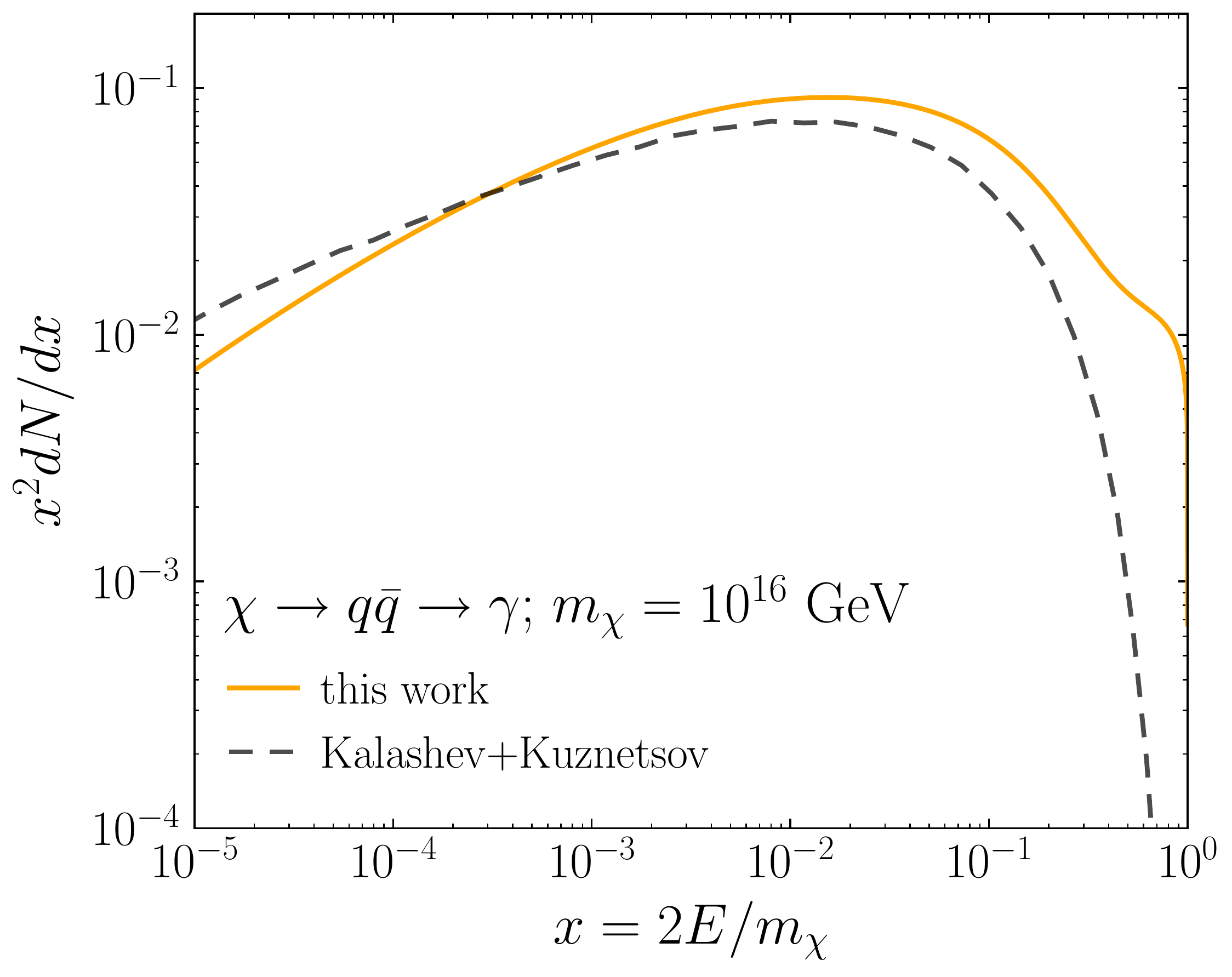}
\end{center}
\vspace{-0.5cm}
\caption{A comparison between the results of the present work and alternative approaches to calculating heavy DM spectra.
The top left figure is an analogue of Fig.~\ref{fig:PythiaComparison} for the photon spectrum, but for a DM mass of $m_{\chi}=2$ TeV.
Even at this lower mass, the additional electroweak effects in our calculation are producing clear differences with \texttt{Pythia}.
We also demonstrate that our spectra differ from those in \texttt{PPPC4DMID}, a point further discussed in the text.
The disagreement is not generic, however, and instead restricted to scenarios where the initial decay involves neutrinos.
We show a single example in the top right, showing significantly improved agreement, and in general for these other channels the agreement is within the uncertainty we have ascribed to our results.
On the bottom we show photon spectra for GUT scale DM ($m_{\chi} = 10^{16}$ GeV) decaying to light quarks, $q=(u+d+s)/3$.
Here our spectrum is compared to results obtained using pure QCD DGLAP of a low energy fragmentation function, taken from Fig. 1 in \Refc{Kalashev:2017ijd}.
Even for this hadronic channel, there is a clear difference in the hard photon contribution associated with electroweak evolution effects.}
\label{fig:Comparison}
\end{figure}

\subsection{Comparison to Existing Approaches}
\label{sec:Comparison}

At present, experimental collaborations are primarily using either \texttt{Pythia} or PPPC4DMID to determine the DM decay spectra at masses above the electroweak scale.
Nonetheless, a number of approaches to calculating these results have been presented in the literature.
Having just expanded upon the basic underpinnings of our approach, here we summarise several alternatives proposed in the literature, highlighting where we differ.

Our first comparison to existing results appeared already in Fig.~\ref{fig:PythiaComparison}.
As emphasized there, the dramatic difference observed between \texttt{Pythia} and our work is driven by the absence of the full electroweak interactions in the former; in particular, \texttt{Pythia} includes $W$ and $Z$ emission off fermions~\cite{Christiansen:2014kba}, but not the electroweak triple gauge couplings $WWZ$ and $WW \gamma$.
These terms are central in the development of electroweak showers at higher energies, and the spectra of particles they produce.
Even at lower masses these differences can be important.
This is shown on the top left of Fig.~\ref{fig:Comparison}: electroweak effects have a visible impact for the hard photon spectrum. 
In particular, we see that \texttt{Pythia} does not predict a hard photon contribution, which arises from polarization effects it does not include.
There we also compare our results to PPPC4DMID,\footnote{The version of PPPC4DMID we used was downloaded in August 2019.} which augmented an earlier version of \texttt{Pythia} with leading order electroweak corrections, and still clear differences are observed.
There are several possible sources for this disagreement, although at present we cannot isolate its exact origin.
Primarily, at TeV-scales, finite electroweak masses become important, and while these were not included in our results (see App.~\ref{sec:Limitations}), they were in PPPC4DMID.
However, we note that the simple expectation would be that including electroweak masses, would suppress the development of electroweak showers and the emission they generate.
This is not realized in Fig.~\ref{fig:Comparison}, where instead the peak of the PPPC4DMID spectrum is roughly a factor of two larger.
We cannot resolve this difference without augmenting our calculation with fixed order corrections, which is beyond the scope of the present work.
For this reason, at TeV-scale masses we do not claim our result is the more accurate, and instead suggest that the difference between the two results can be used to estimate the theoretical uncertainty on the spectra.
The situation will ultimately be resolved by the addition of fixed-order corrections to our calculation.
In spite of the pronounced differences visible in this figure, for non-neutrino channels we find agreement, within the theoretical uncertainties on our spectra, with \texttt{Pythia} and PPPC4DMID at mass scales below 10 TeV, a single example of which is shown on the top right of Fig.~\ref{fig:Comparison}.

In the bottom of Fig.~\ref{fig:Comparison}, we contrast our results to an approach similar in spirit to that taken in the present work.
The idea is to take a FF measured experimentally at low energies (rather than relying on \texttt{Pythia}), and DGLAP evolve this to the desired DM mass.
This approach was put forward in \Refcs{Berezinsky:2000up,Sarkar:2001se,Berezinsky:2002hq,Aloisio:2003xj}, where the DGLAP evolution used either pure QCD or its supersymmetric analogue.\footnote{\Refcs{Barbot:2002ep,Barbot:2002gt} went even further, producing results by evolving low energy FFs including not only QCD, but also electroweak evolution, and further the full splittings in the minimal supersymmetric SM.
The additional SUSY splittings makes a detailed comparison impossible, although a number of features of their results are qualitatively similar to ours.}
These methods have been taken up more recently by the authors of \Refcs{Kalashev:2008dh,Kalashev:2016cre,Kuznetsov:2016fjt,Kalashev:2017ijd,Kachelriess:2018rty,Kalashev:2019xkw,Kalashev:2020hqc} in order to set strong constraints on heavy DM.
For hadronic channels, those works disregard electroweak interactions in the evolution, whereas our formalism includes these effects for all initial states.
A comparison for light quarks is shown in Fig.~\ref{fig:Comparison} to a spectrum taken from \Refc{Kalashev:2017ijd}.
The results are qualitatively similar, although with clear differences.
In particular, the pure QCD evolution misses the hardest photons that result from the electroweak shower, and further there are differences at lower-$x$ likely associated with our inclusion of soft-coherence.
A related approach was taken in \Refc{Ishiwata:2019aet}, where the authors used a hybrid approach of evolving the hadron FFs with QCD DGLAP, and decaying the particles with~\texttt{Pythia}.
As that work did not include electroweak effects, the differences to our results are similar to those shown on the bottom of Fig.~\ref{fig:Comparison}.

\section{Details of the High-Scale Evolution}
\label{sec:HighScale}

Here we expand upon the high-scale evolution from $\mu \sim m_{\chi}$ to $\qW$, describing both the full DGLAP calculation, and our treatment of soft coherence.

\subsection{Review of DGLAP Evolution in the Unbroken Standard Model}
\label{sec:SMFF}

For the DM masses considered in the present work, the starting point for our evolution is a scale far above electroweak symmetry breaking, $\mu \gg \qW$, where the SM can be accurately described by an unbroken ${\rm SU(3)} \times  {\rm SU(2)}_L \times  {\rm U(1)}_Y$ gauge theory.
One can therefore treat the electroweak gauge bosons, as well as all fermions as massless degrees of freedom.
DGLAP evolution in this theory can therefore be used to evolve fragmentation functions
\be
d_i^k(x;\, Q, \mu) = x D_i^k(x;\, Q, \mu)\,,
\ee
where $D_i^k(x;\, Q, \mu)$ gives the distribution at the scale $\mu$ of the momentum fraction $x$ for particle species $k$ in a shower initiated by a parton $i$ (labeled by both type and helicity) produced in a hard process at momentum scale $Q$, and $d_i^k(x;\, Q, \mu)$ denotes the corresponding momentum weighted fragmentation function.
The DGLAP equations take the standard form
\bea
Q\frac{\pd}{\pd Q} d^k_i(x;\, Q, \mu) 
= \sum_I  \frac{\alpha_{I}(Q)}{\pi} &\left[  P^V_{i,I}(Q) \, d^k_i(x;\, Q, \mu) \vphantom{\int_x^{z_{\rm max}^{ji,I}(Q)}} \right. \\
&\left.+ \sum_j  C_{ji,I} 
\int_x^{z_{\rm max}^{ji,I}(Q)} \!\!\! d z \, P^R_{ji, I}(z)\, d^k_j(x/z;\, Q, \mu) \right],
\label{eq:genevol}
\eea
where the two terms in brackets correspond to the virtual and real contributions.
This result can be viewed as the more complete version of the schematic form presented in \Eq{eq:SchematicDGLAP}.
The details of how to solve these evolution equations was presented in~\Refc{Bauer:2018xag}.
We repeat only the salient points and refer the reader to the original work for details.

The sum over $I$ in \Eq{eq:genevol} runs over the different interactions in the SM, and we denote by $I = 1, 2, 3$ the pure ${\rm U(1)}_Y$, ${\rm SU(2)}_L$ and ${\rm SU(3)}$ gauge interactions, and by $I =Y$ the Yukawa interactions.
Besides these contributions to the evolution, there is also a mixed interaction, denoted by $I = M$.
This originates from an interference contributions, where the particle $i$ originates from a ${\rm U(1)}_Y$ gauge boson $B$ in the amplitude and the $W_3$ of the ${\rm SU(2)}_L$ gauge group in the complex conjugate amplitude (or vice versa). 
The coupling of the mixed interaction is therefore proportional to
\be
\alpha_M(Q) = \sqrt{\alpha_1(Q)\, \alpha_2(Q)}\,.
\ee
The maximum cutoff on $z$ in the integration of the real radiation is dependent upon both the splitting and interaction type.
We define
\be
z_{\rm max}^{ji,I}(Q) = \left\{
\begin{array}{ll}
1 - \frac{\mV}{Q} & {\rm for}\, I = 1, 2, \,{\rm and}\, i, j \notin V \,{\rm or}\, i, j \in V\,,
\\
1 & {\rm otherwise}\,,
\end{array}
\right.
\label{eq:zmax}
\ee
where $V$ is the set of vector bosons.
This prescription ensures that an infrared cutoff $\mV$, of the order of the electroweak scale, is applied when a $B$ or $W$ boson is emitted.
To evolve in the full (unbroken) SM, one needs to differentiate the two chiralities of the fermions, the two transverse polarizations of the gauge bosons, the mixed $B/W_3$ state, and include all 4 components of the complex Higgs field (instead of the longitudinal polarizations of the heavy gauge bosons).
The complete set of states required is summarized in Table~\ref{tab:FFcounts}.

\begin{table}[t]
\begin{center}
\begin{tabular}{|c|C{1.2cm}C{1.2cm}C{1.2cm}|C{1.2cm}|}
\hline
\backslashbox{$i$}{$k$} & $f$ & $V$ & $H$& sum \\\hline
$f$ & $42 \times 26$ & $42 \times 11$ & $42$ & $42 \times 38$ \\
$g_\pm$ & $2 \times 26$ & $2 \times 11$ & $2$ & $2 \times 38$ \\
$W_\pm^\pm$ & $4 \times 26$ & $4 \times 11$ & $4$ & $4 \times 38$ \\
$X_\pm^0$ & $6 \times 26$ & $6 \times 11$ &  $6$ & $6 \times 38$ \\
$H^\pm$ & $2 \times 26$ & $2 \times 11$ & $2$ & $2 \times 38$ \\
$H^0$ & $2 \times 26$ & $2 \times 11$ &  $2$ & $2 \times 38$ \\
\hline
sum & $58 \times 26$ & $58\times 11$ & $58$ & $58 \times 38$ \\
\hline
\end{tabular}
\end{center}
\vspace{-0.4cm}
\caption{Total number of FFs, $d_i^k$, required in our high scale evolution.
Here $i$ labels the states required at the high scale $Q \sim m_{\chi}$.
As we work in the full unbroken standard model, $i$ can take 58 values: 42 fermions ($2 \times 3 \times \{e_L,e_R,\nu_L,u_L,u_R,d_L,d_R\}$ for particle/antiparticle and three generations), 2 gluons (helicities), 4 charged electroweak bosons (two helicities and two charges), 6 neutral electroweak bosons (two helicities for each of $B$, $W^3$, and the mixture, collectively labeled $X^0$), and the four degrees of freedom of the SU(2)$_L$ Higgs doublet.
For $k$ at the weak scale, the counting is similar although slightly rearranged.
We now have 38 states as we only distinguish the helicity of the electroweak states (as this information is used in the weak matching), so we now have 26 fermions ($2 \times 3 \times \{e,\nu,d\}+2 \times \{u, c\}+2 \times \{t_L,t_R\}$), 11 vectors ($\{g,\gamma\}+3\times\{W^{\pm},Z\}$ including the three polarizations of the massive bosons), and 1 physical Higgs.
Thus, in general $58 \times 38 = 2204$ FFs are required.}
\label{tab:FFcounts}
\end{table}

Contrary to FF evolution in the strong sector, where the DGLAP equations only give rise to single logarithmic terms, the evolution in the full SM gives rise to double logarithmic sensitivity as well. 
Double logarithmic contributions arise from the limit where radiated particle are simultaneously soft and collinear relative to the particle they were emitted from.
In the strong interaction, these simultaneously soft and collinear contributions cancel between the virtual and real terms in the DGLAP equations.
This occurs as an arbitrarily soft emission of a gluon cannot be observed experimentally, so the divergence associated with this emission must cancel against the virtual contribution.
This is different from the case of the soft emission of a $W$ boson, which can always be observed through the change of flavor (or SU(2) quantum numbers) of the emitting particle.
Thus, as long as a process is sensitive to the SU(2) quantum numbers of the external states, soft radiation of $W$ bosons from these particles leads to an incomplete cancellation of the soft and collinear divergences, which gives rise to double logarithms.
The form of the soft boson cutoff in \Eq{eq:zmax} ensures that these double logarithms have the correct coefficients~\cite{Ciafaloni:2000df,Manohar:2018kfx}.

As discussed in \Refc{Bauer:2018xag}, the set of evolution equations can be decoupled to some degree by switching to a basis of well-defined isospin $\mathbf{T}$ and CP.
The definition of all FFs in this new basis was given in \Refc{Bauer:2018xag}.
To provide examples, for left-handed fermions one can write in a basis $d_i^{\mathbf{T}\mathrm{CP}}$,
\bea
d^{0\pm}_{f_L} &= \frac
                 14\left[\left(d_{u_L}+d_{d_L}\right)\pm\left(d_{{\bar
                 u}_L}+d_{{\bar d}_L}\right)\right],\\
d^{1\pm}_{f_L} &= \frac
                 14\left[\left(d_{u_L}-d_{d_L}\right)\pm\left(d_{{\bar
                 u}_L}-d_{{\bar d}_L}\right)\right],
\label{eq:fLIsospin}
\eea
while for the ${\rm SU(2)}$ bosons we have
\bea
d^{0\pm}_W=\,& \frac 13\left[\left(d_{W_+^+}+d_{W_+^-}+d_{W_+^3}\right)
\pm\left(d_{W_-^+}+d_{W_-^-}+d_{W_-^3}\right)\right],\\
d^{1\pm}_W =\,& \frac 12\left[\left(d_{W_+^+}-d_{W_+^-}\right)
\mp\left(d_{W_-^+}-d_{W_-^-}\right)\right],\\
d^{2\pm}_W=\,& \frac 16\left[\left(d_{W_+^+}+d_{W_+^-}-2d_{W_+^3}\right)
\pm\left(d_{W_-^+}+d_{W_-^-}-2d_{W_-^3}\right)\right].
\eea

Using this isospin basis allows to isolate the double logarithmic dependence.
In particular, for isosinglets (with $\mathbf{T} = 0$) there cannot be any double logarithms generated, since the emission of an isosinglet cannot change the isospin of the emitting particle.
In general, the double logarithmic term is given by a Sudakov factor, which depends on the total isospin, and takes the form
\be
\Delta^{(\mathbf{T})}(Q)\sim \exp\left[-\mathbf{T}(\mathbf{T}+1)\frac{\alpha_2}{2\pi}\ln^2\left(\frac{Q}{\qW}\right)\right].
\label{eq:DeltaTq}
\ee
One can then show that the rescaled FF
\be
\tilde d^{\,\mathbf{T}{\rm CP}}_i(x;\, Q, \mu) = \frac{d^{\,\mathbf{T}{\rm CP}}_i(x;\, Q, \mu)}{\Delta^{(\mathbf{T})}(Q)}\,,
\ee
has only standard single logarithmic evolution.

As mentioned in the main body, rather than perform the DGLAP evolution from $\mu \sim Q = m_{\chi}/2$ down to $\mu \sim \qW$, we instead start at the electroweak scale and evolve upwards.\footnote{This is done as the DGLAP equations have similar properties to the diffusion equation, where $\ln Q$ plays the role of time.
As with diffusion, the evolution can be solved with far greater stability if evolved towards larger time, or here $Q$.}
In detail we start with $\qW = 100$ GeV as the starting point for our evolution.
For quarks and leptons ($k=f$), assuming that the helicity of the fragmentation product is not detected, we take as input
\be
d_{f_L}^f(x;\,\qW,\qW)=d_{f_R}^f(x;\,\qW,\qW)=\delta(1-x)\,,
\ee
setting all other initial FFs to zero.  
The only exception to this is the top quark, where in order to correctly account for its decay at the electroweak scale, we evolve $t_L$ and $t_R$ separately.
Similarly, as neutrinos ($k=\nu$) have no right-handed states, the non-zero initial conditions becomes
\be
d_{\nu_L}^\nu(x;\qW,\qW)=\delta(1-x)\,.
\ee
For fragmentation into a gauge boson $V$ we keep track of helicity (as we did for the top), because for the electroweak scale vectors we will use this information in the matching.
Accordingly, the non-zero initial conditions are
\be
d^V_{V_+}(x;\qW,\qW)=\delta(1-x)\,,\;\;\;{\rm or}\;\;\;d^V_{V_-}(x;\qW,\qW)=\delta(1-x)\,.
\ee

\begin{figure}[t]
\leavevmode
\vspace{-0.2cm}
\begin{center}
\includegraphics[width=.47\textwidth]{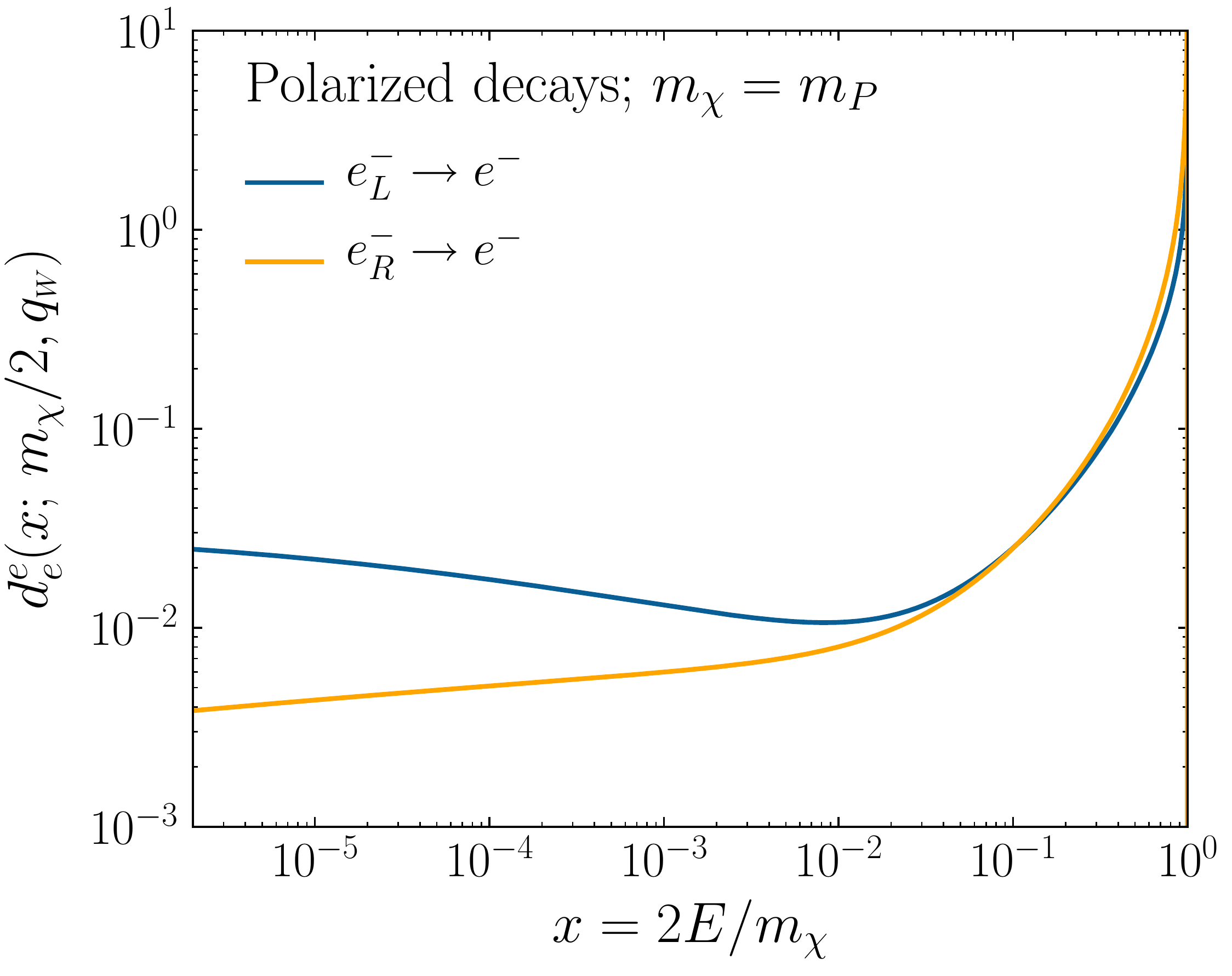}\hspace{0.1cm}
\includegraphics[width=.47\textwidth]{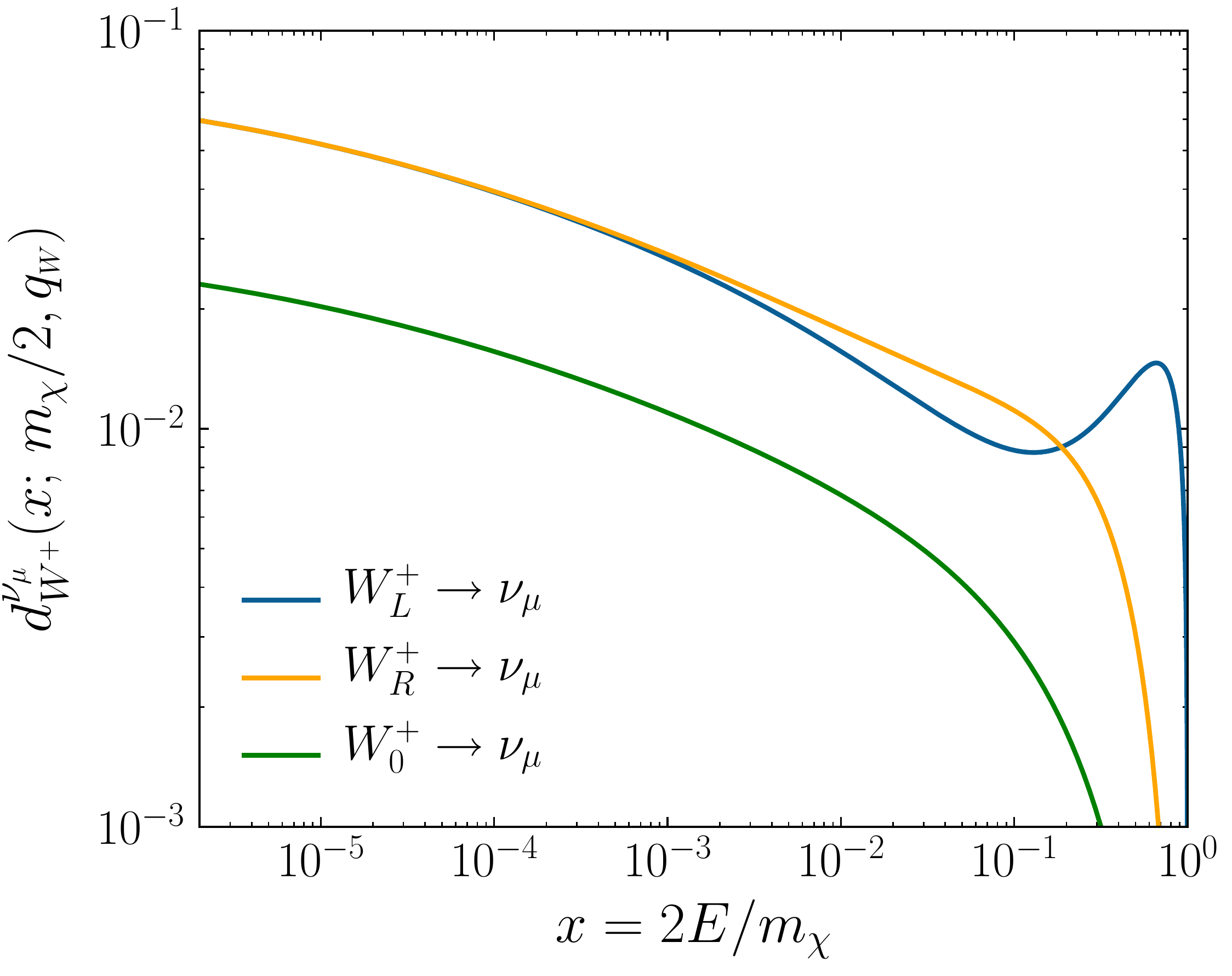}
\end{center}
\vspace{-0.5cm}
\caption{Example FFs obtained using the DGLAP evolution and soft-coherence correction outlined in App.~\ref{sec:HighScale}.
In this figure we have chosen to highlight the generation of polarization due to the chiral nature of the SM.
In particular, on the left we see that left and right handed electrons evolve to produce a significantly different distribution of $e = e_L + e_R$.
Similarly, on the right we see that all three polarizations of the $W$ evolve differently.
In both cases the evolution is performed for a Planck scale mass DM particle down to the weak scale.
}
\label{fig:Polarization}
\end{figure}

Ultimately, the DGLAP evolution will provide expressions for 
\be
D_i^k(x;\, Q, \qW^+) \qquad {\rm with} \qquad Q > \qW
\,.
\ee
Several example outputs at this stage are shown in Fig.~\ref{fig:Polarization} (including soft coherence), where we have chosen to highlight the generation of polarization effects through the evolution.
In particular, due to the chiral nature of the SM, the chiral fermion states and boson polarizations do not evolve identically.
See \Refc{Bauer:2018arx} for an extended discussion of this effect.
Because we are neglecting any SU(2) breaking effects in this evolution, one expects a break down in this description for $Q \sim \qW$.
We discuss this in more detail in App.~\ref{sec:Limitations}.
To continue the evolution below $\qW$ we will first need to remove particles with masses $m \sim \qW$, and we will describe how to do so in App.~\ref{sec:Matching}.
Before doing so, however, we next outline how to incorporate a partial treatment of soft coherence effects into the DGLAP evolution.

\subsection{Incorporating the Soft-Coherence of Real Radiation}
\label{sec:Smallx}

At low energy fractions the simple DGLAP evolution described above is not reliable: there are large double logarithms of $x$ that need to be resummed.
The physical origin of these logarithms is the soft-coherence or angular-ordering effect described in the main body.
In this section we will demonstrate that taking the output of our DGLAP evolution and applying the substitution $m_{\chi} \to x\, m_{\chi}$ incorporates a partial treatment of these effects.
This treatment is manifestly incomplete: in particular no accounting for the virtual effects of soft-coherence will be included.
One manifestation of this shortcoming will be incomplete momentum sums, which we discuss in App.~\ref{sec:AdditionalResults}.
A full treatment of these effects is left as an open problem.

To begin with, consider the soft coherence of gluons in the context of QCD, as reviewed in~\Refc{Bassetto:1984ik}.
The resummation of the gluon-to-gluon FF at small $x$ is given in leading-logarithmic approximation (LLA) by Eq.~(5.21) of~\cite{Bassetto:1984ik} as\footnote{We emphasize that the single scale appearing in the FF in \Eq{eq:BCM} is the high scale -- the low scale, $\qW$, has been suppressed.
This is not the same as \Eq{eq:SchematicDGLAP}, where the single scale that appeared there, $\mu$, was instead the low scale.}
\be
xD_g(x;\,Q)= \delta(1-x)+\sum_{n=1}^\infty\frac{(C_A\alpha_3/\pi)^n}{n!(n-1)!}
\left(\ln\frac{Q^2x^2}{Q_0^2}\right)^n \left(\ln\frac 1x\right)^{n-1}.
\label{eq:BCM}
\ee
This result provides a FF with soft-coherence included, at least as far as it impacts real radiation.
As we will now show, this FF can be determined as the solution to the DGLAP equation, but in $q=xQ$ rather than $Q$.

In order to expose this, first we introduce $xD_g(x;\,Q) = x \bar{D}_g(x;\,q) \equiv d_g(x;\,q)$ -- we will find that $\bar{D}_g$ satisfies the unmodified DGLAP equations.
Introducing this notation to rewrite \Eq{eq:BCM}, and the substitution $Q=q/x$, after differentiating with respect to $\ln q^2$, we find
\bea
q^2\frac{\pd}{\pd q^2} d_g(x;\,q) =\,& \sum_{n=1}^\infty\frac{(C_A\alpha_3/\pi)^n}{[(n-1)!]^2}
\left(\ln\frac{q^2}{Q_0^2}\right)^{n-1} \left(\ln\frac{1}{x}\right)^{n-1} \\
=\,& \frac{C_A \alpha_3}{\pi} + \sum_{n=1}^\infty\frac{(C_A\alpha_3/\pi)^{n+1}}{[n!]^2}
\left(\ln\frac{q^2}{Q_0^2}\right)^n\left(\ln\frac{1}{x}\right)^n.
\eea
We can rewrite this using the identity
\be
\frac{1}{n!}\left(\ln\frac{1}{x}\right)^n = \frac{1}{(n-1)!}\int_x^1\frac{dz}{z} \left(\ln\frac {z}{x}\right)^{n-1},
\ee
which holds for $n>0$.
Accordingly,
\bea
q^2\frac{\pd}{\pd q^2} d_g(x;\,q)
=\,& \frac{C_A \alpha_3}{\pi} \int_x^1 \frac{dz}{z}\, \left[ \delta(1-x/z) + \sum_{n=1}^\infty\frac{(C_A\alpha_3/\pi)^n}{n!(n-1)!}
\left(\ln\frac{q^2}{Q_0^2}\right)^n \left(\ln\frac {z}{x}\right)^{n-1} \right] \\
=\,& \frac{C_A \alpha_3}{\pi} \int_x^1 \frac{dz}{z} d_g(x/z;\,q) \\
=\,& \frac{\alpha_3}{2\pi} \int_x^1 dz\, P_{gg}(z)\, d_g(x/z;\,q)\,.
\eea
In the second step we used \Eq{eq:BCM}, and in the final step we used the fact that in the small-$x$ limit $P_{gg}(x) = 2 C_A/x$.
Accordingly, $d_g(x;\,q)$ satisfies the DGLAP equation to LLA.
Therefore to obtain the correct small-$x$ gluon FF we should solve the DGLAP equations for evolution in $q$ and then set $q=xQ$ where $Q$ is the hard process scale.
At large $x$, and for those terms in the evolution equations that do not give rise to extra small-$x$ logarithms (i.e. those without a $1/z$ singularity in the splitting function), this gives rise to unenhanced NLO contributions, which are in any event beyond the precision of our treatment.
Therefore the same procedure can be applied to include the small-$x$ suppression of all the QCD FFs.
Note for $x < \qW/Q$ the substitution samples $D_g(x;\,Q)$ for $Q < \qW$.
As we begin our evolution at $\qW$, these results simply vanish, and thus at the high scale there is an artificial cut in the distribution at small-$x$.
This effect is washed out after convolution with weak matching and \texttt{Pythia}, but is unphysical and a manifestation of our incomplete treatment of soft coherence.

The same large double logarithms of $x$ arise whenever a sequence of emissions with $1/z$ singularities in the splitting functions can occur, which is also the case for $W$ boson fragmentation in the unbroken SM.
We expect a similar small-$x$ behavior of the SU(2) evolution equations, with $\alpha_3$ replaced by the appropriate gauge coupling $\alpha_2$.
There is, however, an additional complication in this case, namely the double-logarithmic evolution of FFs with non-zero weak isospin.
As discussed above, these FFs are suppressed by the Sudakov factor in \Eq{eq:DeltaTq}, due to a mismatch of real and virtual contributions, which is not
relevant to small $x$.
As such, the substitution $q=xQ$ does not apply to the isospin suppression factor, and the formula for small-$x$ resummation becomes
\be
D^{\mathbf{T}}(x,Q) = \tilde{D}^{\mathbf{T}}(x,xQ)\,\Delta^{(\mathbf{T})}(Q) = \bar{D}^{\mathbf{T}}(x,xQ)
\frac{\Delta^{(\mathbf{T})}(Q)}{\Delta^{(\mathbf{T})}(xQ)}.
\ee
We apply this prescription to the FFs above the EW breaking scale, $D_h^{h'}(x;\,q>\qW^+)$, before the matching corrections discussed below.
The boundary conditions for $D$ and $\bar D$ at $\qW^+$ are identical, as they are proportional to $\delta(1-x)$.
We do not apply the prescription to the FFs at lower scales $q<\qW$ provided by \texttt{Pythia}, since these already take coherent emission into account and are tuned to experimental data at such scales.

\begin{figure}[t]
\leavevmode
\vspace{-0.2cm}
\begin{center}
\includegraphics[width=.47\textwidth]{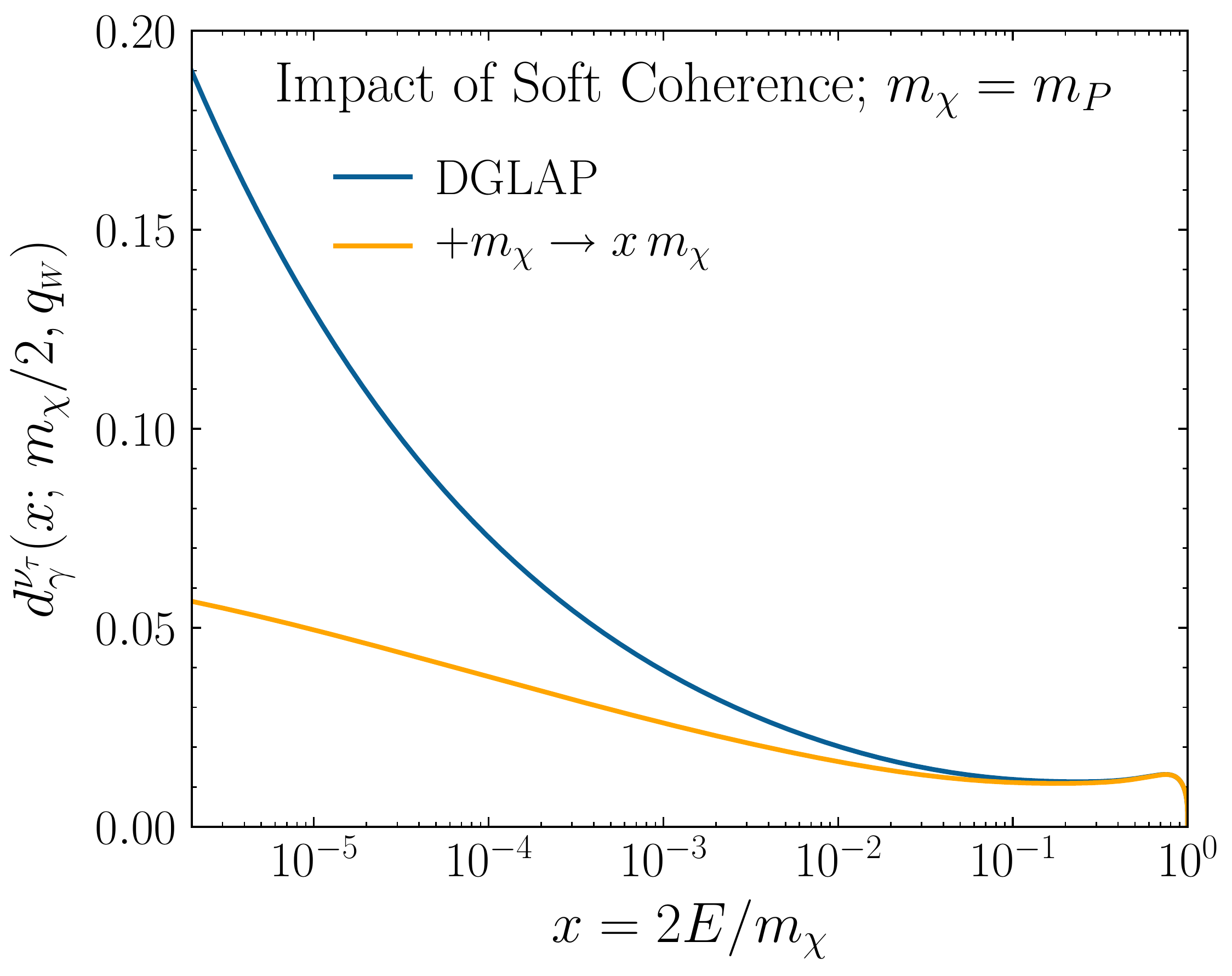}
\end{center}
\vspace{-0.5cm}
\caption{An example of the impact of the $m_{\chi} \to x\, m_{\chi}$ substitution on our high scale FFs.
This particular example shows the evolution of a Planck scale $\gamma$ to an electroweak scale $\nu_{\tau}$ with and without the substitution.
}
\label{fig:SoftCoherence}
\end{figure}

The above expands upon the $m_{\chi} \to x\, m_{\chi}$ substitution mentioned in the main text.
In practice, this effect generically suppresses small-$x$ contributions to FFs, as the states that the suppressed bosons would have split into are also removed.
An example is given in Fig.~\ref{fig:SoftCoherence} for the example of a photon evolving to a tau neutrino.

\section{Matching at the Electroweak Scale}
\label{sec:Matching}

Evolution through the electroweak scale is handled analytically.
As already emphasized, the chiral nature of the SM ensures that quite generically the spectra of states resulting from the high scale evolution will be significantly polarized.
Through a matching procedure outlined in this section, we ensure the polarization information of the electroweak states is not discarded.
In detail, we will compute FFs
\be
d_i^j(x;\,\qW^+,\qW^-)\,,
\ee
where $\qW^{\pm} = \qW(1\pm \epsilon)$, i.e. we evolve the states through a threshold of differential width at the electroweak scale.
We choose $\epsilon \ll 1$ in order to ensure no large logarithms associated with the evolution can be generated at this step.

For states that do not have electroweak masses, the threshold is uneventful.
For fermions, we combine chiralities at this stage.
For example, we only track the evolution of $e$ below $\qW$, whereas $e_L$ and $e_R$ are evolved separately at the high scale.
For a fermion $f$ we take\footnote{As $x \delta(1-x) = \delta(1-x)$, an identical equation to \Eq{eq:fermionmatching} holds with $d(x) \to D(x)$.}
\be
d_{f_{L/R}}^j(x;\,\qW^+,\qW^-) = \delta^j_f\, \delta(1-x)\,.
\label{eq:fermionmatching}
\ee
Similarly, the helicities of the photon and gluon are combined into a single unpolarized state.
Note this procedure washes out any remaining information about the polarization: we explicitly assume that the experiments are searching for unpolarized states.
Even then, this assumption is associated with an imprecision in that muon and tau decays do depend on helicity.
Extending our formalism to evolve the polarization down to the scale of leptonic decays is left to future work.

For the Higgs, $Z$, $W$, and top, $\qW$ marks the end of their evolution.
In the following subsections we outline how their momentum is redistributed amongst their decay products.
For each, the strategy is as follows.
We begin by calculating the differential energy and, where relevant, angular spectrum of the decay products for all polarizations or spins in the rest frame of the electroweak state.
At this stage we account for all electroweak scale masses, although particles that will continue their evolution below $\qW$ are left massless.
Then, in order to match these results onto the high scale distributions where all particles were treated as massless, we perform an infinite boost, i.e. boosting the electroweak states to an energy $E \gg \qW$.
At this stage angular differences in the rest frame are transformed into energy differences in the boosted frame, indicating why this information was retained.
The distribution of the energy fractions amongst the boosted decay products then gives us exactly $D(x)$.
We now implement this procedure case by case.

\subsection{Higgs Decays}

As a scalar, the Higgs carries no polarization information: there are no initial polarizations or spins to account for.
For this reason, we can extract the $D(x)$ simply by generating the spectrum of boosted Higgs decay products in \texttt{Pythia}.
To do so, we generate $e^+ e^- \to HH$ events at $\sqrt{s} = 200$ TeV.\footnote{At this energy, the minimum energy fraction for two-body decay product is $x \sim \mH^2/4\EH^2 \lesssim 10^{-6}$, and therefore below the smallest values considered in this work.}
We forbid initial and final state showering of any kind, and turn off hadronization.
The only states we allow to decay are the $W$ and $Z$.
The energy distribution of leptons, neutrino, quarks, gluons, and photons associated with each Higgs is collected, and used to form
\be
d_h^j(x;\,\qW^+,\qW^-)\,.
\ee

\subsection{$Z$ Decays}

We begin by computing the differential decay spectrum of $f$ in $Z \to f \bar{f}$, working in the $Z$ rest frame.
Here, and throughout, we work in unitary gauge.
In order to establish our conventions, the coupling between $f$ and $Z$ is determined by
\be
\mathcal{L} \supset Z_{\mu} \frac{\gW}{\cW} \bar{f} \gamma^{\mu} (c_L P_L + c_R P_R) f\,.
\ee
Here $c_L = \IW^3 - Q \sW^2$ and $c_R = - Q \sW^2$, with $\IW^3$ weak isospin and $Q$ electric charge, whereas $\cW$ and $\sW$ are the cosine and sine of the Weinberg angle, respectively.
In the broken phase of the SM, for each generation, $\nu$ and $u$ carry $\IW^3=1/2$, whilst $e$ and $d$ have $\IW^3=-1/2$.
The charges are, of course, $Q=0,-1,2/3,-1/3$ for $\nu$, $e$, $u$, and $d$.
For the moment, we will perform the calculation for the case of a single $f$ with arbitrary couplings, and then we can ultimately weight this result by the appropriate branching fractions for the specific states in the SM.
Continuing with our conventions, we define our coordinates such that the $f \bar{f}$ are produced in the $x$-$z$ plane, with $f$ produced at an angle $\theta$ from the $z$-axis.
When we eventually boost the $Z$, we will do so in the $z$ direction, and accordingly in the rest frame we choose our polarization vectors as follows,
\be
\epsilon_{\pm}^{\mu} = \frac{1}{\sqrt{2}} (0,1,\pm i,0)\,,\hspace{0.5cm}
\epsilon_{0}^{\mu} = (0,0,0,1)\,.
\ee

In terms of these quantities, we can now compute $\Gamma(Z_{\pm 0} \to f \bar{f})$, and thereby determine the following distribution of angles,
\bea
\frac{1}{\Gamma} \frac{d \Gamma}{d\cos \theta} (Z_+ \to f \bar{f}) &= \frac{3}{8} \frac{c_L^2(1-\cos \theta)^2+c_R^2(1+\cos \theta)^2}{c_L^2+c_R^2}\,, \\
\frac{1}{\Gamma} \frac{d \Gamma}{d\cos \theta} (Z_- \to f \bar{f}) &= \frac{3}{8} \frac{c_L^2(1+\cos \theta)^2+c_R^2(1-\cos \theta)^2}{c_L^2+c_R^2}\,, \\
\frac{1}{\Gamma} \frac{d \Gamma}{d\cos \theta} (Z_0 \to f \bar{f}) &= \frac{3}{4} (1-\cos^2 \theta)\,.
\label{eq:ZdecayDists}
\eea
Observe that each expression is a normalized probability distribution for $\cos \theta$, and thus we will label each of these expressions as $p_{\pm 0}^Z(\cos \theta)$, denoting the distribution the $f$ emission angle is drawn from in the $Z$ rest frame, for a given polarization.
Note, these distributions also contain the information on the emission angles of $\bar{f}$.
Defining $\bar{\theta}$ to be the angle $\bar{f}$ makes with the $z$-axis, we have $\bar{p}_{\pm 0}^Z(\cos \bar{\theta}) = p_{\pm 0}^Z(-\cos \theta)$.

Armed with these results, we will now determine the energy distribution of $f$ and $\bar{f}$ in the boosted $Z$ frame.
We wish to boost the $Z$ to energy $\EZ$, propagating along the $+z$ direction, which we can achieve through the boost
\be
\gamma = \frac{\EZ}{\mZ}\,,\hspace{0.5cm}
\beta_z = - \sqrt{1-\frac{\mZ^2}{\EZ^2}}\,.
\ee
After this boost, the energy of $f$, which was $\mZ/2$ in the rest frame, is now
\be
E_f = \frac{\EZ}{2} \left( 1 + \cos \theta \sqrt{1 - \frac{\mZ^2}{\EZ^2}} \right) \approx \frac{\EZ}{2} \left( 1 + \cos \theta \right),
\ee
where in the last step we implemented the large boost approximation.
In this limit, the energy fraction carried by $f$ is given by $x_f = E_f/\EZ = (1+\cos \theta)/2 \in [0,1]$, with $\cos \theta$ drawn from the appropriate distribution given in \Eq{eq:ZdecayDists}.
We then determine the distribution of $x_f$ by the following change of variables,
\be
p(x_f) 
= \int_{-1}^1 d (\cos \theta)\, p_{\pm 0}^Z(\cos \theta)\, \delta \left[ x_f - (1+\cos \theta)/2 \right]
= 2 p_{\pm 0}^Z(2 x_f - 1)\, \Theta \left[x_f (1-x_f) \right]\,,
\ee
where $\Theta$ is the Heaviside step-function.
Similarly,
\be
p(x_{\bar{f}}) 
= 2 p_{\pm 0}^Z(1-2 x_{\bar{f}})\, \Theta \left[x_{\bar{f}} (1-x_{\bar{f}}) \right]\,.
\ee

In terms of these expressions, we can calculate the expected energy fractions for an arbitrary $p(\cos \theta)$ as follows,
\bea
\langle x_f \rangle = \int_0^1 d x_f\, x_f\, p(x_f) = \frac{1}{2} \int_{-1}^1 d(\cos \theta)\,(1+\cos \theta) p(\cos \theta) = \frac{1+\langle \cos \theta \rangle}{2}\,,
\eea
and an identical calculation yields $\langle x_{\bar{f}} \rangle = (1 - \langle \cos \theta \rangle)/2$.
Accordingly, independent of the exact form of $p(\cos \theta)$, we have $\langle x_f \rangle + \langle x_{\bar{f}} \rangle = 1$, consistent with momentum conservation.
This also indicates that the momentum weighted fragmentation functions should be associated with $x \,p(x)$, and so for the case of a single fermion we have
\bea
d_{Z_{\pm 0}}^f(x;\,\qW^+,\qW^-) = 2x\, p_{\pm 0}^Z(2x-1)\,, \hspace{0.5cm}
d_{Z_{\pm 0}}^{\bar{f}}(x;\,\qW^+,\qW^-) = 2x\, p_{\pm 0}^Z(1-2x)\,,
\eea
where again, the explicit $p_{\pm 0}^Z(\cos \theta)$ are given in \Eq{eq:ZdecayDists}.
In detail,
\bea
d_{Z_+}^f(x;\,\qW^+,\qW^-) &= 3x \left[ \frac{c_L^2(1-x)^2+c_R^2 x^2}{c_L^2 + c_R^2} \right], \\
d_{Z_-}^f(x;\,\qW^+,\qW^-) &= 3x \left[ \frac{c_L^2x^2+c_R^2 (1-x)^2}{c_L^2 + c_R^2} \right], \\
d_{Z_0}^f(x;\,\qW^+,\qW^-) &= 6 (1-x) x^2\,.
\label{eq:dforZ}
\eea
The equivalent results for $\bar{f}$ follow by taking $c_L \leftrightarrow c_R$ in each case.
Observe that for each polarization, we have explicit momentum conservation independent of the values of $c_L$ and $c_R$,
\be
\sum_{j=f,\bar{f}} \int_0^1 dx\, d_{Z_{\pm 0}}^j(x;\,\qW^+,\qW^-) = 1\,.
\ee
We can extend the result immediately to all the states in the SM by weighting these distributions by the appropriate $Z$ branching fractions, and in all cases inserting the appropriate values of $c_L$ and $c_R$.

\subsection{$W$ Decays}

Having determined in detail the spectrum for the $Z$ decay products, the result for $W$ decays follows almost immediately.
As the $W$ couples only to left-handed fermions, we simply take \Eq{eq:dforZ} (and the analogous result for $\bar{f}$) and set $c_L=1$ and $c_R=0$.
Again weighting the different final states by the appropriate branching fractions, this specifies all relevant fragmentation functions.

\subsection{Top Decays}

As the top decay is a three-body process, the distribution of energy fractions amongst the decay products is a more involved calculation than for the states already considered.
We take $|V_{tb}|=1$, considering only $t \to b W$, and to be explicit let us take the decay $W \to f \bar{f}$, so that the full process is
\begin{center}
\includegraphics[width=.25\textwidth]{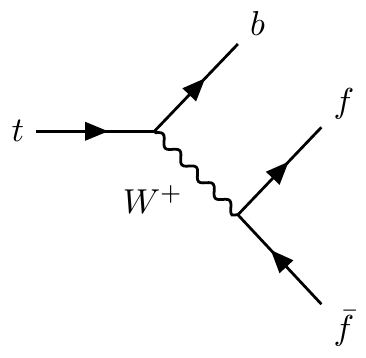}
\end{center}
The translation of the final results to all relevant $W$-decay modes is straightforward, and each channel can then be weighted by the appropriate $W$ branching fraction.

Once again, we will begin the calculation in the top rest-frame.
We measure the spin of the top along the $z$-axis, denoting spin up and down by $t_{\pm}$.
Ultimately we will boost the top in the $z$ direction, so that in the infinite boost limit, spin up and down will be associated with positive and negative helicity, or right and left handed chirality.
In that sense, while the sign of $t_{\pm}$ will represent spin in the first part of the calculation, it will later be translated directly to helicity.

\begin{figure}[t]
\leavevmode
\vspace{-0.2cm}
\begin{center}
\includegraphics[width=.47\textwidth]{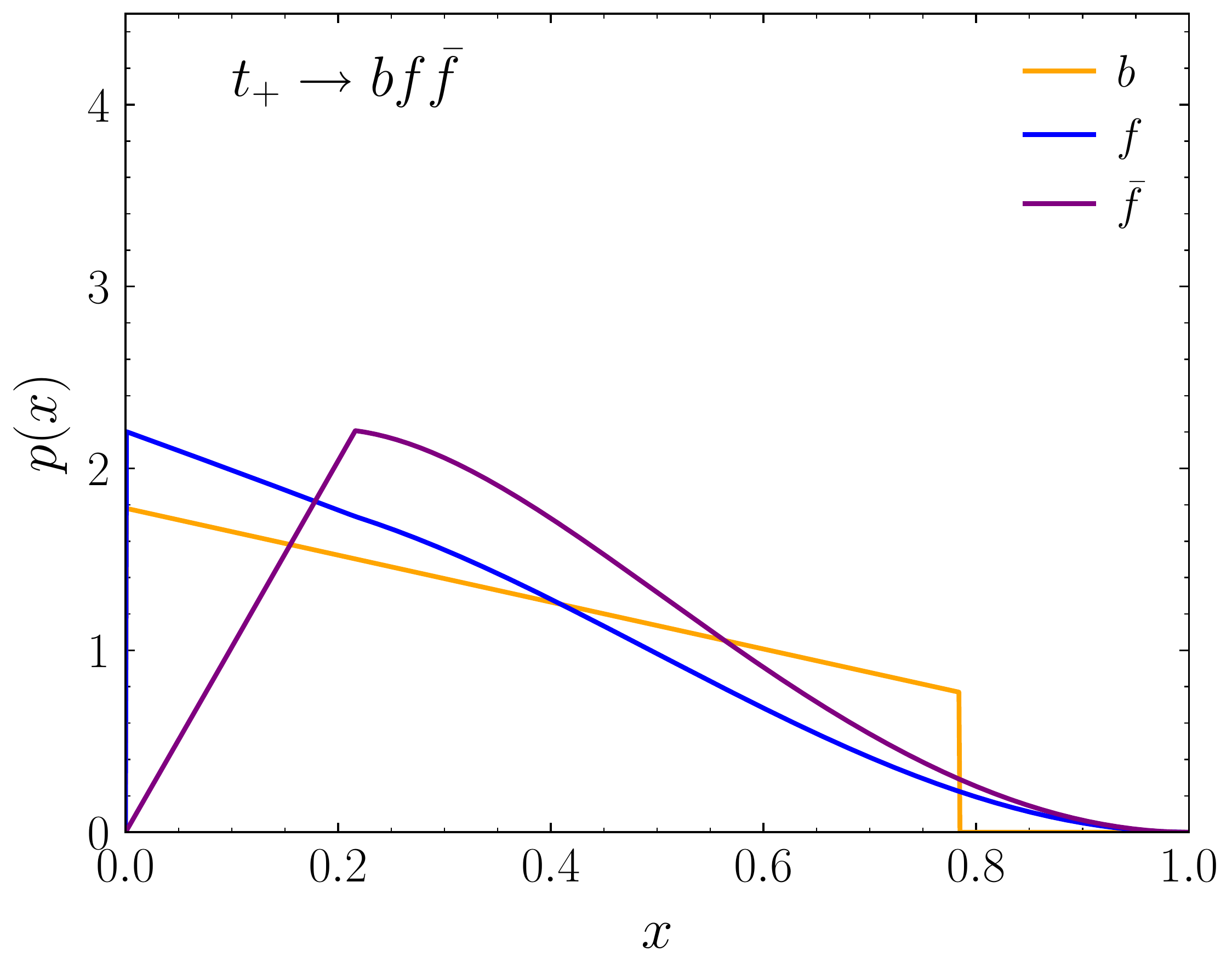}
\hspace{0.5cm}
\includegraphics[width=.47\textwidth]{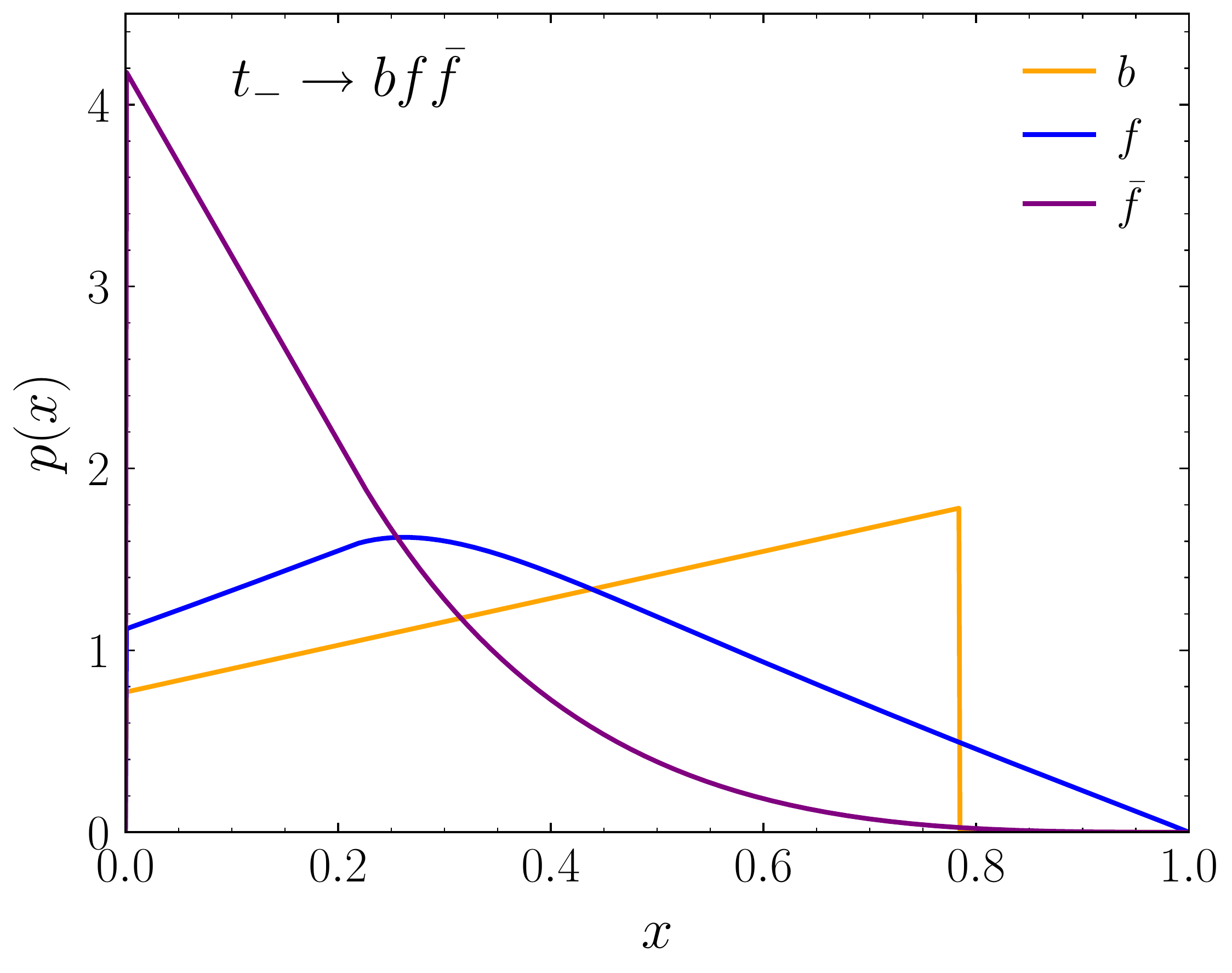}
\end{center}
\vspace{-0.5cm}
\caption{The distribution of energy fractions carried by polarized top-decay products, as described by \Eq{eq:top2b} and \eqref{eq:top2f}.
Here $f \bar{f}$ represent the particle and antiparticle that result from the decay of the intermediate on-shell $W$.
}
\label{fig:PolarizedTop}
\end{figure}

Consider first the spectrum of $b$-quarks produced from $t_{\pm} \to W b$ (ignoring any contribution from $W$ decays at this stage).
The intermediate $W$ is produced on-shell, and thus the calculation proceeds similarly to that of the $Z$.
A key difference, however, is that here and throughout we will retain both the top and $W$ mass, defining the ratio $\rho = \mW^2/m_t^2 \sim 0.2$.
All other fermions, including the $b$-quark, are left massless.
Further, we need to account for the three possible polarizations of the $W$.
Due to the chiral nature of the weak interaction, the resulting $b$-quark must be left handed.
This implies in the limit where $m_b=0$, only a longitudinal and negative helicity $W$ can participate in the process.
The ratio of the branching fractions is
\be
\frac{{\rm Br}(t \to W_0\, b)}{{\rm Br}(t \to W_-\, b)} = \frac{1}{2\rho} \sim 2\,,
\ee
so that the longitudinal polarized $W$s are produced twice as often.
We can use this to gain a rough intuition for the resulting spectrum of $b$-quarks.
For a longitudinal $W$, the left-handed $b$ will be preferentially emitted in the opposite direction to the spin of the top.
After boosting, we then expect a softer spectrum for $t_+$ than $t_-$.
Performing the calculation, we find
\bea
d_{t_+}^b(x;\,\qW^+,\qW^-) &= \frac{2x(1-\rho) + 2x^2(2\rho-1)}{(1+2\rho)(1-\rho)^2}\, \Theta(1-\rho-x)\,, \\
d_{t_-}^b(x;\,\qW^+,\qW^-) &= \frac{4x(1-\rho) \rho - 2x^2(2\rho-1)}{(1+2\rho)(1-\rho)^2}\, \Theta(1-\rho-x)\,.
\label{eq:top2b}
\eea
The step-function enforces the condition $x_b \in [0,1-\rho] \sim [0,0.8]$: the finite $\mW$ ensures the $b$ cannot carry away all the energy.
The distribution of energy fractions is presented in \Fig{fig:PolarizedTop}.

We now turn to the distribution of $W$ decay products.
As the $W$ is produced on shell, we can start in the $W$ rest frame, where the energy spectrum is a two-body $\delta$-function.
Yet, there will be a non-trivial angular dependence associated with the $W$ polarization, which will be converted into energy differences when we boost to the top rest frame, and again when we boost the top itself, at which point we need to account for the angular distribution of the $W$ as a function of boson polarization and fermion spin.
Further, we must account for the fact that as the $W$ is an intermediate state, we will have interference between the different polarizations.

We begin by establishing our coordinates.
In the $W$ rest frame, we represent $p_f$ with a general lightlike four-vector
\be
p_f^{\mu} = \frac{1}{2} \mW (1,\,\sin \theta \cos \phi,\, \sin \theta \sin \phi,\, \cos \theta)\,.
\label{eq:pfWrest}
\ee
The equivalent expression for $p_{\bar{f}}$ can be obtained by sending $\theta \to \pi - \theta$ and $\phi \to \pi + \phi$.
In the top rest frame, we choose coordinates such that the $b$ and $W$ are emitted in the $x$-$z$, with the $W$ at an angle $\theta'$ from the $z$-axis, i.e.
\be
\pW^{\mu} = (\EW,\,\pW \sin \theta',\,0,\, \pW \cos \theta')\,,
\ee
where $\EW = m_t(1+\rho)/2$ and $\pW = m_t(1-\rho)/2$.
In terms of this, to transform $p_f$ to the top rest frame, we boost in the $W$ direction by an amount
\be
\gamma = \frac{\EW}{\mW} = \frac{1+\rho}{2 \sqrt{\rho}}\,,\hspace{0.5cm}
\beta = -\frac{\pW}{\EW} = -\frac{1-\rho}{1+\rho}\,,
\ee
after which \Eq{eq:pfWrest} becomes
\bea
p_f^{\mu} = 
\frac{1}{2} \mW &\big(
\gamma[1+\beta \cos \theta],\,
\gamma[\beta+\cos \theta] \sin \theta' + \cos \theta' \sin \theta \cos \phi, \\
&\hspace{0.1cm}\sin \theta \sin \phi,\,
\gamma[\beta+\cos \theta] \cos \theta' - \sin \theta' \sin \theta \cos \phi
\big).
\label{eq:pftoprest}
\eea

In terms of these coordinates, the fully differential spectrum in the top rest frame is
\bea
\frac{1}{\Gamma} \frac{d\Gamma(t_{\pm} \to b f \bar{f})}{d\phi\, d \cos \theta\, d \cos \theta'}
= \frac{3}{16 \pi} \frac{1+\cos \theta}{1 + 2 \rho} 
&\left[ (1 \pm \cos \theta')(1-\cos \theta) + \rho(1 \mp \cos \theta')(1+\cos \theta) \right.\\
&\left. \pm 2 \sqrt{\rho} \sin \theta' \sin \theta \cos \phi \right].
\eea
For both spins this is a normalized probability distribution for the three relevant angles, which we denote $p_{\pm}^t(\phi,\,\cos \theta,\,\cos \theta')$.
We can now use this to determine the energy fraction carried by $f$ in the boosted top frame.
To determine this, we boost \Eq{eq:pftoprest} in the $-z$ direction by an amount $\gamma = E_t/m_t$ and $\beta \approx 1$, to obtain
\be
x_f = \frac{1}{4} \left( [1+\cos \theta'][1 + \cos \theta] + \rho[1 - \cos \theta'][1 - \cos \theta] - 2 \sqrt{\rho} \sin \theta' \sin \theta \cos \phi \right).
\ee
Note that $x_f \in [0,1]$.
At this stage, we know the energy fraction as a function of the angles, and the distribution from which the angles are drawn, so we can formally write down the relevant FF,
\bea
D_{t_{\pm}}^f(x;\,\qW^+,\qW^-) = &\int d\phi\,d (\cos \theta) d (\cos \theta') p_{\pm}^t(\phi,\,\cos \theta,\,\cos \theta') \\
\times &\delta \left[ x - \frac{1}{4} \left( [1+\cos \theta'][1 + \cos \theta] + \rho[1 - \cos \theta'][1 - \cos \theta] \right. \right. \\
&\left.\left.\hspace{1.6cm}- 2 \sqrt{\rho} \sin \theta' \sin \theta \cos \phi \right) \vphantom{\frac{1}{4}} \right].
\label{eq:top2f}
\eea
The equivalent expression for $\bar{f}$ follows by sending $\theta \to \pi - \theta$ and $\phi \to \pi + \phi$ in the argument of the $\delta$-function.

These expressions can be readily computed numerically, and are depicted in \Fig{fig:PolarizedTop}.
To facilitate rapid evaluation, we further determined piecewise polynomial fitting functions.
Recall that $x_b < 1-\rho$.
Momentum conservation then requires $x_f + x_{\bar{f}} > \rho$, and thus both distributions display a discontinuous derivative at $x=\rho$.
As such, we determine fitting functions of the form
\be
g(x) = \left\{
\begin{array}{ll}
(1-x)^p \sum_{n=0}^N a_n(1-x)^n & x > \rho\,, \\
(1-\rho)^p \sum_{n=0}^N a_n(1-\rho)^n + \sum_{n=1}^{N'} b_n(\rho-x)^n & x < \rho\,.
\end{array}
\right.
\label{eq:topfitfunc}
\ee
For each spectrum, $p$ is fixed by the asymptotics as $x \to 1$, whereas $N$ and $N'$ were chosen such that $\{a_n,b_n\}$, determined by a least squares fit, provided a satisfactory description.
Explicit values are provided in Table~\ref{tab:topfit}.

We do not need to repeat any calculations in order to obtain the equivalent spectra for anti-top decays, instead we obtain the result by a CP transformation.
In detail, CP flips the helicity off all states, and also interchanges particles and antiparticles.
Recalling that helicity and chirality are identified for a massless particle, but opposite for anti-particles, we have
\be
d \Gamma(t_{\pm} \to b f \bar{f}) = d \Gamma(\bar{t}_{\pm} \to \bar{b} \bar{f} f)\,.
\ee
Consequently, the spectrum of $\bar{b}$ is given directly by \Eq{eq:top2b}.
The distribution for $\bar{f}$ is given now by \Eq{eq:top2f}, and that for $f$ can be obtained by taking the same equation, but with $\theta \to \pi - \theta$ and $\phi \to \pi + \phi$ in the $\delta$-function argument.

\begin{table}[t]
\begin{center}
\begin{tabular}{|C{1.25cm}|C{0.2cm}|C{1.2cm}C{1.2cm}C{1.2cm}C{1.2cm}C{1.2cm}C{1.2cm}C{1.2cm}C{1.2cm}|}
  \hline
  Decay & $p$ & $a_0$ &  $a_1$ &  $a_2$ &  $a_3$ & $a_4$ & $a_5$ & $b_1$ & $b_2$ \\
 \hline
  $t_+\to f$ & 2 & 5.278 & -1.732 & -2.231 & 0.571 & -- & -- & 2.223 & -0.253 \\
  $t_-\to f$ & 1 & 2.380 & -1.786 & 11.450 & -32.748 & 47.768 & -27.897 & -2.235 & 0.337 \\ 
  $t_+\to \bar{f}$ & 2 & 6.943 & -3.541 & 2.764 & -4.717 & -- & -- & -10.266 & 0.257 \\
  $t_-\to \bar{f}$ & 3 & 2.121 & 2.504 & -3.135 & 4.063 & -- & -- &  10.249 & -0.192 \\
 \hline
\end{tabular}
\caption{Parameters of the fitting function \Eq{eq:topfitfunc} used to provide an adequate description of the $W$-decay products resulting from polarized top decays.
The spectra themselves are depicted in \Fig{fig:PolarizedTop}.}
\label{tab:topfit}
\end{center}
\end{table}

\section{Low Scale Evolution with \texttt{Pythia}}
\label{sec:LowScale}

The evolution of our FFs below $\qW$ is computed with \texttt{Pythia} v8.235~\cite{Sjostrand:2006za,Sjostrand:2007gs,Sjostrand:2014zea}.
In this section we expand upon this, outlining the options used in running \texttt{Pythia}, and also the modifications we made for our purposes such as an improved treatment of FSR, and how we incorporate the proton mass.

Firstly, let us outline the basic details of how we used the program.
In order to calculate $d_X^S(x;\,\qW,0)$ appearing in \Eq{eq:threesteps}, we simulate events with a hard interaction $e^+ e^- \to X \bar{X}$ at $\sqrt{s} = 2 \qW$, with initial state radiation switched off so that the $e^+ e^-$ operates as an energy injection, starting the $X$ at $\mu \sim \qW$.
We then determine $d_X^S(x;\,\qW,0)$ from the spectrum of stable $S$ particles in the hemisphere of the initial $X$.
This is all handled using a modified version of the example script \texttt{main07}.
All long lived particles that may not necessarily decay on collider scales are forced to decay, in particular muons, pions, kaons, and neutrons.
We leave off all electroweak radiation effects, as these were in the higher steps of our evolution.
This means that neutrinos do not evolve at all at this lower stage, although their spectra can still receive contributions from various decays.
Finally, a number of details of photon emission are modified; the choices and motivations here are discussed next.

\subsection{Improved Treatment of FSR for DM}
\label{sec:FSRPythia}

Below the weak scale, charged fermions will contribute to the photon spectrum via FSR, in the form $\chi \to f \bar{f} \gamma$.
The expression for this contribution is known analytically, as given in \Eq{eq:photonFSR}.
We can arrive at this result by, for example, calculating the photon spectrum from a three-body decay of a vector, $V$, via the process $V \to f \bar{f} \gamma$.
Giving the vector a mass $\mV = 2\qW$, and defining $\epsilon = m_f^2/\mV^2$, we can extract the leading order result from the analogous QCD calculation in \Refc{Ioffe:1978dc},
\bea
\frac{dN}{dx} 
=\frac{\alpha e_f^2}{\pi} &\left[ \frac{1+(1-x)^2-4\epsilon (x+2\epsilon)}{x \left( 1 + 2 \epsilon \right) \sqrt{1-4\epsilon}} \ln \left( \frac{1+\sqrt{1-4\epsilon/(1-x)}}{1-\sqrt{1-4\epsilon/(1-x)}} \right) \right. \\
&\left.
- \frac{1+(1-x)^2+4\epsilon(1-x)}{x \left( 1 + 2 \epsilon \right) \sqrt{1-4\epsilon}} \sqrt{1-\frac{4\epsilon}{1-x}} \right]\,,
\eea
where $e_f$ is the charge of the fermion, and $\alpha = \alpha_{\rm EM}$.
In the limit $\qW \gg m_f$, we have $\epsilon \ll 1$ and we can expand the above result to arrive at twice the result in \Eq{eq:photonFSR} (the factor of two arises as this is the spectrum of $f + \bar{f}$).\footnote{If we calculated the decay of a scalar instead of a vector, we obtain the same result after expanding in the limit $\epsilon \to 0$~\cite{Coogan:2019qpu}.}
The first correction to \Eq{eq:photonFSR} is $\mathcal{O}(\epsilon)$, and so this is an excellent approximation for all remaining SM fermions below $\qW$.

\begin{figure}[t]
\leavevmode
\vspace{-0.2cm}
\begin{center}
\includegraphics[width=.47\textwidth]{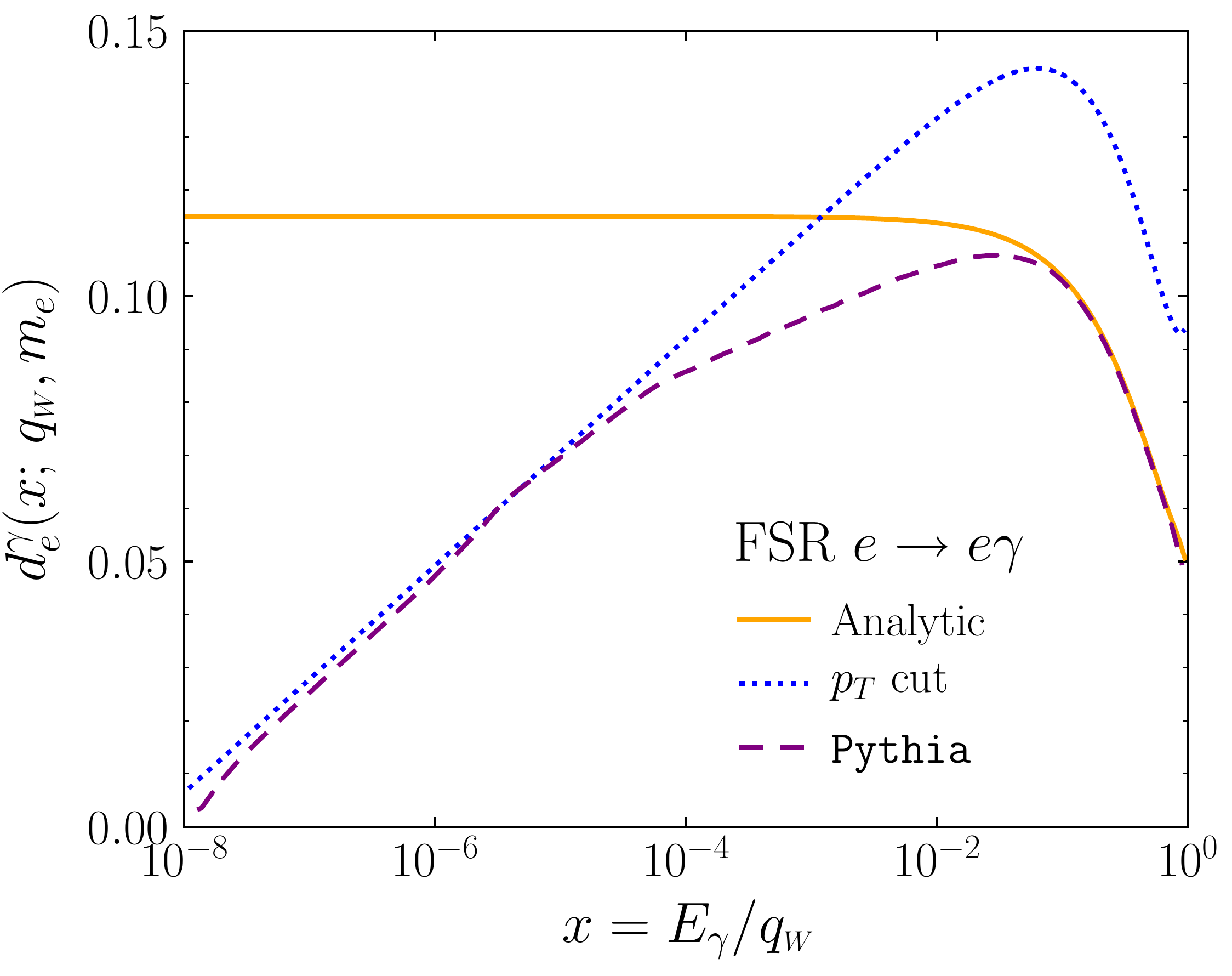}
\end{center}
\vspace{-0.5cm}
\caption{A comparison of the electron FSR spectrum to various analytic calculations.
In solid orange we show the expected analytic result from \Eq{eq:photonFSR}, in dashed purple the corresponding result from \texttt{Pythia}, and in dotted blue the analytic spectrum in \Eq{eq:photonFSR-pT} with a $p_T^{\rm cut} = 1$ keV.
The deviation of \texttt{Pythia} from the expected analytic result to the $p_T$ spectrum motivated an alternative treatment of this effect.
}
\label{fig:FSRSpec}
\end{figure}

With the analytic expression in hand, we can compare this to the output of \texttt{Pythia} for an example state, $f = e$.
The result of doing so is shown in \Fig{fig:FSRSpec}.
At lower $x$ values \texttt{Pythia} is under-predicting the number of photons.
The origin of this mismatch is that \texttt{Pythia} imposes a $p_T$ cutoff on the QED shower: photons that would carry a $p_T < 1$ keV by default are not produced in the shower (this is controlled by \texttt{TimeShower:pTminChgL}).
In order to validate this interpretation, note that the analogue of \Eq{eq:photonFSR} in the presence of a $p_T$ cut is,
\be
\frac{dN}{dx} = \frac{\alpha e_f^2}{\pi} \left[ \frac{1+(1-x)^2}{x} \ln \left( \frac{4 \qW^2 x^2}{(p_T^{\rm cut})^2(1-x)} \right) - x \right]\,,
\label{eq:photonFSR-pT}
\ee
after expanding in $(p_T^{\rm cut})^2/\qW^2 \ll 1$.
This expression is also plotted in \Fig{fig:FSRSpec}, and we see the \texttt{Pythia} result trending towards it at low energy fractions.

As discussed, given concerns about the accuracy of our results at small-$x$, we do not produce spectra below $x=10^{-6}$.
Further, for most channels, due to showering into other states that can produce photons through hadronic channels, FSR is often a subdominant contribution to the photon spectrum at low-$x$, particularly as we move above $\qW$ and the shower can develop.
Nevertheless, at low masses certain states such as right handed electrons still receive a large contribution from FSR and thus we outline a procedure we used to correct this issue.
We note this treatment may also be useful for studies of DM at lower masses also.

Our approach is to turn off FSR in \texttt{Pythia} and instead deal with it analytically.
Showering for all fermions is switched off, although we leave on $\gamma \to f \bar{f}$, as this is not subject to the issue described above.
The full photon spectrum for a given initial state can then be determined using \Eq{eq:photonFSR} and,
\bea
D_X^{\gamma}(x;\,m_{\chi}/2,0) &= D_X^{\gamma}(x;\,m_{\chi}/2,\qW) \\
&+ \sum_f \int_x^1 \frac{dz}{z} D_X^f(x/z;\,m_{\chi}/2,\qW)\,[D_f^{\gamma}(z;\,\qW,0)]_{\rm FSR}\,,
\label{eq:AddFSR}
\eea
so that the low scale FSR is handled analytically instead of with \texttt{Pythia}.

Adding \Eq{eq:AddFSR} alone is inconsistent with momentum conservation.
The charged particles that would have emitted photons also need to have their distributions corrected to ensure the momentum is correctly subtracted.
In particular, there will be both real and virtual corrections to the charged fermions' FFs.
For real emission, if we have a process where the high scale evolution produced a charged fermion, $f$, from an initial particle $X$, then we need to account for the fact the $f$ would have lost momentum to radiated photons via $f \to f \gamma$ in the evolution below $\qW$, which we now exclude.
In particular, if the fermion has an initial momentum fraction $z$ and the photon carries away a fraction $1-u$, then the fermion FF receives a correction,
\begin{align}
\left[D_X^f(x;\,m_{\chi}/2,\qW)\right]_{\rm Real} 
&= \int_0^1 dz\, \int_0^1 du\, D_X^f(z;\,m_{\chi}/2,\qW)\,[D_f^{\gamma}(1-u;\,\qW,0)]_{\rm FSR} \delta(x-zu) \nn
&= \int_x^1 \frac{dz}{z} D_X^M(x/z;\,m_{\chi}/2,\qW)\,[D_M^{\gamma}(1-z;\,\qW,0)]_{\rm FSR}\,.
\label{eq:RealEmissionFSR}
\end{align}
This same fermion would also have received virtual corrections associated with the probability of no photon being emitted, which can be computed from one minus the probability there was an emission, i.e.
\begin{align}
\left[D_X^f(x;\,m_{\chi}/2,\qW)\right]_{\rm Virtual} 
&= D_X^f(x;\,m_{\chi}/2,\qW) \int_0^1 dz\, \left( 1 - [D_f^{\gamma}(z;\,\qW,0)]_{\rm FSR} \right) \\
&= D_X^f(x;\,m_{\chi}/2,\qW) - D_X^f(x;\,m_{\chi}/2,\qW) \int_0^1 dz\, [D_f^{\gamma}(1-z;\,\qW,0)]_{\rm FSR}\,. \nonumber
\end{align}
From the combination of these two terms we can see the appropriate correction is given by,
\bea
\Delta D_X^f(x;\,m_{\chi}/2,\qW)
= \int_0^1 dz\, &\left[ \frac{1}{z} D_X^f(x/z;\,m_{\chi}/2,\qW) - D_X^f(x;\,m_{\chi}/2,\qW) \right] \\
\times &[D_f^{\gamma}(1-z;\,\qW,0)]_{\rm FSR}\,.
\label{eq:ChargeFermionCorrection}
\eea
Note $D(z) = 0$ for $z > 1$, explaining the difference in integration limits between this expression and \Eq{eq:RealEmissionFSR}.

In order to confirm the correctness of this result, we can explicitly confirm momentum conservation.
Firstly, the momentum partitioned into photons via \Eq{eq:AddFSR} is given by
\bea
&\int_0^1 dx\, x\,\sum_f \int_x^1 \frac{dz}{z} D_X^f(x/z;\,m_{\chi}/2,\qW)\,[D_f^{\gamma}(z;\,\qW,0)]_{\rm FSR} \\
=&\sum_f \int_0^1 dz\,[D_f^{\gamma}(z;\,\qW,0)]_{\rm FSR} \int_0^z dx\, \frac{x}{z} D_X^f(x/z;\,m_{\chi}/2,\qW) \\
= &\sum_f \left[ \int_0^1 dz\,z\,[D_f^{\gamma}(z;\,\qW,0)]_{\rm FSR} \right] \left[ \int_0^1 du\, u\, D_X^f(u;\,m_{\chi}/2,\qW) \right]\,,
\label{eq:FSRmomentumsum}
\eea
where in the final step we changed variables to $u=x/z$.
This is then equal to the momentum lost by the sum of all charged fermions.
Looking at one in particular, we have
\begin{align}
&\int_0^1 dx\, x \Delta D_X^f(x;\,m_{\chi}/2,\qW) \nn
= &\int_0^1 dx\, x \int_0^1 dz\, \left[ \frac{1}{z} D_X^f(x/z;\,m_{\chi}/2,\qW) - D_X^f(x;\,m_{\chi}/2,\qW) \right] [D_f^{\gamma}(1-z;\,\qW,0)]_{\rm FSR} \nn
= &\int_0^1 dz\,z  [D_f^{\gamma}(1-z;\,\qW,0)]_{\rm FSR} \int_0^1 du\, u D_X^f(u;\,m_{\chi}/2,\qW) \nn
&- \int_0^1 dz [D_f^{\gamma}(1-z;\,\qW,0)]_{\rm FSR} \int_0^1 du\, u D_a^i(u;\,m_{\chi}/2,\qW) \\
= &-\left[ \int_0^1 dz\,(1-z)  [D_f^{\gamma}(1-z;\,\qW,0)]_{\rm FSR} \right] \left[ \int_0^1 du\, u D_X^f(u;\,m_{\chi}/2,\qW) \right] \nn
= &-\left[ \int_0^1 dz\,z\,  [D_f^{\gamma}(z;\,\qW,0)]_{\rm FSR} \right] \left[ \int_0^1 du\, u\, D_X^f(u;\,m_{\chi}/2,\qW) \right]\,, \nonumber
\end{align}
Summing over $f$, this is exactly the negative of \Eq{eq:FSRmomentumsum} as claimed, demonstrating using both terms restores momentum conservation.

In summary, we implement the corrected FSR treatment with a combination of \Eq{eq:AddFSR} and \Eq{eq:ChargeFermionCorrection}.
The latter needs to be computed carefully, as $[D_f^{\gamma}(1-z;\,\qW,0)]_{\rm FSR}$ diverges as $z \to 1$, although that divergence is regulated by the vanishing of the two terms in square brackets in the same limit.

\subsection{Incorporating the Proton Mass}
\label{sec:ProtonMass}

\begin{figure}[t]
\leavevmode
\vspace{-0.2cm}
\begin{center}
\includegraphics[width=.47\textwidth]{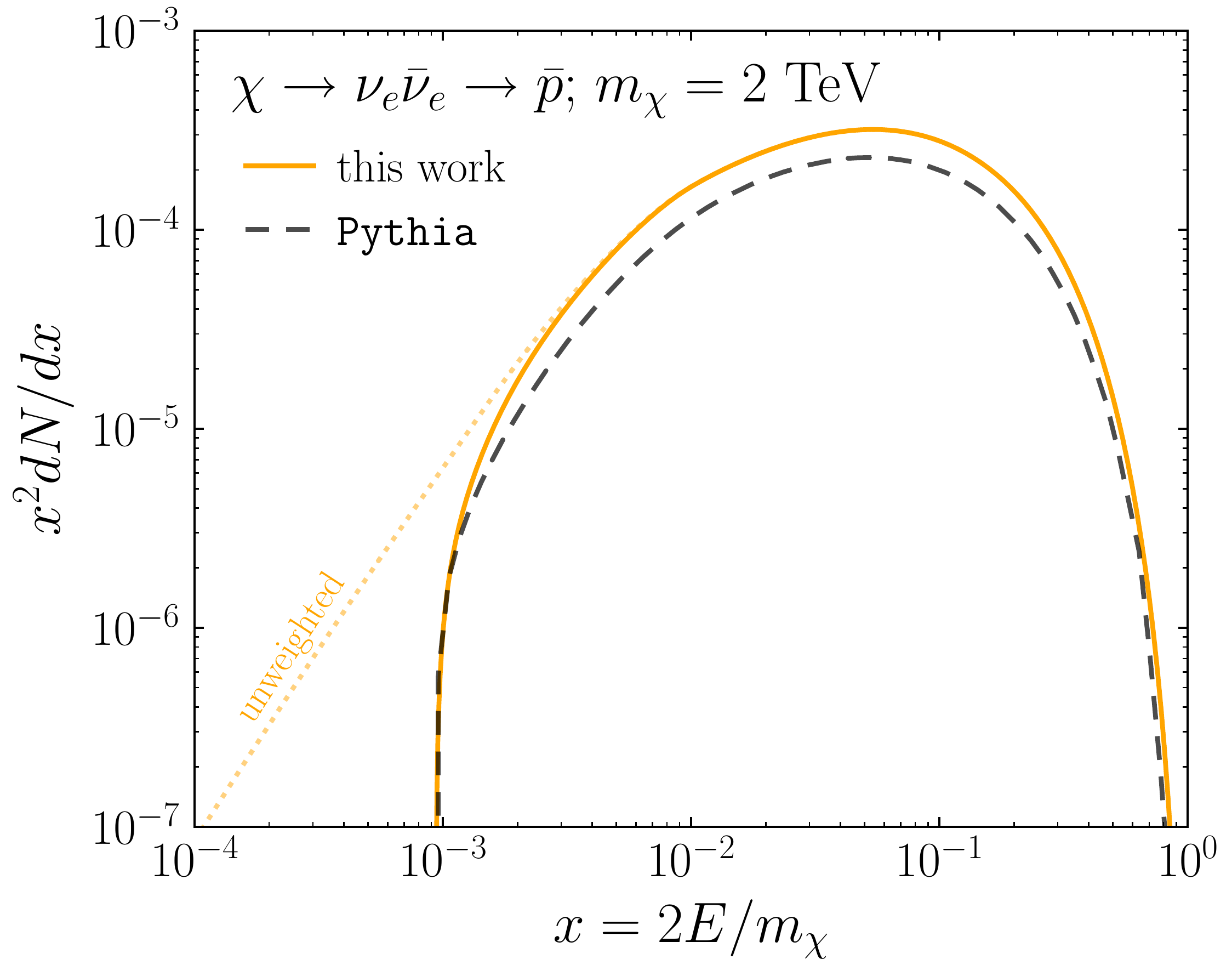}
\hspace{0.5cm}
\includegraphics[width=.47\textwidth]{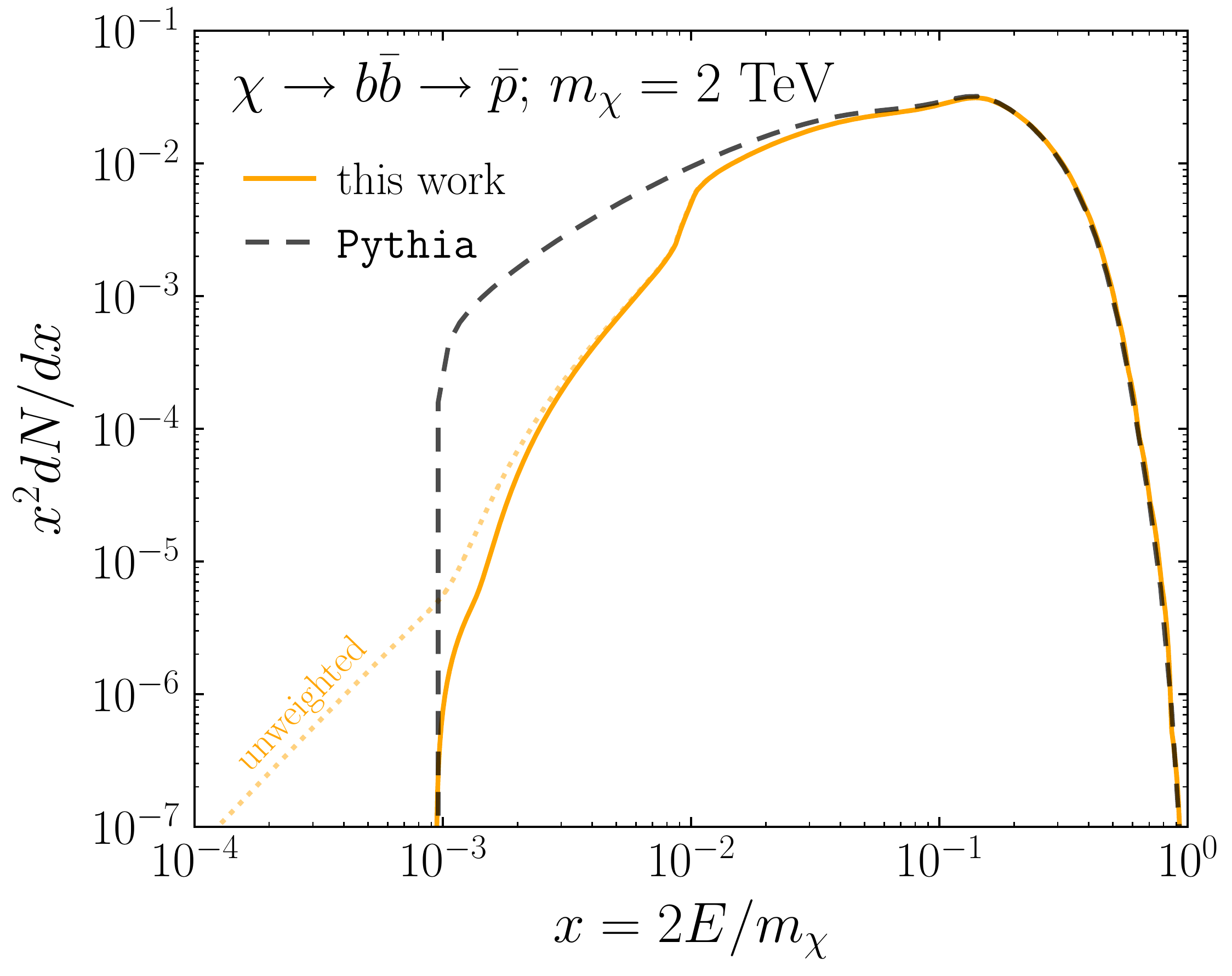}
\end{center}
\vspace{-0.5cm}
\caption{Spectrum of antiprotons for two different initial states, electron neutrinos (left) and $b$-quarks (right), obtained for $m_{\chi} = 2$ TeV.
Our default results, labelled ``unweighted", continue to unphysical $x$-values.
In App.~\ref{sec:ProtonMass} we outline a weighting procedure to correct these to the solid orange curves.
Even then, residual features at $x \sim m_p/\qW$ remain, although they are washed out for larger masses.
This feature is a result of the incomplete treatment of the proton mass in this work, see text for details.
}
\label{fig:ProtonSpec}
\end{figure}

The finite proton mass provides a physical cutoff in the energy fraction the $p/\bar{p}$ final states can carry of $x \geq 2m_p/m_{\chi}$.
In our calculation at the high scale, however, all states are treated as massless, allowing for in principle arbitrarily small energy fractions.
At the low-scale, however, the evolution in \texttt{Pythia} explicitly includes $m_p \neq 0$, and therefore will have a sharp cut-off at $x = m_p/\qW \sim 10^{-2}$.
In summary, the finite mass of the stable hadrons we are interested in is not treated consistently throughout the calculation.
By default, this inconsistency will manifest in two ways.
Firstly, as at the high scale we allow arbitrarily small energy -- or, as our particles are massless, equivalently momentum -- fractions, after convolving this with \texttt{Pythia}, by default we will have a non-zero spectrum for $x < 2m_p/m_{\chi}$.
Secondly, as the proton mass is only treated in one part of our calculation, a feature at $x \sim 10^{-2}$ can generically be expected to appear.
This occurs because, if the high scale spectrum has a sharp feature near $x=1$, a residual bump at $10^{-2}$ can remain after convolution with \texttt{Pythia}.

Both of these problems are on display in \Fig{fig:ProtonSpec}, focusing on the comparison between the unweighted spectrum and \texttt{Pythia} for the moment.
On the right, we see a clear feature at $10^{-2}$.
This results from the fact that for a relatively light DM mass of $m_{\chi} = 2$ TeV, the $b$-quarks will not always undergo significant evolution by $\qW$, and there will still be a large contribution to the fragmentation functions near $x=1$.
When convolved with \texttt{Pythia}, the result is a clear bump.
Fixing this particular problem in detail is left to future work, however we note the feature rapidly becomes less pronounced as we increase in mass, and for states that do not have a large colored component near $x=1$, as is the case for neutrinos viewed on the left, the issue is far less apparent.

The second issue, of a non-zero spectrum for unphysical energy fractions, we will resolve.
Our default output at this stage is the ``unweighted" dotted distributions, exhibiting this exact behaviour.
We will reweight these results as follows.
To begin with, we treat all $x$ values as momentum fractions (these are interchangeable at the high scale as discussed).
So for a given $x$ and DM mass $m_{\chi}$, a proton would have a momentum $p = x_p\, m_{\chi}/2$, and hence an energy fraction
\be
x_E = \sqrt{x_p^2 + \frac{4 m_p^2}{m_{\chi}^2}}\,.
\ee
Importantly, as $x_p \to 0$, $x_E \geq 2m_p/ m_{\chi}$, as required.
This allows us to transform to the relevant energy fractions, and further the spectrum as
\be
\frac{dN}{dx_E} = \frac{x_E}{x_p} \frac{dN}{dx_p}\,.
\ee

The above change of variables from $x_p \to x_E$ will ensure that the spectrum is cut-off at the correct $x$ value, however, it does not ensure the spectrum exhibits the correct asymptotics as $x_p \to 2m_p/m_{\chi}$.
In general we expect there will be power corrections to our result in this limit, which manifest here in the form of a threshold factor $(x_p/x_E)^k$, so that
\be
\frac{dN}{dx_E} = \left(\frac{x_p}{x_E} \right)^{k-1} \frac{dN}{dx_p}\,.
\ee
By comparing our results near the threshold with \texttt{Pythia}, we found good agreement for $k=3$.
The combined result of the rescalings is depicted as the solid orange curves in \Fig{fig:ProtonSpec}.

\section{Estimating the Accuracy of our Results}
\label{sec:Limitations}

As discussed in the main text, there is considerable scope for systematic improvement of the results presented in this work.
In this section we will present a more quantitative discussion of this point, outlining the formal accuracy of our calculation and estimating the size of neglected terms.
As we will show, at high DM masses our spectra for $x \in [10^{-3},1]$ are accurate to $\mathcal{O}(10\%)$.
For $x \lesssim 10^{-3}$, the uncertainty increases (particularly due to an incomplete treatment of soft coherence) to the $\mathcal{O}(1)$ or even order-of-magnitude level.
Nevertheless, in light of the pronounced differences with existing results as shown in Figs.~\ref{fig:PythiaComparison} and \ref{fig:Comparison}, our results represent a significant improvement.
Obtaining a full NLL calculation with theoretical uncertainty bands and $\mathcal{O}(5\%)$ errors at the level that has been achieved for specific DM spectra, see e.g.~\cite{Baumgart:2018yed}, represents a clear target for future work.

Our treatment of fragmentation function evolution above the electroweak scale $\qW\sim 100$ GeV is according to the leading-order DGLAP evolution equations, with neglect of all particle masses.
The evolution code uses Heun's method for direct solution of the integro-differential equations, with typical precision of a few parts per mille.
However, the evolution equations only treat terms enhanced by
logarithms of $m_{\chi}/\qW$ and therefore results are only
reliable well above the electroweak scale.  
While mass corrections are expected to be of order $\qW^2/m_{\chi}^2$, other terms that are not logarithmically enhanced can be important for TeV-scale DM masses.
These could be included in future by matching to fixed-order matrix elements, as discussed in \Refc{Bauer:2017bnh}.
Meanwhile, for DM decay or annihilation at or below the TeV scale, a direct parton shower simulation taking account of masses, such as \texttt{Pythia}, would be more reliable for the majority of final states.
The exception would be final states where the spectra is dominated by electroweak showers, in particular neutrinos as highlighted in the left of \Fig{fig:Comparison}.

\begin{figure}[t]
\leavevmode
\vspace{-0.2cm}
\begin{center}
\includegraphics[width=.47\textwidth]{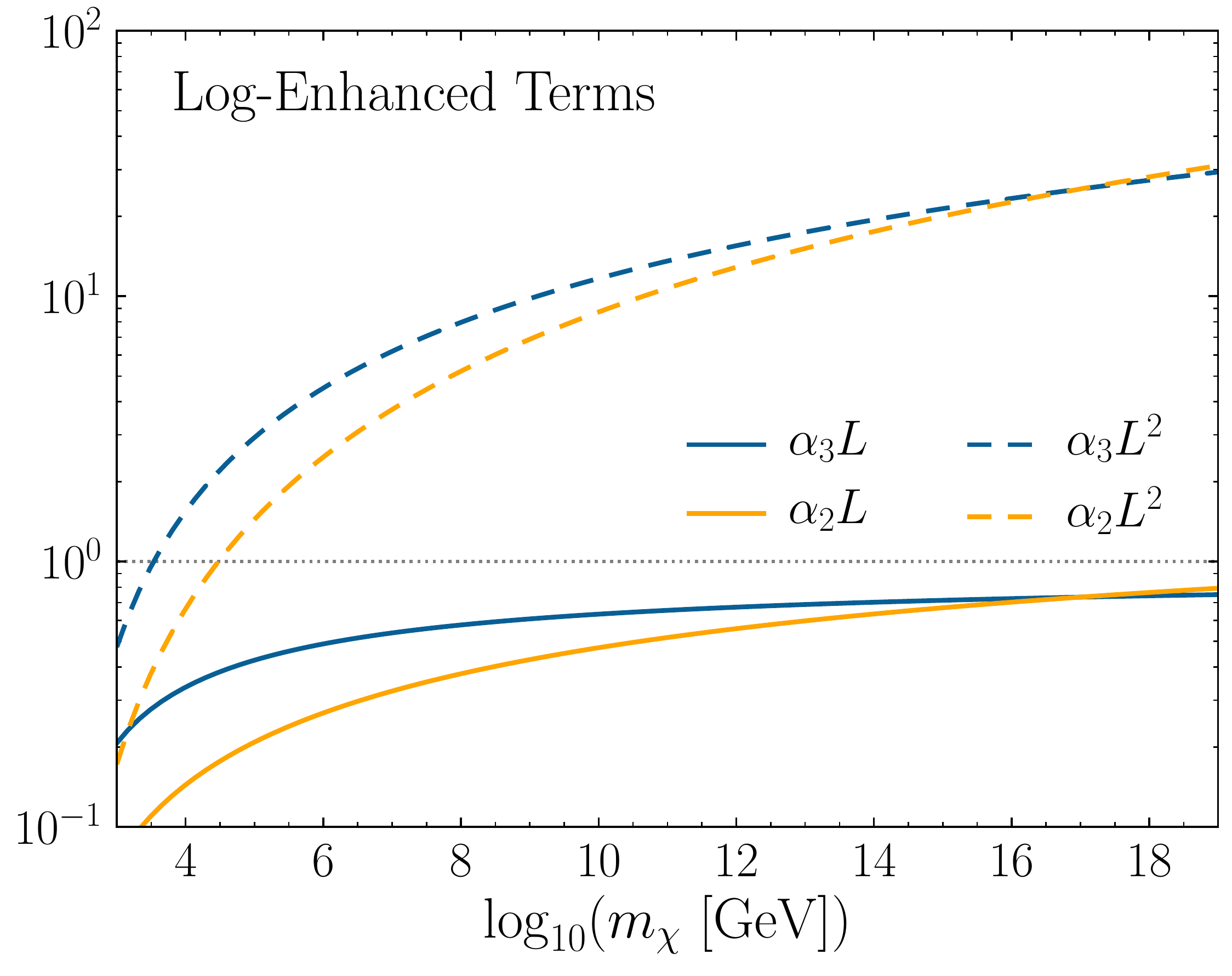}
\end{center}
\vspace{-0.5cm}
\caption{Qualitative size of double and single logarithmic terms as a function of DM mass $m_{\chi}$.
Here $L = \ln[m_{\chi}/\qW]$, and results are shown for SU(3) and SU(2)$_L$ couplings.
A LL calculation resums the double logarithmic terms, whereas a full NLL result would resum also the single logarithmic terms.
Clearly the $\alpha L^2$ terms are rapidly larger than one, and thus must be resummed for a reliable result for most of the mass range considered in this work, and our calculation achieves this aim.
The $\alpha L$ terms, which we do not fully resum, remain less than unity across the entire mass range.
In both cases we account for the running couplings.
}
\label{fig:MissingLogs}
\end{figure}

For all except the SU(2)$_L$ electroweak interaction, terms resummed by leading-logarithmic (LL) DGLAP evolution from the electroweak scale to scale $\sim m_{\chi}$ are of order $\alpha^n_IL^n$ where $L=\ln(m_{\chi}/\qW)$.
However, our calculation fails to resum all $\alpha L$ terms related to soft emissions.
Nevertheless, $\alpha L$ remains less than 1 to the Planck scale, as shown in \Fig{fig:MissingLogs}. 
In detail $\alpha_3L\sim 0.75$ and $\alpha_2L\sim 0.80$, taking into account the running of both couplings.

In the case of SU(2)$_L$, fragmentation functions in general have non-isosinglet contributions of order $\alpha^n_2L^{2n}$.
These, together with a class of terms down to order $\alpha^n_2L^{n+1}$, sum up to yield leading-logarithmic Sudakov factors, of the form $\exp[-Lg_1(\alpha_2L)]$ where $g_1$ is a known function.
The missing NLL terms are of order $\alpha_2 L$, just as in QCD.
As shown in \Fig{fig:MissingLogs}, these terms are smaller for masses up to $m_{\chi} \sim$ GUT scale, and at higher masses remain comparable to $\alpha_3 L$ and crucially less than unity.
Failure to fully resum these terms induces an error of size $\mathcal{O}(10\%)$ up to the EeV scale.
A full NLL calculation would suppress these effects by an additional $\alpha_I \sim 0.1$, and reduce these errors further to the $1\%$ scale.

\begin{figure}[t]
\leavevmode
\vspace{-0.2cm}
\begin{center}
\includegraphics[width=.47\textwidth]{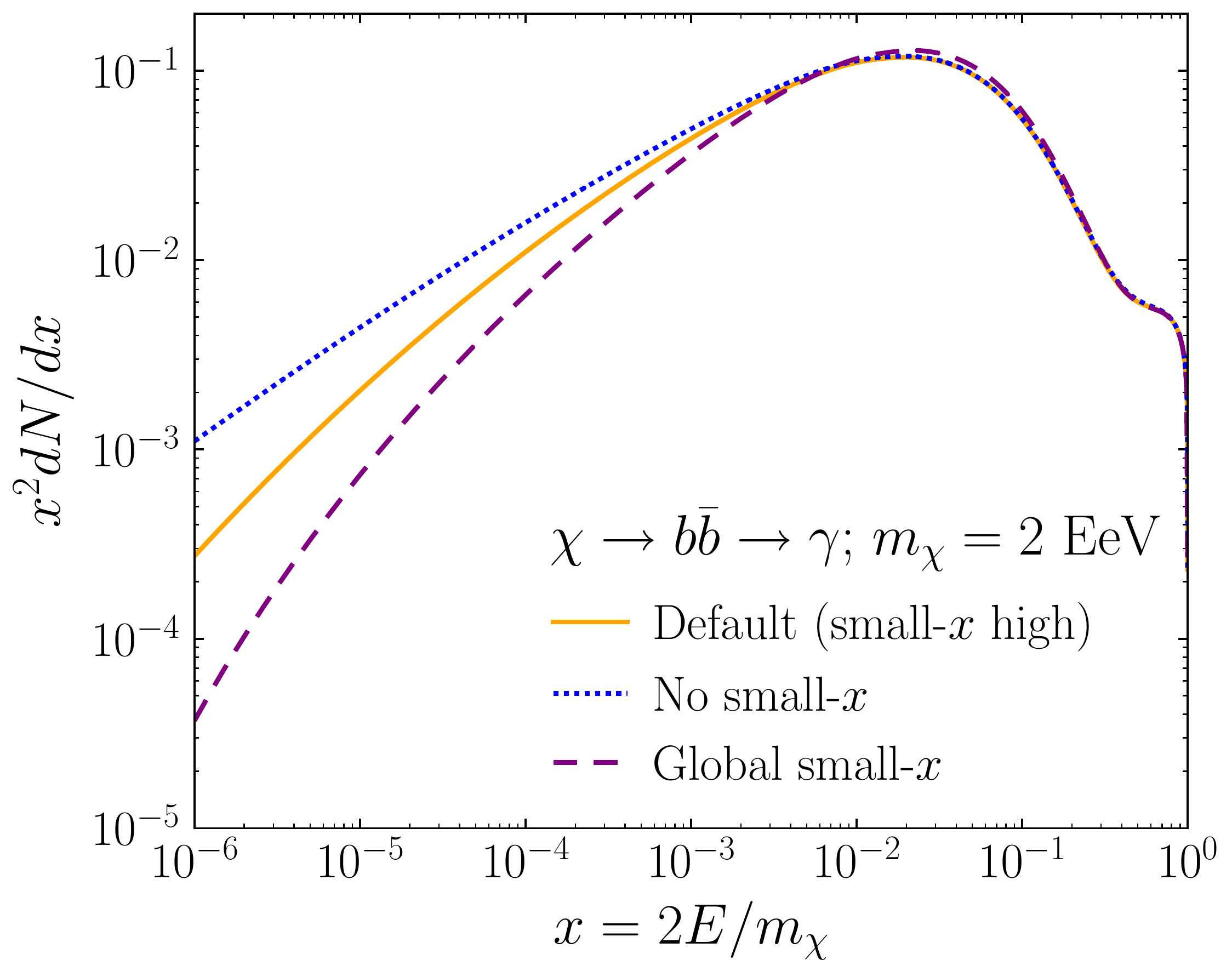}
\includegraphics[width=.47\textwidth]{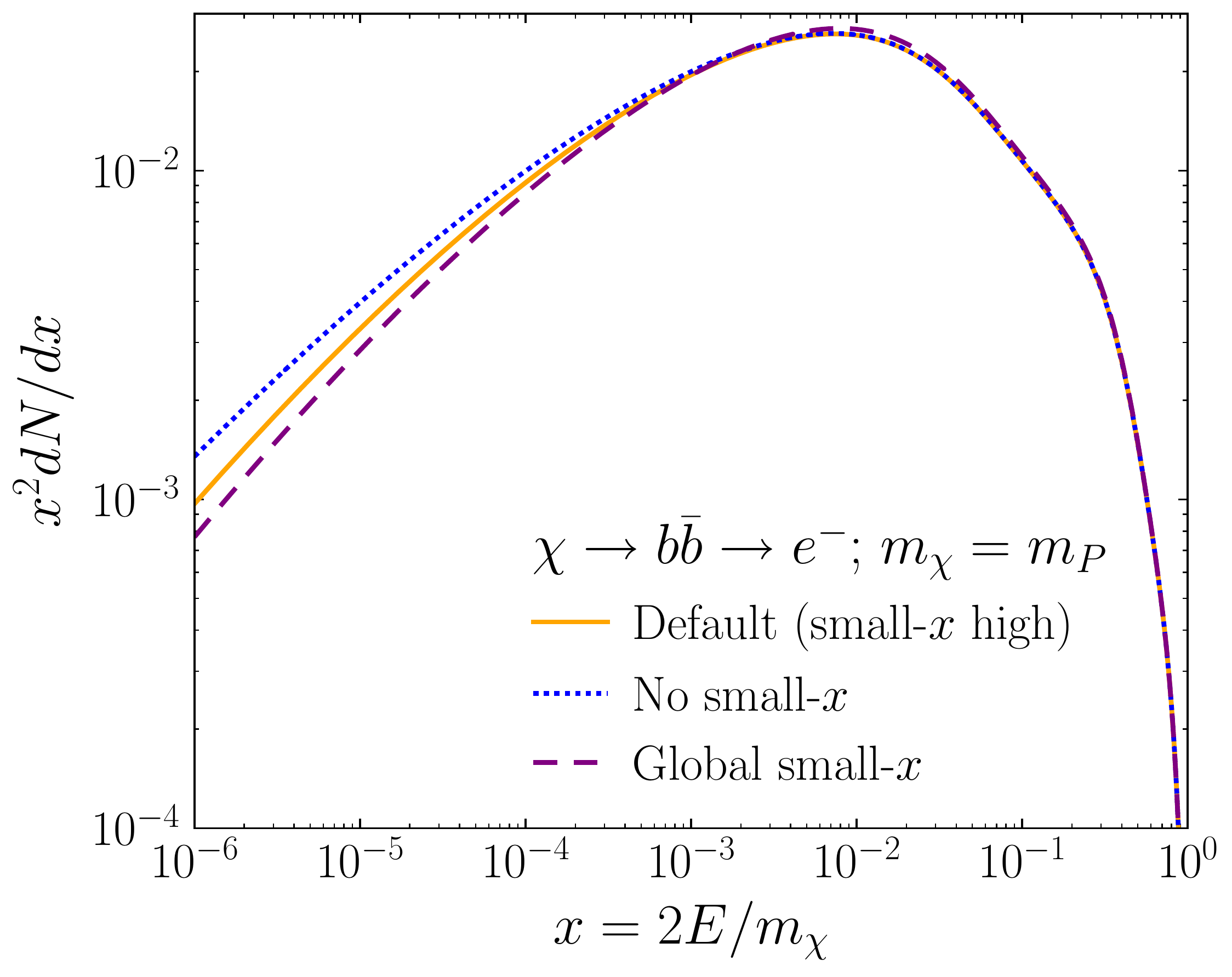}
\end{center}
\vspace{-0.5cm}
\caption{An estimate of the uncertainty associated with our treatment of the small-$x$ physics.
We achieve this by showing results for three different treatments of soft coherence, for the spectra resulting from $\chi \to b \bar{b}$ for two different masses and final states: the photon spectrum for $m_{\chi} = 2$ EeV (left), and the electron spectrum for $m_{\chi} = m_P$ (right).
In dotted blue, we show the result of applying no corrections at all, a result which represents an undoubted overestimate of the small-$x$ flux.
In solid orange we show our default procedure of applying a substitution $m_{\chi} \to x m_{\chi}$ in the high scale FF, as outlined in App.~\ref{sec:Smallx}.
Finally, in dashed purple we show the result of applying that same substitution to our full FF that results from the convolution of all three steps.
This final result will double-count soft-coherence from the low-scale evolution, and therefore sits below the other approaches.
We see that at an EeV, and for $x \sim 10^{-6}$ the uncertainty is at the order-of-magnitude level.
For $x \gtrsim 10^{-3}$ or at higher masses, the result reduces to a $\mathcal{O}(1)$ error.
}
\label{fig:Smallx-3ways}
\end{figure}

All the above relates to evolution of fragmentation functions due to collinear enhancements in the relevant matrix elements, giving rise to large logarithms of the energy scale ratio $m_{\chi}/\qW$.
At small values of the energy fraction $x$ there are also large (double) logarithms of $x$ that need to be resummed.
As discussed above, these have the effect of strongly suppressing fragmentation at small $x$, essentially due to destructive interference between different amplitudes involving soft gauge bosons.
These soft coherence effects are well understood at the leading-logarithmic level in QCD, but less so at NLL in QCD and even at LL in the electroweak sector.
Our treatment takes account of soft coherence at LL level in QCD, with a plausible extension of the same effects to the full SM.
However, it is difficult to estimate the quantitative uncertainty of this procedure.
A qualitative estimate of the uncertainties of this procedure is provided in \Fig{fig:Smallx-3ways}.
The figure depicts spectra arising from $\chi \to b \bar{b}$ for three different treatments of the small-$x$ physics.
Firstly, we show the result if you use pure DGLAP evolution with no treatment of soft-coherence.
This is certainly an overestimate of the true soft multiplicity.
On the other extreme, we show a result where we apply our $m_{\chi} \to x m_{\chi}$ correction to the final FF, i.e. the substitution is applied to the left-hand side of \Eq{eq:threesteps}, rather than just to the high-scale or DGLAP FF, as we do by default.
This results in most likely an overly aggressive suppression.
\texttt{Pythia} already has a partial accounting for soft-coherence, so applying the substitution globally we are double counting the effect at this stage.
Further, as mentioned in App.~\ref{sec:Smallx} this substitution should not be applied to the electroweak double logs, however we cannot factor out their contribution to the final FF.
Finally, we show our default procedure where the correction is only applied to the high-scale evolution.
We see that our approach sits in between the two extreme alternatives.
For $x \gtrsim 10^{-3}$ the differences are not that pronounced, but by $x \sim 10^{-6}$ there is now more than an order of magnitude separation between the approaches.
This can be taken as a rough estimate for the uncertainty we have at these scales, although \Fig{fig:Smallx-3ways} also demonstrates that by the Planck scale the differences are reduced to a factor of $\sim 2$ (we emphasize that the difference between the three approaches shown in the two plots is driven entirely by the mass scale).
The results also clarify that in order to reduce these uncertainties, a serious study of coherence effects in the whole SM is required.

Turning to the matching at the electroweak scale, the calculation was performed at leading order.
Next-to-leading QCD corrections, of order $\alpha_3\sim 10\%$, could be included to improve precision, although this would require care to avoid double counting in the subsequent \texttt{Pythia} shower.
Electroweak corrections would contribute at the few percent level.
Finally, below the electroweak scale, the \texttt{Pythia} parton shower and hadronization generator generally agrees with collider data at such energies at the $10\%$ level.
This would be difficult to improve significantly without major advances in event generator technology

In summary, our predictions at high DM masses and moderate to high $x$ values are subject to uncertainties at the few times 10\% level, which could be reduced somewhat by inclusion of higher-order corrections at the evolution and matching stages.
Uncertainties increase markedly at lower energy fractions, due to a lack of precise understanding of soft coherence effects in the full SM.

\section{Additional Results}
\label{sec:AdditionalResults}

\begin{figure}[t]
\leavevmode
\vspace{-0.2cm}
\begin{center}
\includegraphics[width=.47\textwidth]{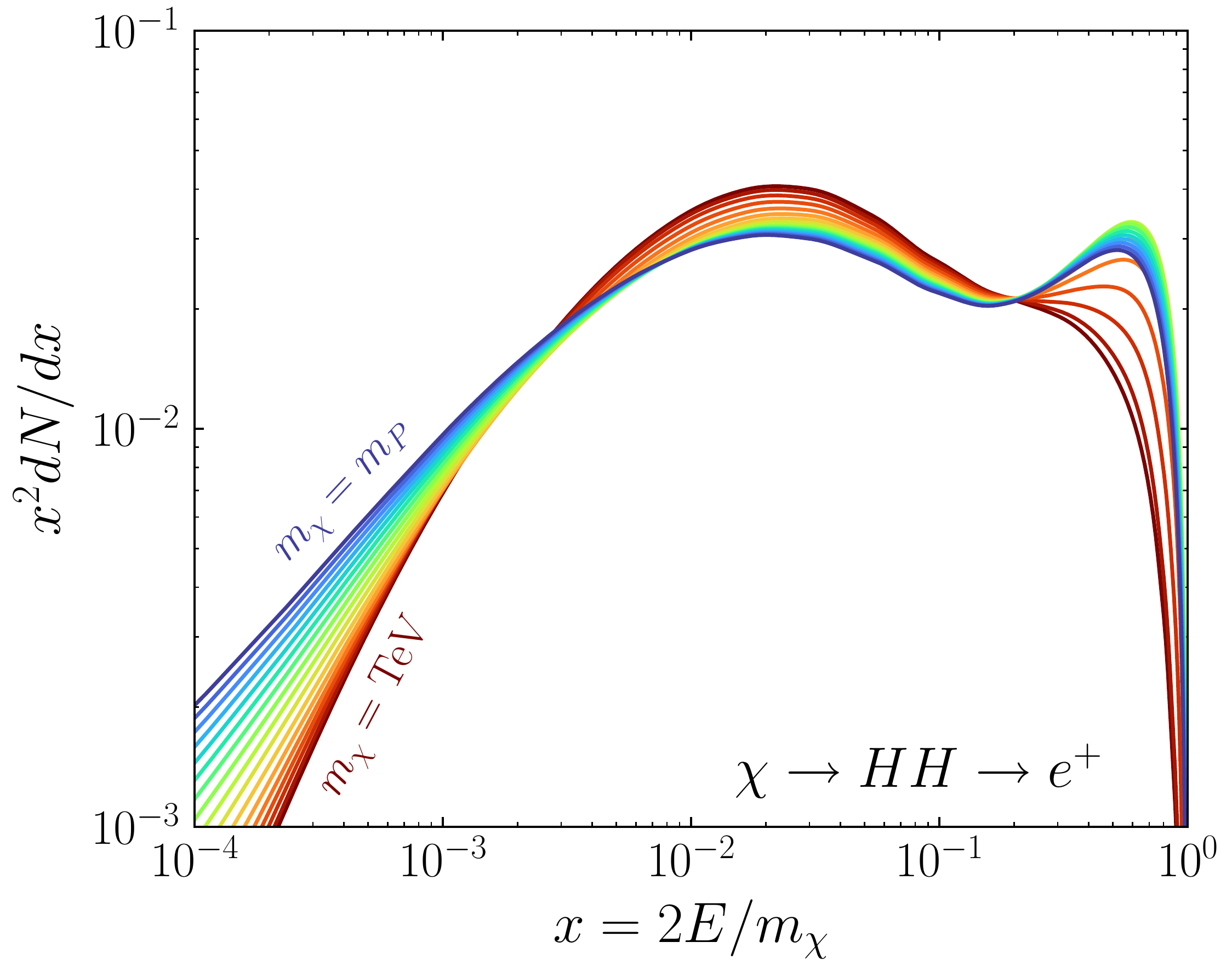}
\hspace{0.5cm}
\includegraphics[width=.47\textwidth]{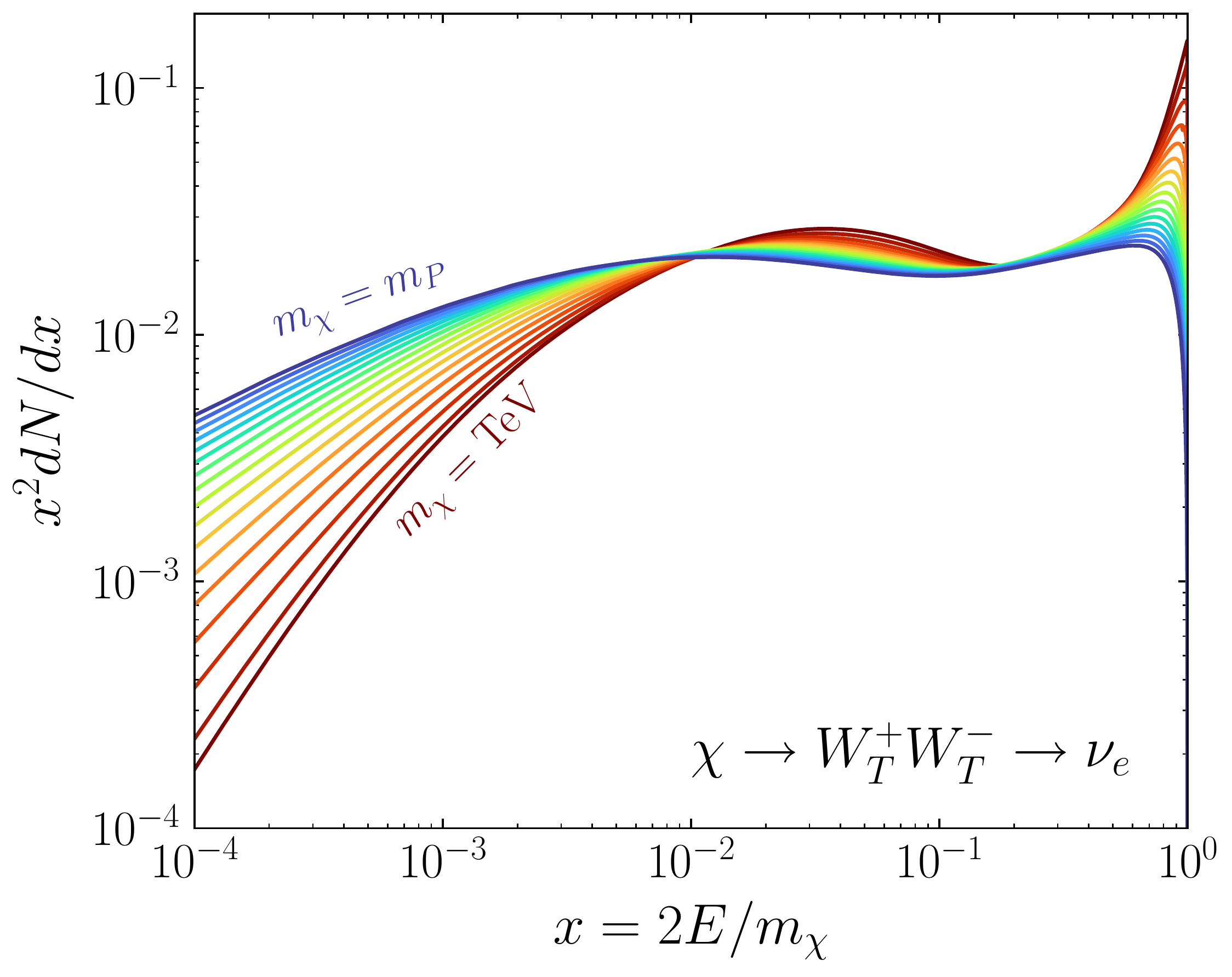} \vspace{0.5cm}\\
\includegraphics[width=.47\textwidth]{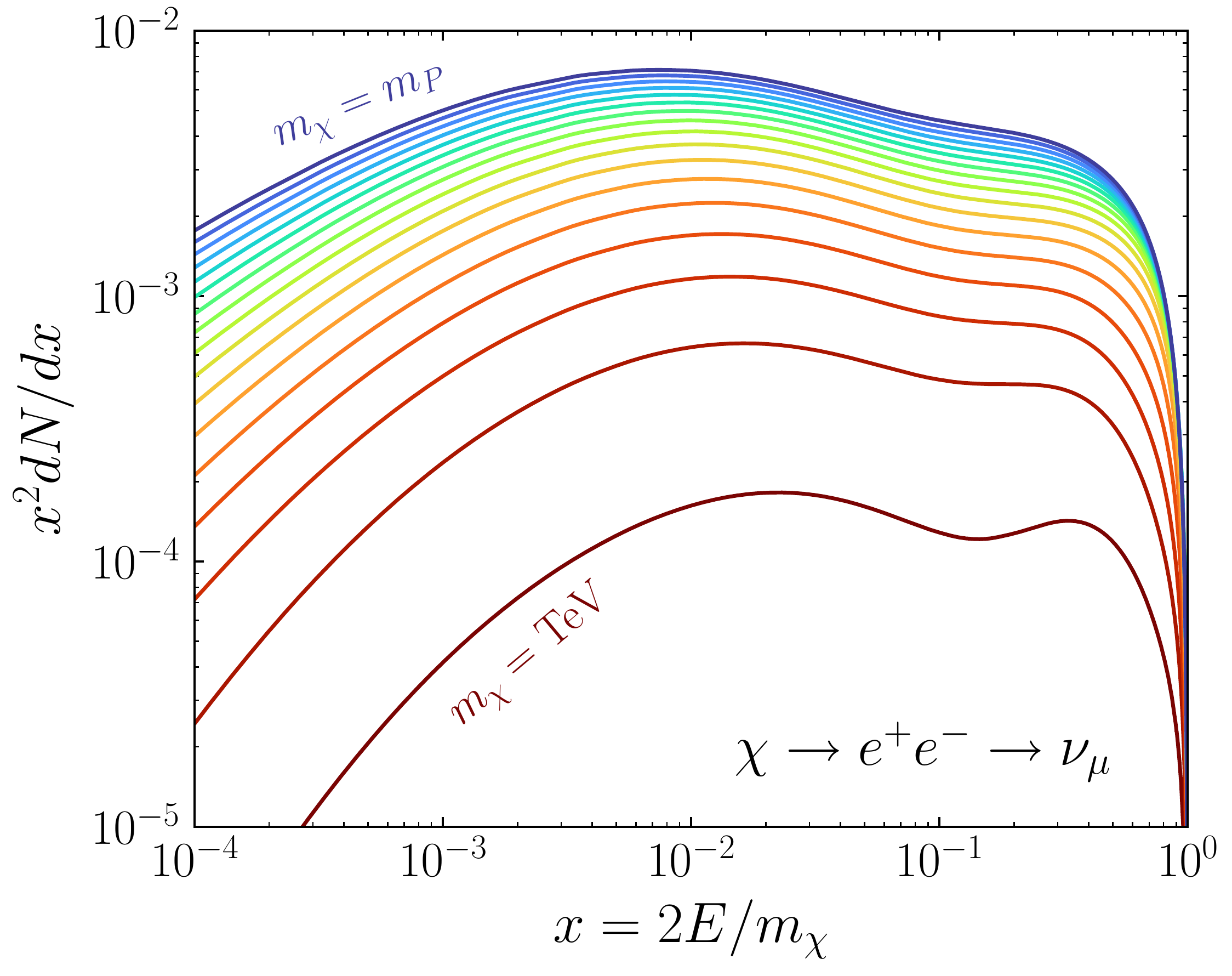}
\hspace{0.5cm}
\includegraphics[width=.47\textwidth]{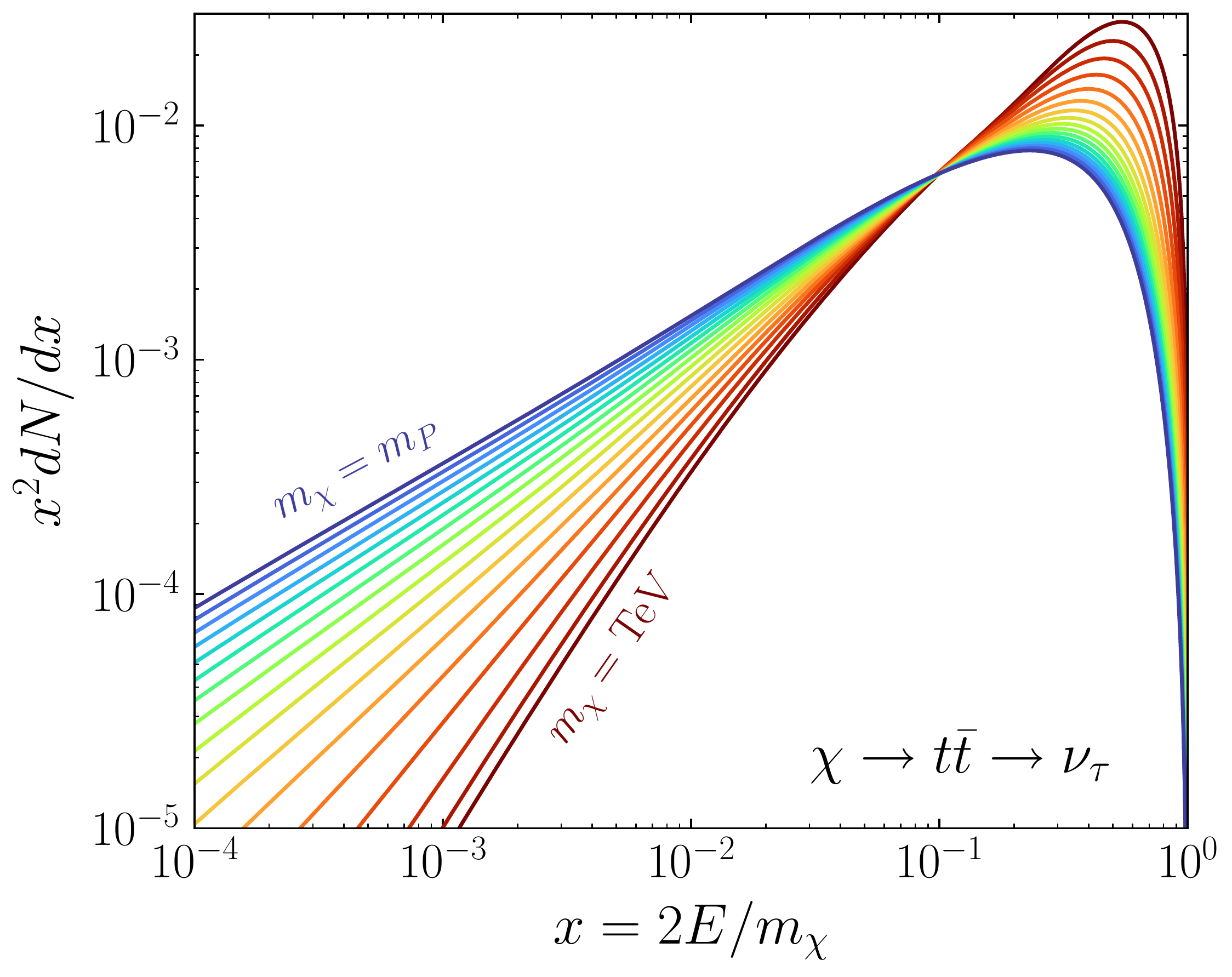}
\end{center}
\vspace{-0.5cm}
\caption{Analogs of \Fig{fig:ballM} for four additional DM decay processes: higgs to positrons (top left), transverse $W$-bosons to electron neutrinos (top right), electrons to muon neutrinos (bottom left), and top-quarks to tau neutrinos (bottom right).
Each figure shows the spectra for each decade of DM mass between $m_{\chi} = 1$ TeV and $10^{19}$ GeV.
Further, initial helicities or polarizations are averaged over, where relevant. 
}
\label{fig:ExtraSpec}
\end{figure}

Having outlined the details of our calculation and discussed their accuracy, we now present a number of additional outputs from our formalism.
Firstly, we present a number of additional spectra in the same spirit as \Fig{fig:ballM}, highlighting additional physics inherent in our results.
Afterwards we present a non-trivial cross-check on our results, demonstrating the extent to which momentum is conserved as it is repartitioned among the various states through our evolution.
In a similar vein, we will then show how the momentum is distributed between the various states for selected processes, demonstrating how this varies as the mass is increased.

\subsection{Additional Spectra}
\label{sec:additionalespectra}

\begin{figure}[t]
\leavevmode
\vspace{-0.2cm}
\begin{center}
\includegraphics[width=.47\textwidth]{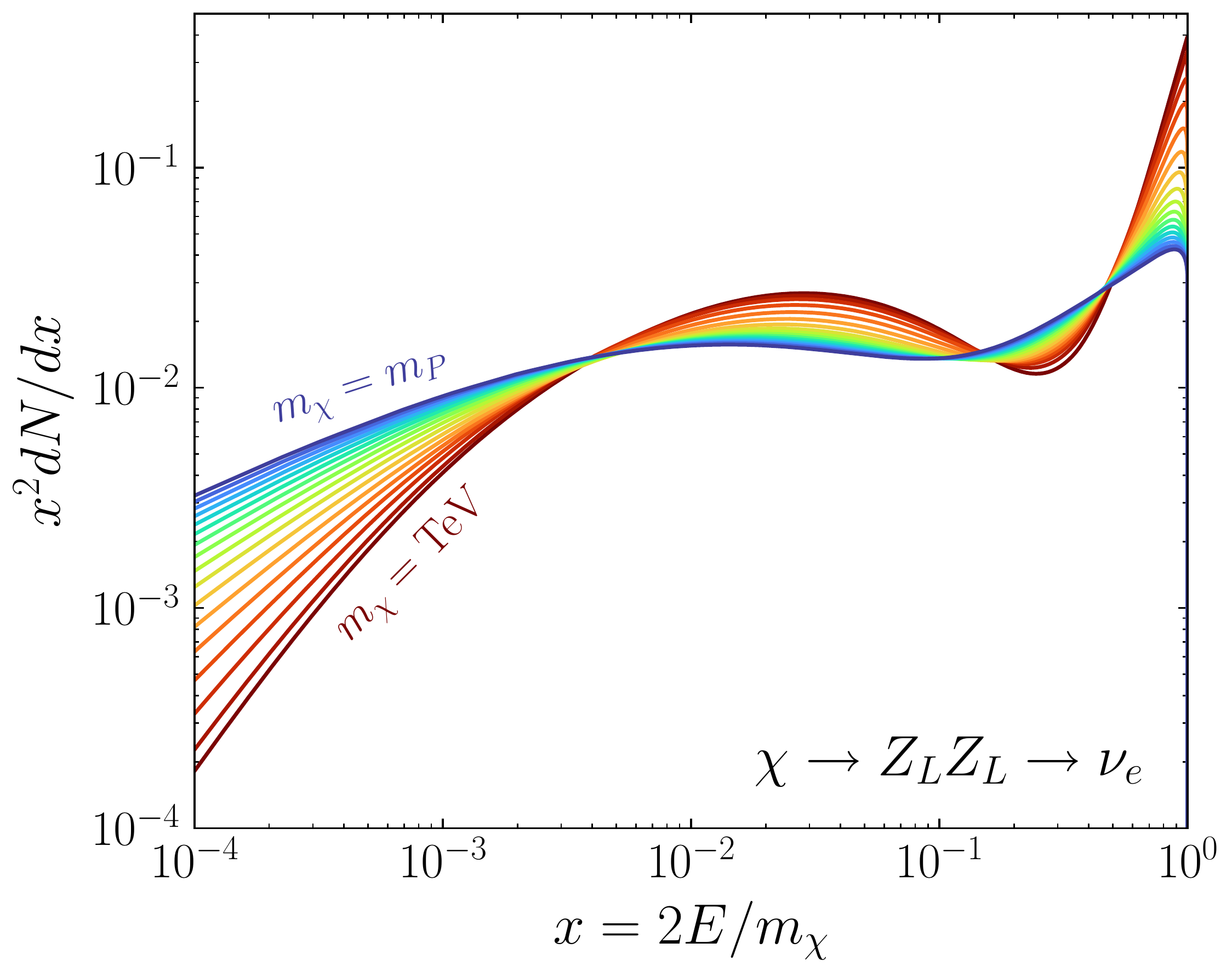}
\hspace{0.5cm}
\includegraphics[width=.47\textwidth]{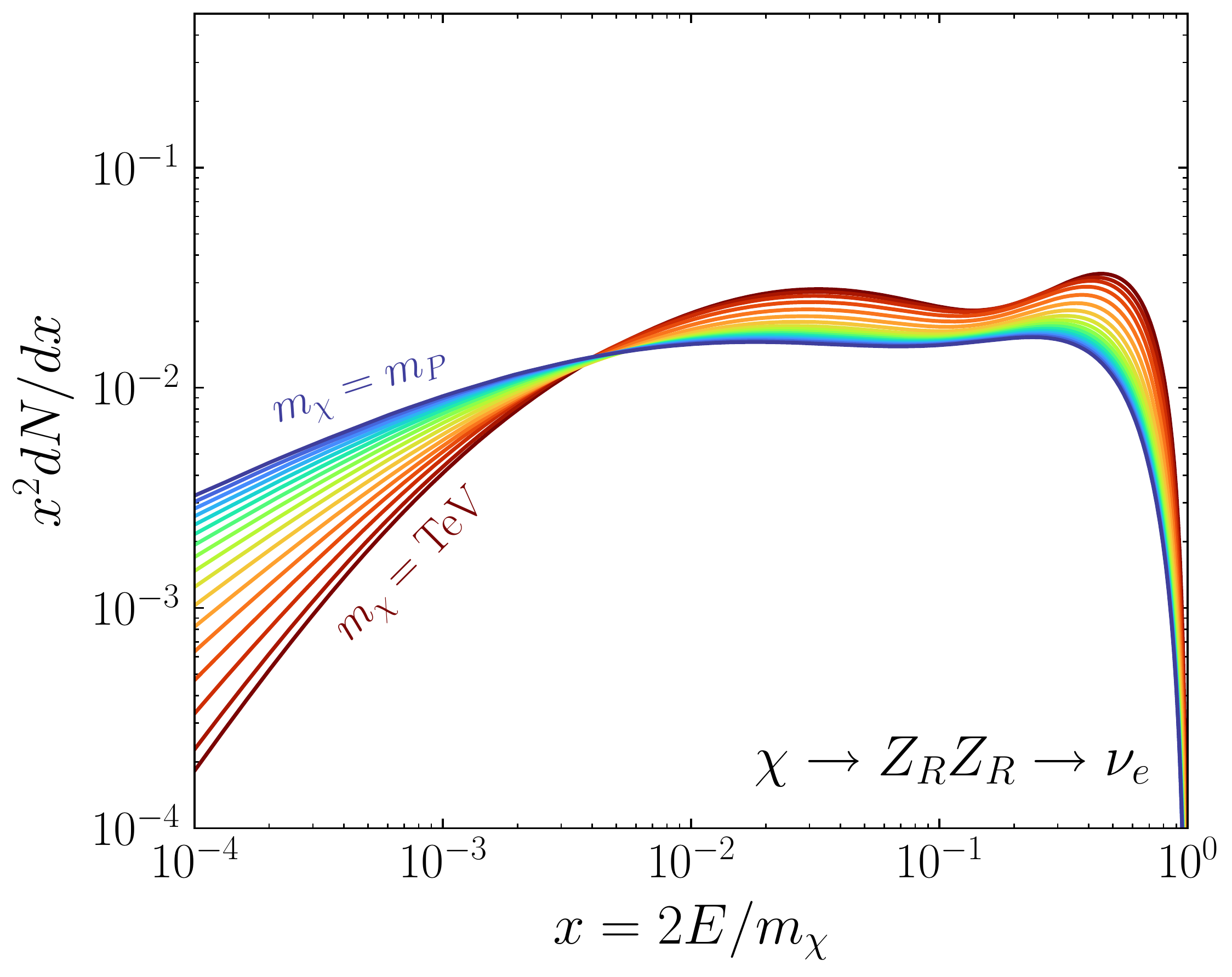} \vspace{0.5cm}\\
\includegraphics[width=.47\textwidth]{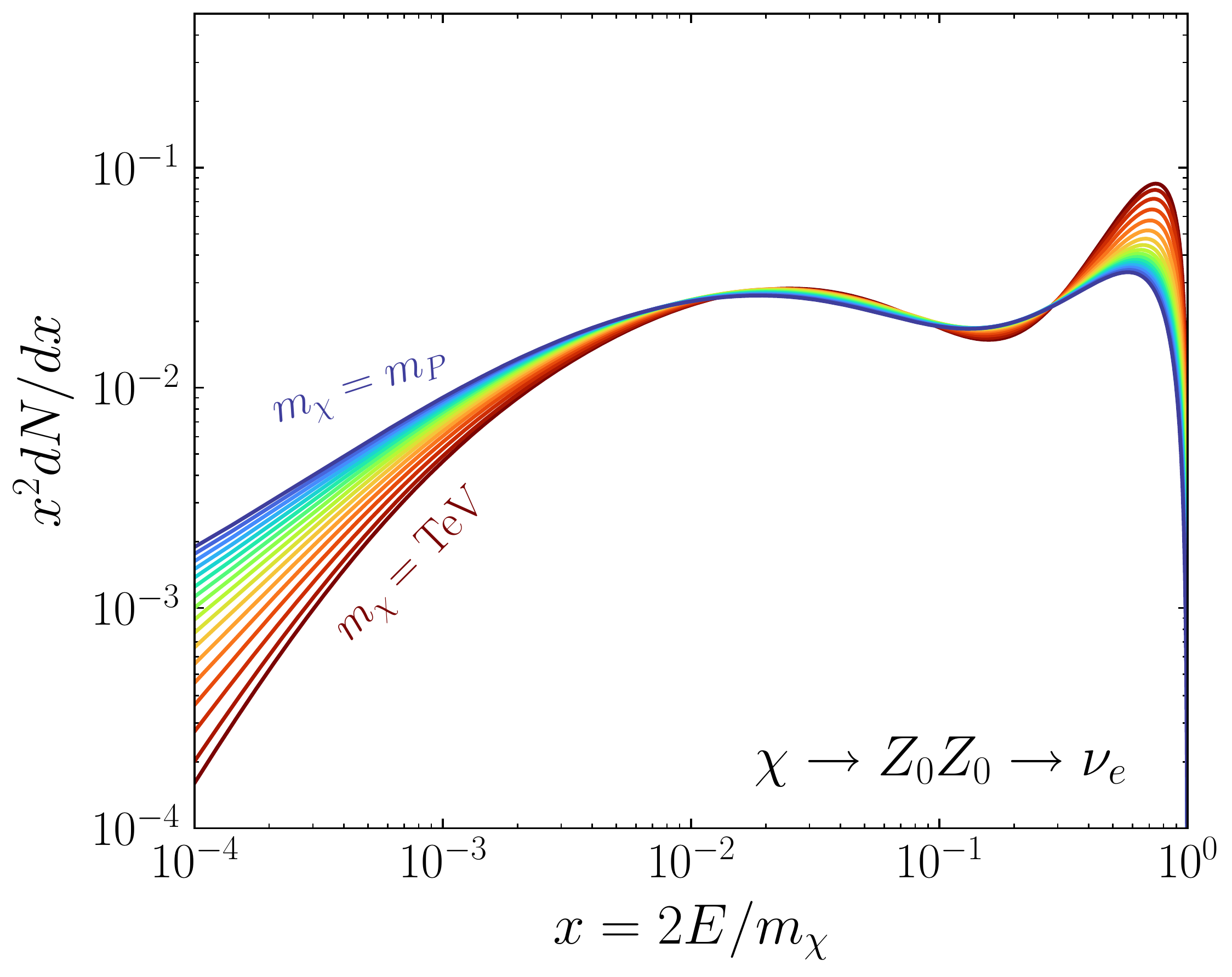}
\end{center}
\vspace{-0.5cm}
\caption{A further analog of \Fig{fig:ballM}, here for the electron neutrino spectrum from DM decay to $Z$-bosons.
In this case, instead of averaging over the three available initial polarizations, we depict each polarization separately: left (top left) and right (top right) handed initial states (top left), and longitudinally polarized bosons (bottom).
Once more, a separate curve is shown for each decade in DM mass between a TeV and the Planck scale.
}
\label{fig:ExtraSpecPolarized}
\end{figure}

Figure~\ref{fig:ExtraSpec} and \Fig{fig:ExtraSpecPolarized} furnish additional examples of the output of the formalism introduced in this paper.
In all cases, we show the spectrum of stable SM final states for DM masses between 1 TeV and $m_P \sim 10^{19}$ GeV, with a spectrum shown at each decade in mass.

Quite generically, we see that when plotted in dimensionless variables, the spectra vary most rapidly for scales around the TeV scale, often slowing as the mass approaches the Planck scale.
Just above the electroweak scale, new channels become kinematically available through the emission of states with $m \sim \qW$, and as the evolution heads into the full unbroken SM.
This is clear in the case of the muon neutrino spectrum from $\chi \to e^+ e^-$ shown in \Fig{fig:ExtraSpec}, which can only arise from electroweak boson or hadronic decays, both of which are primarily accessed through electroweak states.
This can be contrasted with the positron spectrum resulting from $\chi \to HH$.
The rich decay pattern of the SM Higgs already involves many SM states, explaining the lack of significant evolution in the spectrum.
Nevertheless, the hardest emissions near $x=1$ do evolve considerably, growing rapidly as multiple electroweak emissions become available, and then softening again as the size of electroweak showers develops.
In all cases, the evolution eventually slows down as the various channels become well mixed.
This is perhaps unsurprising, given the connections between the DGLAP equation and diffusion.
We reiterate that our calculation assumes that there are no new-physics thresholds crossed between $\qW$ and $m_P$.
If in fact there were, rapid variations could again be observed as momenta is redistributed amongst the newly available channels.
For results in this direction, see e.g.~\cite{Berezinsky:2000up,Berezinsky:2002hq,Aloisio:2003xj,Barbot:2002ep,Barbot:2002gt}.

As emphasized a number of times already, the chiral nature of the SM plays a central role in the high-scale evolution.
This point is further emphasized in \Fig{fig:ExtraSpecPolarized}, where we depict the electron neutrino spectrum resulting from the three polarizations of the massive $Z$-boson.
In general the softest emissions, dominated by QCD hadron decays, is comparable between all three states.
The hard emissions, however, differ dramatically, particularly when considering the spectrum of a purely chiral state.

\subsection{Confirming Momentum Sums}
\label{sec:MomentumSums}

Conservation of momentum implies that the momentum weighted FFs must satisfy the following consistency condition,
\be
\sum_b \int_0^1 dx\,d_a^b(x;\,Q,\mu_0) = 1\,.
\label{eq:MomSum}
\ee
Starting with $d_a^b(x;\,Q,\mu_Q) = \delta_a^b \delta(1-x)$, this equation is satisfied trivially.
However, it must also remain true as the momentum is repartitioned amongst different states through the evolution in virtuality, and the result becomes an important check on the evolution.
In this section we discuss how well our results satisfy \Eq{eq:MomSum}, taking various values of $Q$, and $\mu_0 \sim 0$ appropriate for the end of our evolution.

The results are provided in Table~\ref{tab:MomSum}.
The table shows a select set of values for $a$ and $Q$ in \Eq{eq:MomSum}.
In particular, it is clear that near the electroweak thresholds, the deviation from perfect momentum conservation is $\mathcal{O}(5-20\%)$, whereas by the highest scales considered in this work, they have shrunk to $\mathcal{O}(2-4\%)$.
These results represent an irrefutable uncertainty in our results.
Nevertheless, although there are a many steps in our calculation where momentum is redistributed, each of which could contribute to the errors shown in the table, the uncertainty is almost exclusively due to a single source: our procedure for incorporating soft coherence described in App.~\ref{sec:Smallx}.
Indeed, when we do not implement the corrections described in that section, we find momentum conservation is obeyed to better than $1\%$ in all cases, limited by the numerical precision used in our calculation.

Our procedure for implementing soft-coherence removes the real-radiation associated with destructive color-interference.
However, we do not account for the associated virtual corrections: the suppression of soft emission also increases the probability for a state to not emit and thereby retain a larger fraction of its momentum.
Accordingly, the presence of an offset in Table~\ref{tab:MomSum} is unsurprising, although also representative of a clear target for improving the treatment of color-coherence in our results.

\begin{table}[t]
\begin{center}
\begin{tabular}{|c|C{1.4cm}C{1.4cm}C{1.4cm}|}
  \hline
\backslashbox{$a$}{$Q$} & TeV & EeV &  $m_P$ \\
 \hline
$d_L$ & 0.890 & 0.932 & 0.963 \\
$d_R$ & 0.903 & 0.939 & 0.967 \\
$c_L$ & 0.890 & 0.932 & 0.963 \\
$c_R$ & 0.900 & 0.938 & 0.968 \\
$e_L$ & 0.953 & 0.973 & 0.979 \\
$e_R$ & 0.964 & 0.975 & 0.985 \\
$\nu_{\tau}$ & 0.981 & 0.977 & 0.978 \\
$g_L$ & 0.811 & 0.918 & 0.961 \\
$g_R$ & 0.811 & 0.918 & 0.961 \\
$Z_L$ & 0.961 & 0.964 & 0.981 \\
$Z_R$ & 0.961 & 0.964 & 0.981 \\
$Z_0$ & 0.981 & 0.975 & 0.977 \\
$H$ & 0.964 & 0.973 & 0.977 \\ \hline
\end{tabular}
\caption{Momentum sums determined from \Eq{eq:MomSum} for a representative subset of initial states and $Q$ values.
In all cases, disagreement from unity primarily results from our treatment of soft coherence, outlined in App.~\ref{sec:Smallx}.
}
\label{tab:MomSum}
\end{center}
\end{table}

\subsection{Momentum Distributions Amongst Final States}
\label{sec:MomDist}

\begin{figure}[t]
\leavevmode
\vspace{-0.2cm}
\begin{center}
\includegraphics[width=.47\textwidth]{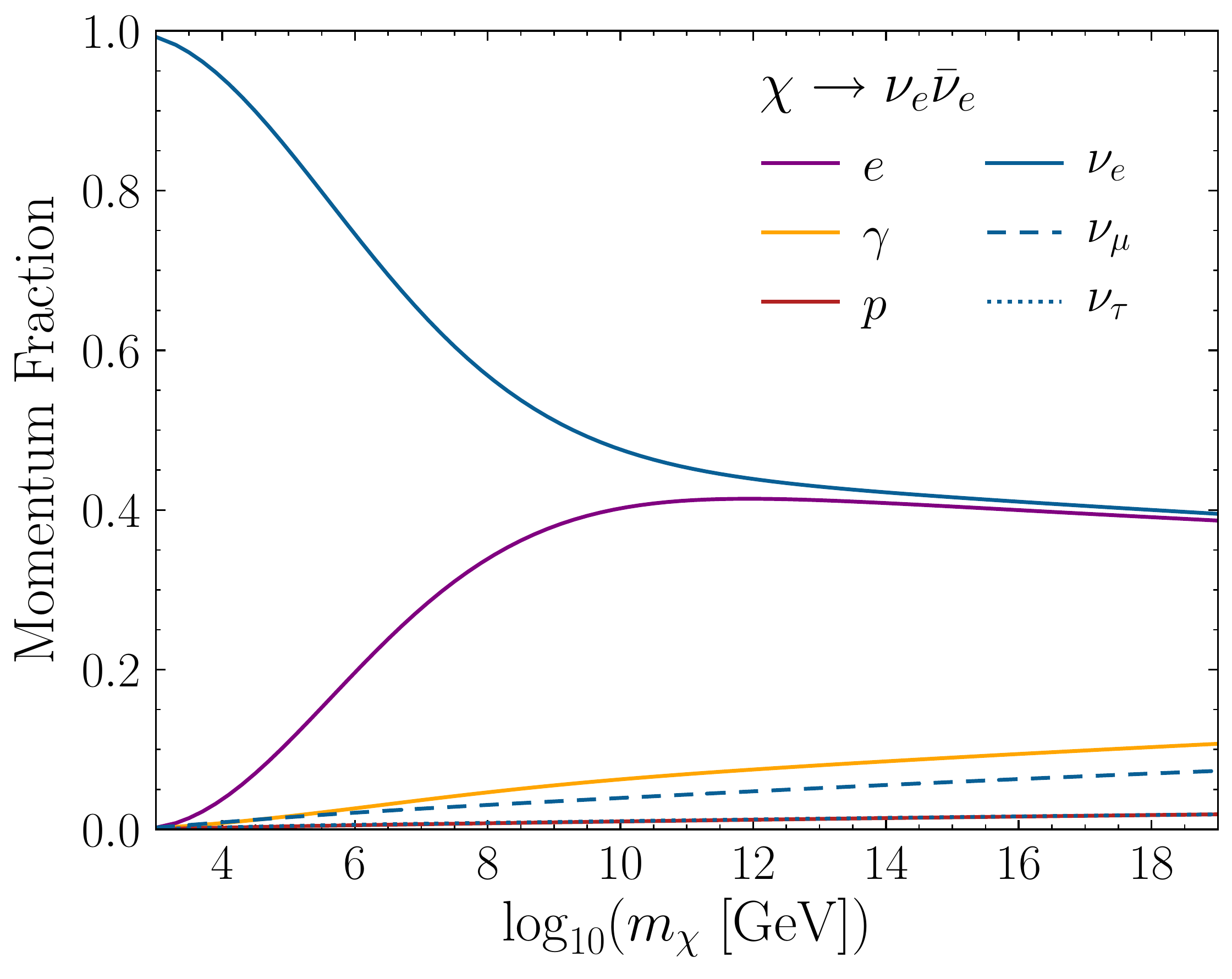}\hspace{0.5cm}
\includegraphics[width=.47\textwidth]{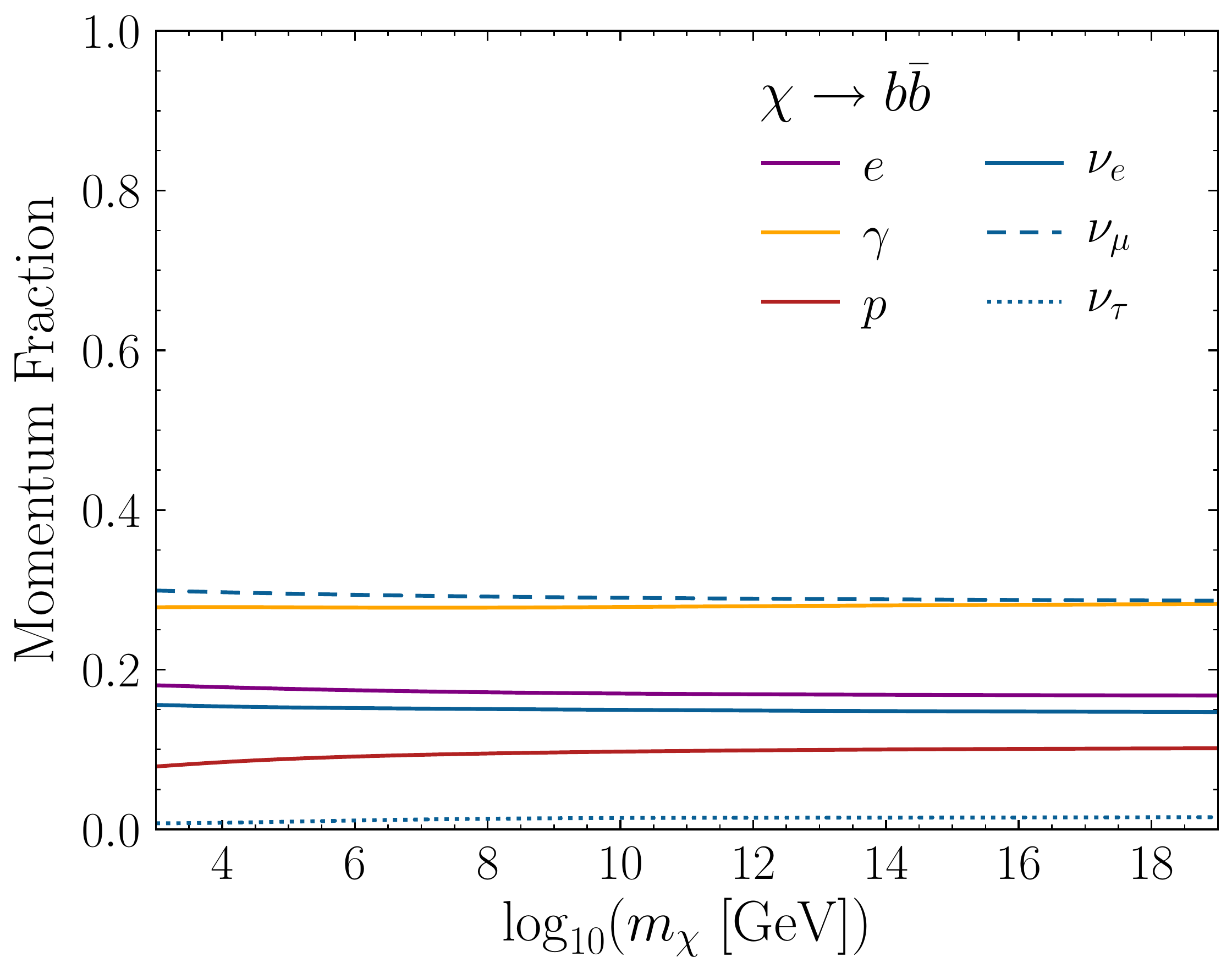}
\end{center}
\vspace{-0.5cm}
\caption{Momentum fraction carried by all stable SM final states considered in this work for two example decays, $\chi \to \nu_e \bar{\nu}_e$ (left) and $\chi \to b \bar{b}$ (right).
In both cases we show how the fractions evolve as a function of DM mass.
In the legend, particle labels are a proxy for the contribution of both particles and antiparticles, so $\nu_{\mu}$ labels the momentum fraction carried by both $\nu_{\mu}$ and $\bar{\nu}_{\mu}$.
}
\label{fig:MomFrac}
\end{figure}

In addition to checking the overall conservation of momentum, we can also consider how that momentum is redistributed amongst the stable SM final states.
In \Fig{fig:MomFrac} we show exactly that, plotting the momentum fractions for all final states, for two example spectra, $\chi \to \nu_e \bar{\nu}_e$ and $\chi \to b \bar{b}$.
In both cases, particles represent the momentum carried by the state and its conjugate (where applicable), so $p$ represents the momentum carried by $p+\bar{p}$.

For the neutrino initial state, we see considerable variation as a function of mass.
The difference between the $e$ and $\nu_e$ FFs, being an isovector quantity, evolves double logarithmically according to \Eq{eq:DeltaTq}, so that their fractions become almost equal at high enough masses.
At high enough masses, the fractions are identical for $e$ and $\nu_e$, as a result of their connection in the unbroken SM.
For the representative hadronic channel, the distributions are highly stable as a function of mass.
Of course, even if the total momentum fraction deposited into a given SM state is constant, how that momentum is divided between individual states can still evolve considerably, as demonstrated in \Fig{fig:ballM}. 
The tight correlation between the photon and muon neutrino can be understood as follows.
Strong isospin implies charged and neutral pions will be produced in these decays at a ratio of two to one.
The decays of $\pi^0$ will produce two hard photons, each carrying a large momentum fraction, whereas the $\pi^{\pm}$ decays will produce only a single hard muon neutrino (and a softer one which will not generally carry a large momentum fraction).

\section{Details of the Public Code}
\label{sec:PubCode}

\begin{figure}[t]
\leavevmode
\vspace{-0.2cm}
\begin{center}
\includegraphics[width=.47\textwidth]{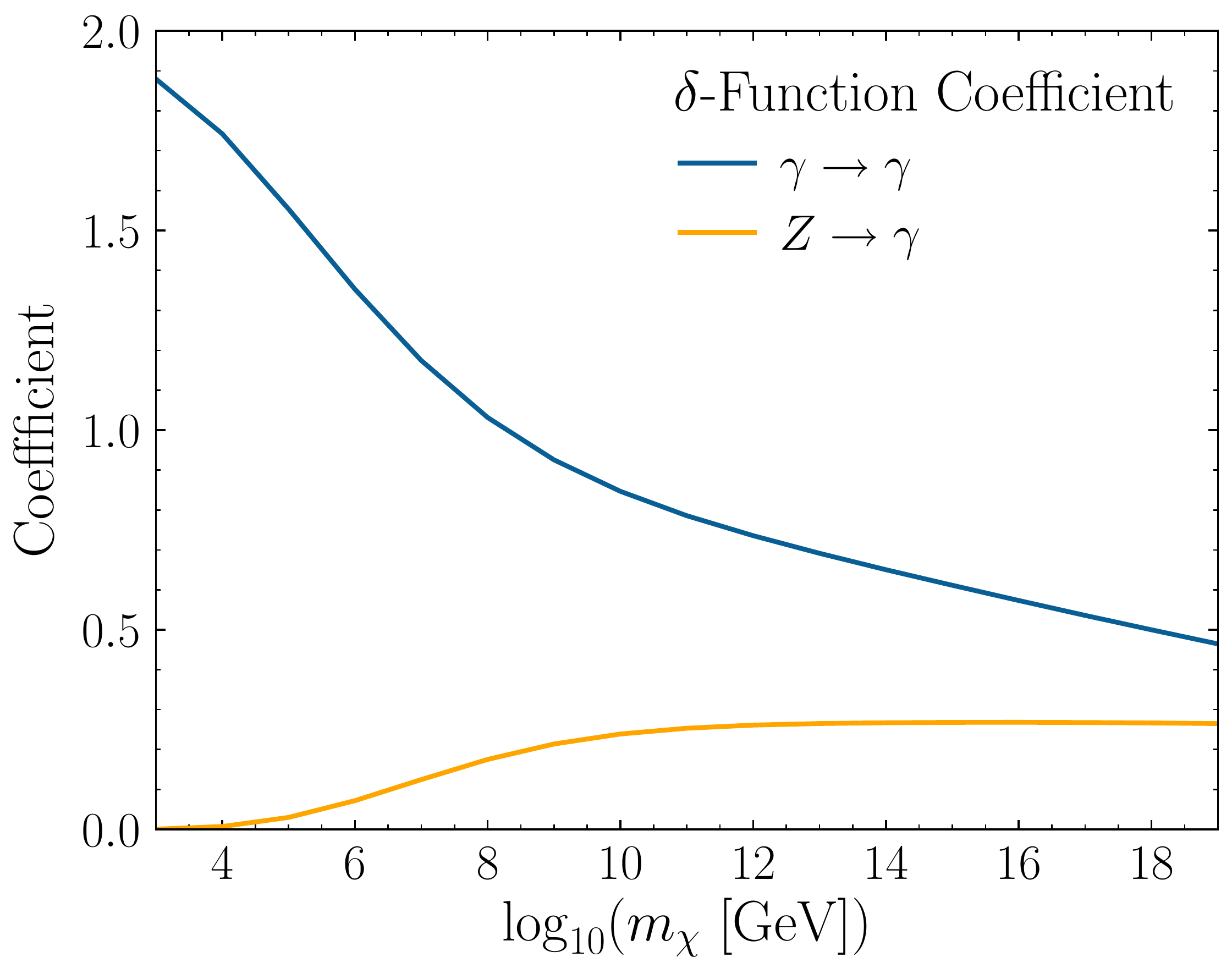}
\end{center}
\vspace{-0.5cm}
\caption{The coefficient of $\delta(1-x)$ associated with $d_i^{\gamma}(x;\,m_{\chi}/2,0)$  or $i=\gamma$ and $Z$.
More generally, a number of $\delta$-function coefficients can be accessed in the public code, in addition to the continuum results.
}
\label{fig:DeltaCoefficient}
\end{figure}

The spectra generated in this work are publicly available at \href{https://github.com/nickrodd/HDMSpectra}{github.com/nickrodd/HDMSpectra}.
Examples of how to generate spectra for arbitrary initial states or even individual FFs is provided.
For several cases, the code can also be used to extract the coefficient of $\delta(1-x)$ in the spectrum, an example output is shown in \Fig{fig:DeltaCoefficient}.
Further, the repository contains the details of how to reproduce many of the figures in this work.
In this section we outline several additional details of how those results were computed, but for details of how to use them we refer to the repository.
We emphasize once more that all spectra provided in the repository are prompt: no propagation effects are included.

With the exception of the weak matching, all details of our calculation are computed numerically.
For the high scale evolution, we solve the DGLAP equations using the procedure outlined in~\cite{Bauer:2018xag,Bauer:2018arx}.
We note that in this stage of the calculation we made use of \texttt{LHAPDF}~\cite{Buckley:2014ana}.
At the low scale, we evolve our results using \texttt{Pythia}.
In each case, we determine the FFs $d(x)$ as $\ln(x)$ spaced histograms, and then perform the convolution using the approach in App.~\ref{sec:Convolution}.

At the end of the procedure, we have a collection of 616 FFs evaluated at a set of $Q$ values between 500 and $10^{19}$ GeV.
We then implement a reduction algorithm to reduce this to a minimal set of points necessary for retaining the details of the spectra at the level of accuracy of our calculation for $500\,  {\rm GeV} < Q < 10^{19}\,  {\rm GeV}$ and $10^{-6} < x < 1$. 
In this reduction of points we ensure that all spectra are unchanged to within 1\% in the region $10^{-4} < x < 0.99$, while we allow for larger deviations in regions where the precision of our calculation is expected to be worse.
To be precise, the accuracy as a function of $x$  we use is
\begin{align}
{\rm acc}(x) = \left\{
\begin{tabular}{ll}
$10^{-4} \sqrt{\frac{10^{-4}}{x}}$ & $x < 10^{-4}\,,$\\
$10^{-4}$ & $10^{-4} < x < 0.99\,,$\\
$10^{-4} \, \frac{0.99}{1-x}$ & $x > 0.99\,.$
\end{tabular}
\right.
\end{align}
We then discard as many points as possible, while maintaining this accuracy, in order to compress the output dataset.
This data is then packaged into a single file which is accessed via the public code.

\subsection{Computing the Convolution of Binned Fragmentation Function}
\label{sec:Convolution}

A central tool in the present work was the use of the convolution expression satisfied by FFs,
\be
d_a^c(x;\,\mu_1,\mu_3) = \sum_b \int_x^1 \frac{dz}{z}\,
d_a^b(z;\,\mu_1,\mu_2)\,
d_b^c(x/z;\,\mu_2,\mu_3)\,.
\label{eq:convolution}
\ee
This result was exploited to simplify our calculation into steps as shown in \Eq{eq:threesteps}, and appears frequently in our calculation of the DM spectra.
A key ingredient in the final result is the spectra obtained from \texttt{Pythia}, which are inherently binned.
It is convenient to have a form of \Eq{eq:convolution} appropriate for binned FFs.
Such a result is presented in this section.

Before doing so, let us briefly provide some intuition for \Eq{eq:convolution}.
The result quantifies that the probability a particle $a$ at a scale $\mu_1$ produces a particle $c$ at $\mu_3$ carrying momentum fraction $x$, is given by the combination of the probability of $a \to b$ at an intermediate scale $\mu_2$, and then $b \to c$, summing over the allowed states and momenta for $b$.
One may worry about only keeping track of a single state across the threshold $\mu_2$, rather than all details of the evolution.
For example, if $b$ is quark or gluon, it will be color connected to other objects, and this information is lost across the threshold.
Here we invoke Amati-Veneziano preconfinement~\cite{Amati:1979fg}: as long as the various scales are sufficiently separated, what we color connect $b$ to is asymptotically irrelevant.
As our non-matched evolution always satisfy $\mu_1/\mu_2 \gg 1$, we are always in this regime.
For this reason, when simulating our \texttt{Pythia} results to compute the low scale FFs, we always initiate these at $\sqrt{s} = 2 \qW$ for states $X \bar{X}$, with the $X$ and $\bar{X}$ color connected to form a singlet.
There will be corrections to this picture at higher order, although this is sufficient for the level of accuracy of the present work.

In practice we evaluate the FFs over a wide dynamic range, and thus it is convenient to change variables to $l_x = \ln x$, yielding
\be
d_a^c(l_x;\,\mu_1,\mu_3) = \sum_b \int_{l_x}^0 \frac{dz}{z}\,
d_a^b(l_x-l;\,\mu_1,\mu_2)\,
d_b^c(l;\,\mu_2,\mu_3)\,.
\label{eq:convolutionlog}
\ee
Now, our FFs will be a logarithmically binned histogram with $N$ bins, and bin edges $l_1,\ldots,l_{N+1}$, where $l_{N+1} = 0$. Then we have
\bea
d_a^b(l;\,\mu_1,\mu_2) &= \sum_{n=1}^N \tilde{d}^b_{a,n} \Theta(l_{n+1}-l)\Theta(l-l_n)\,, \\
d_b^c(l;\,\mu_2,\mu_3) &= \sum_{n=1}^N d^c_{b,n} \Theta(l_{n+1}-l)\Theta(l-l_n)\,,
\eea
i.e. the high-scale values are described by $\tilde{d}$, and the low scale by $d$.

To determine the convolution for a given $l_x$, it is convenient to define three additional quantities. 
The first is simply the bin width $l_{n+1}-l_n = \Delta$.
The next two specify the location of $l_x$.
We introduce an integer $m$ defining the bin $l_x$ falls in, in detail $l_m < l_x < l_{m+1}$.
From here, the fractional bin width to the point $l_x$, is defined as $\delta = (l_x-l_m)/\Delta \in [0,1]$.
In terms of these auxiliary quantities, we can then evaluate \Eq{eq:convolutionlog},
\bea
d_a^c(l_x;\, \mu_1, \mu_3) 
= \Delta \sum_b &\left[ d^c_{b,m} \tilde{d}^b_{a,N} (1-\delta) \vphantom{\sum_{n=1}^{N-m}} \right. \\
&\left.+ \sum_{n=1}^{N-m} d^c_{b,n+m} \left( \tilde{d}^b_{a,N+1-n} \delta + \tilde{d}^b_{a,N-n} (1-\delta) \right)\right]\,,
\eea
which can be readily evaluated numerically.
Note if $m=N$ the summation within square brackets above expression vanishes.

As a simple check, imagine $\mu_2=\mu_1$, so that $\tilde{d}^b_{a,n} = \delta_a^b \delta_n^N$.
Then the above becomes,
\bea
d_a^c(l_x;\, \mu_1, \mu_3) 
&= \Delta \left[ d^c_{a,m} (1-\delta)
+ d^c_{a,m+1} \delta \right]\,,
\eea
which is an appropriately weighted sum.

\addcontentsline{toc}{section}{\protect\numberline{}References}%
\bibliography{HDMSpectra}
\bibliographystyle{JHEP}

\end{document}